\documentclass[twocolumn,showpacs,preprintnumbers,pra,aps,superscriptaddress,longbibliography,footinbib]{revtex4-1}

\usepackage{graphicx}
\usepackage{multirow}
\usepackage{amsmath}
\usepackage{mathtools}
\usepackage{amssymb}
\usepackage{amsthm}
\usepackage{soul}
\usepackage{verbatim}
\usepackage{bm}
\usepackage[colorlinks=true, linkcolor=blue, citecolor=gray]{hyperref}
\usepackage[table]{xcolor}
\usepackage{tikz}
\usepackage{xspace}
\usepackage{import}
\usepackage[caption=false]{subfig}
\usepackage{bbold}

\makeatletter
\newcommand\footnoteref[1]{\protected@xdef\@thefnmark{\ref{#1}}\@footnotemark}
\makeatother

\newcommand{\Hh}{\mathcal{H}}
\newcommand{\RR}{\mathbb{R}}
\newcommand{\PP}{\mathbb{P}}
\newcommand{\Nn}{\mathcal{N}}
\newcommand{\Tt}{\mathcal{T}}

\newcommand{\One}{\mathbb{1}}

\newcommand{\psis}{\psi_{\mathrm{s}}}
\newcommand{\Psis}{\Psi_{\mathrm{s}}}

\newcommand{\Psir}{\Psi_{\mathrm{r}}}
\newcommand{\rhoc}{\rho_{\mathrm{c}}}

\newtheoremstyle{break}
  {\topsep}{\topsep}%
  {\itshape}{}%
  {\bfseries}{}%
  {\newline}{}%
\theoremstyle{break}
\newtheorem{alg}{Algorithm}

\begin{document}

\title{Error mitigation via verified phase estimation}

\author{Thomas E.~O'Brien}
\email{corresponding author: teobrien@google.com}
\affiliation{Google Research, Venice, CA 90291, United States}
\affiliation{Instituut-Lorentz, Universiteit Leiden, 2300 RA Leiden, The Netherlands}
\author{Stefano Polla}
\affiliation{Instituut-Lorentz, Universiteit Leiden, 2300 RA Leiden, The Netherlands}
\author{Nicholas C.~Rubin}
\affiliation{Google Research, Venice, CA 90291, United States}
\author{William J.~Huggins}
\affiliation{Google Research, Venice, CA 90291, United States}
\author{Sam McArdle}
\affiliation{Google Research, Venice, CA 90291, United States}
\affiliation{Department of Materials, University of Oxford, Parks Road, Oxford OX1 3PH, United Kingdom}
\author{Sergio Boixo}
\affiliation{Google Research, Venice, CA 90291, United States}
\author{Jarrod R.~McClean}
\affiliation{Google Research, Venice, CA 90291, United States}
\author{Ryan Babbush}
\email{corresponding author: babbush@google.com}
\affiliation{Google Research, Venice, CA 90291, United States}

\begin{abstract}
The accumulation of noise in quantum computers is the dominant issue stymieing the push of quantum algorithms beyond their classical counterparts.
We do not expect to be able to afford the overhead required for quantum error correction in the next decade, so in the meantime we must rely on low-cost, unscalable error mitigation techniques to bring quantum computing to its full potential.
This paper presents a new error mitigation technique based on quantum phase estimation that can also reduce errors in expectation value estimation (e.g., for variational algorithms).
The general idea is to apply phase estimation while effectively post-selecting for the system register to be in the starting state, which allows us to catch and discard errors which knock us away from there.
We refer to this technique as ``verified phase estimation'' (VPE) and show that it can be adapted to function without the use of control qubits in order to simplify the control circuitry for near-term implementations.
Using VPE, we demonstrate the estimation of expectation values on numerical simulations of intermediate scale quantum circuits with multiple orders of magnitude improvement over unmitigated estimation at near-term error rates (even after accounting for the additional complexity of phase estimation).
Our numerical results suggest that VPE can mitigate against any single errors that might occur; i.e., the error in the estimated expectation values often scale as ${\cal O}(p^2)$, where $p$ is the probability of an error occurring at any point in the circuit.
This property, combined with robustness to sampling noise reveal VPE as a practical technique for mitigating errors in near-term quantum experiments.
\end{abstract}

\maketitle

\section{Introduction}

Error mitigation is likely essential for near-term quantum computations to realize valuable applications.
State-of-the-art technology in superconducting qubits has recently pushed quantum computers beyond the capability of their classical counterparts~\cite{Google19Quantum} and enabled intermediate scale demonstrations of quantum algorithms for optimization~\cite{Google20Quantum,Pagano19Quantum}, quantum chemistry~\cite{Google20Hartree,Kandala17Hardware, Hempel18Quantum}, and machine learning~\cite{Otterbach17Unsupervised}, with tens of qubits and hundreds of quantum gates.
However, these experiments clearly reveal a noise barrier that needs to be overcome if such applications will ever scale to the classically intractable regime.
In the long-term, a path towards this goal is known through quantum error correction~\cite{Gottesman97Theory,Kitaev97Fault,Fowler12Surface}.
Yet, the requirements to successfully error correct large-scale quantum applications~\cite{Berry19Qubitization, Gidney19How,Burg20Quantum, Sanders20Compilation,Campbell18Applying} are still a few orders of magnitude above the current state-of-the-art, and will likely require many years to achieve.
In the meantime, quantum applications research has focused on finding the elusive beyond-classical NISQ (noisy, intermediate-scale quantum) application~\cite{Preskill18Quantum}, with the hope to accelerate the path to practical quantum computing.
However without the resources to correct errors, one must develop strategies to mitigate the aforementioned noise barrier.
Otherwise, the output of NISQ devices will be corrupted beyond usefulness for algorithms significantly more complex than those already attempted.

Much of the attention in the NISQ era has been directed towards variational algorithms, with applications in optimization~\cite{Farhi14Quantum}, chemistry and materials science~\cite{Mcclean16Theory}, and machine learning~\cite{Silva16Quantum,Chen18Universal}.
These shift much of the complexity of the algorithm to a classical outer loop involving many circuit repetitions, leaving the quantum computer with the task only of preparing quantum states and estimating expectation values of operators on said states.
However, preparation circuits need to have significant depth to avoid being classically simulated~\cite{Bravyi19Classical}.
Errors accumulated over this circuit quickly distort the prepared state to one different than was targeted.
This has meant that most quantum experiments to date have had difficulty achieving standard accuracy benchmarks prior to applying error mitigation techniques~\cite{Kandala17Hardware,Google20Hartree,Google20Quantum,Sagastizabal19Error,Hempel18Quantum}.
However, accuracy improvements of orders of magnitude have been achieved with error mitigation in these experiments, suggesting there may yet be hope for NISQ.

The zoo of error mitigation techniques is large and varied.
One may first attempt to design algorithms that are naturally noise-robust.
For example, the optimization procedure in a variational algorithm makes the algorithm robust against control errors (e.g. over- or under-rotations when gates are applied)~\cite{Mcclean16Theory}.
Also, subspace expansions of the variational quantum eigensolver (VQE) in materials science and chemistry correct errors that keep one within the desired subspace considered~\cite{Mcclean17Hybrid} or more generally through by approximate symmetry projection~\cite{mcclean2020decoding}.
Given the ability to artificially introduce additional noise into a device, one can extrapolate from multiple experiments at different noise levels to a hypothetical noiseless experiment~\cite{Temme16Error}, which has shown promising results on real devices~\cite{Kandala19Error}.
One may alternatively probabilistically compile circuits by inserting additional gates to average out or cancel out noise, given sufficient knowledge of the error model of the device~\cite{Temme16Error,Endo17Practical}.
When classically post-processing partial state tomography data from an experiment, one may attempt to regularize the obtained results using reduced density matrix constraints~\cite{Rubin18Application}.
Finally, one may mitigate errors that take a state outside of a symmetry-conserving subspace of a quantum problem, either by direct post-selection, or artificial projection of the estimated density matrix in post-processing, producing a `symmetry-verified' state~\cite{Mcardle19Error,Bonet18Low,mcclean2020decoding,Huggins19Efficient}.
Recent efforts have extended this protocol by introducing symmetries into problems to increase the range of errors that may be detected~\cite{Jiang19Majorana}, which is analogous to the way quantum error correcting codes introduce engineered symmetries.

Ideally, we would prefer to go beyond verifying that a system's state remains within a target subspace and instead directly verify that the system's state is the one we desire.
This would result in reaching the information theoretic optimal limit of post-selected error mitigation in which one could completely mitigate the effect of all errors by repeating the experiment a number of times scaling inversely with the circuit fidelity (equivalent to the ability to perfectly detect errors).
The fact that the circuit fidelity is expected to decrease exponentially in the gate complexity indicates that eventually we will still need error-correction; however, moving closer to this limit is certain to enable more powerful NISQ experiments.

In this work we develop a method for error mitigation of quantum phase estimation experiments, by verifying that the system returns to its initial state after the phase estimation step.
We show that the set of experiments that pass this condition contain all the necessary information to perform quantum phase estimation.
This yields a powerful error mitigation technique, as in most cases errors will not return the system to this initial state.
Our techniques apply to variants of phase estimation which might involve post-processing on a single control qubit~\cite{Obrien19Quantum,Somma19Quantum}, or when performing recently-developed control-free variants~\cite{Lu20Algorithms,Russo20Evaluating}.
We further develop it into a simple scheme for verified expectation value estimation by dividing a target Hamiltonian into a sum of fast-forwardable terms.
This yields a simple, low-cost scheme for the measurement of expectation values, which may be immediately incorporated into the quantum step of a variational quantum algorithm.
We study the mitigation power of this protocol in numerical simulations of small-scale experiments of free-fermion, transverse Ising, and electronic structure Hamiltonians.
Verification is observed to mitigate all single (and even all double) errors throughout many of these simulations, as evidenced by a clear second (or third)-order sensitivity in our results to the underlying gate error rate.
We observe in the best-case scenario case an up to $10,000$-fold suppression of error at physical error rates; this is not achieved for all systems studied, but verification is found to improve experimental error in all simulations performed.
We find the error mitigation power to be highly system-, circuit-, and noise model-dependent.
Finally, we study the measurement cost of this protocol in the presence of sampling noise, finding that it is comparable to standard partial state tomography techniques for energy estimation.

The outline of this paper is as follows.
In Sec.~\ref{sec:outline}, we give a pedagogical example of how one might verify the estimation of expectation values of an arbitrary Hamiltonian, by writing it as a sum of Pauli operators and performing (fast-forwarded) verified phase estimation on each individual term.
In Sec.~\ref{sec:theory} we then derive the theory behind verified phase estimation itself, outline how it can mitigate errors, and give algorithms for performing verified phase estimation with a single control qubit, or with access to a reference state.
In Sec.~\ref{sec:VEVE}, we extend these ideas to give algorithms for verified expectation value estimation, and derive the conditions under which one may perform verified estimation of multiple expectation values in parallel (i.e. using the same system register).
In Sec.~\ref{sec:results}, we then implement these ideas, studying the mitigation power of verified expectation value estimation a variety of systems and implementations developed earlier in the text, on a set of small numerical systems under various noise models.

\section{Pedagogical example of verification protocol for expectation value estimation}\label{sec:outline}

\begin{figure*}
    \centering
    \includegraphics[width=\textwidth]{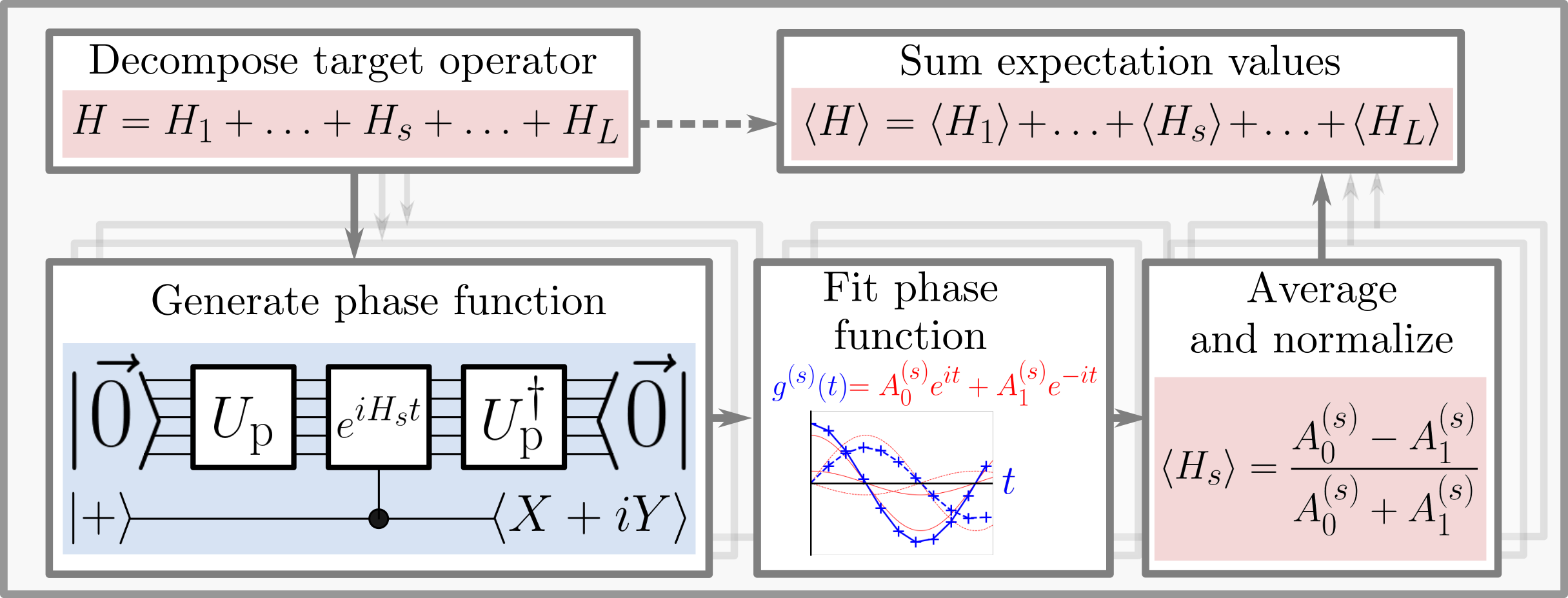}
    \caption{Process diagram of the protocol for verified estimation of the expectation value of a Hamiltonian on a state $|\psi\rangle=U_{\mathrm{p}}|\vec{0}\rangle$. Blue denotes circuits to be executed or data to be extracted from a quantum computer, red denotes signal details to be estimated via classical post-processing. The protocol proceeds as follows: (top-left) a complex Hamiltonian $H$ is split into a number of fast-forwardable summands $H_s$. The spectral function $g(t)$ of $|\Psi\rangle$ under time evolution of each piece is obtained (bottom-left) via verified, fast-forwarded phase estimation. In this example, a control qubit is used to extract the phase function via phase kickback. The resulting data is a weighted sum of oscillations with frequencies equal to the eigenvalues $E_j^{(s)}$ of the corresponding factor (bottom middle). This may be decomposed in a variety of classical post processing techniques to obtain approximations for the expectation values $\langle H_s\rangle$ depending on the type of $H_s$ chosen (bottom-right). Regardless of the method used, the expectation values must be normalized to obey Eq.~\ref{eq:normalization}, the last step in the verification process. As the expectation value is linear, the verified estimates of $\langle H_s\rangle$ obtained may be immediately summed together to give a verified estimate for $\langle H\rangle$ (top-right).}
    \label{fig:process_diagram}
\end{figure*}

In this section we outline a simple implementation of verified expectation value estimation of a target operator $H$ on a state $|\psi\rangle$, as a practical example of the more complicated methods to be found later in the text.
The idea behind all verification protocols is to prepare $|\psi\rangle=U_{\mathrm{p}}|0\rangle$, indirectly estimate $\langle H\rangle$ via phase estimation, and then verify that we remain in $|\psi\rangle$ by uncomputing $|0\rangle=U^{\dag}_\mathrm{p}|\psi\rangle$ and measuring in the computational basis.
If $|\psi\rangle$ is not an eigenstate of $H$, the system may by shifted away from this state by the QPE unitary --- i.e. even in the absence of error we do not expect the system to always pass verification.
However, as we will show later in this work, the data required for phase estimation is contained entirely within the set of experiments that pass verification; we may effectively ignore any experiments that fail.
This in turn allows us to ignore any errors that knock the system away from $|\psi\rangle$, making this a potent error mitigation scheme.
We have constructed various implementations of this idea, which we will expand on in Sec.~\ref{sec:theory} and Sec.~\ref{sec:VEVE}, and compare in Sec.~\ref{sec:results}.
However, the most general protocols require relatively complicated circuits and classical post-processing.
For clarity of exposition, in this section we focus on stepping through a simple protocol for the verification of expectation values, which avoids complex signal processing and circuity requirements.
The protocol we describe will work for arbitrary $H$ and $|\psi\rangle$, and may often be a desirable choice for a real experiment.
However, depending on the choice of $H$ and $|\psi\rangle$ and the noise model, other protocols described later in the text may be more optimal in terms of their mitigation power.

A process diagram for a simplified verified phase estimation protocol is given in Fig.~\ref{fig:process_diagram}.
To begin, we write $H$ as a sum of fast-forwardable terms
\begin{equation}
H=\sum_sH_s.
\end{equation}
Here, by fast-forwardable, we mean that each $H_s$ is chosen such that time evolution $e^{iH_st}$ may be implemented on a quantum register with gate complexity that is ${\cal O}(1)$ with respect to $t$.
Although fast-forwarding is forbidden for arbitrary $H$~\cite{Berry07Efficient}, decomposition of any sparse, row-computable $H$ into a linear combination of polynomially many fast-forwardable Hamiltonians is always possible \cite{Aharonov03Adiabatic}.
For example, the $N$-qubit Pauli operators $P_i\in \PP^N=\{\One,X,Y,Z\}^{\otimes N}$ form a basis for the set of all $N$-qubit operators and are themselves fast-forwardable; we take this decomposition for our simple example.

We then implement verified phase estimation (with a single control qubit) to estimate the expectation values $\langle\psi|H_s|\psi\rangle$.
This involves evolving the system by $H_s$ conditional on a control qubit.
(Circuits to implement this are well-known, see e.g. Ref.~\cite{Whitfield11Simulation}.)
The conditional evolution encodes a phase function on the the control qubit.
That is, if we write $X_{\mathrm{c}}$ and $Y_{\mathrm{c}}$ for the $X$ and $Y$ Pauli operators on this control qubit, we have following the conditional evolution that
\begin{equation}
    \langle X_{\mathrm{c}}\rangle+i\langle Y_{\mathrm{c}}\rangle = A_0e^{it}+A_1e^{-it}=:g(t).\label{eq:phase_function_simple}
\end{equation}
Here, $A_0$ and $A_1$ are the squared amplitudes of $|\psi\rangle$ in the eigenbasis of $H_s$ (which has known eigenvalues $\pm 1$).
The expectation value $\langle X_{\mathrm{c}}\rangle$ may be estimated by measuring the control qubit $M$ times in the $x$-basis, counting the number of times $m_{x,0}$ or $m_{x,1}$ a $0$ or $1$ was seen, and approximating
\begin{equation}
    \langle X_{\mathrm{c}}\rangle \approx \frac{m_{x,0}-m_{x,1}}{M}.\label{eq:expectation_value}
\end{equation}
(A similar procedure may be performed for $Y$.)
To verify this estimate, we uncompute the preparation of the system, and count the number $m^{(v)}_{x,0}$ ($m^{(v)}_{x,1}$) of measurements of $0$ ($1$) on the control qubit when the uncomputed state on the system is returned to the initial $|0\rangle$ state.
We then replace our approximation by
\begin{equation}
    \langle X_{\mathrm{c}}\rangle \approx \frac{m^{(v)}_{x,0}-m^{(v)}_{x,1}}{M}.\label{eq:expectation_value_verified}
\end{equation}
(Note that we only replace the numerator, and not the denominator, of Eq.~\ref{eq:expectation_value}, which makes this not strictly post-selection.)
The expectation value $\langle H_s\rangle$ is encoded within the phase function $g(t)$, and must be inferred from these estimates above.
In our example protocol, this requires inferring the amplitudes $A_0$ and $A_1$ (as the eigenvalues $\pm 1$ are already known).
These may be simply estimated by a two-parameter fit of Eq.~\ref{eq:phase_function_simple} to the extracted values of $g(t)$.

As we show later in the text, in the absence of error Eq.~\ref{eq:expectation_value} and Eq.~\ref{eq:expectation_value_verified} yield the same result (in the large $M$ limit).
Errors tend to scatter the system into a state that fails verification.
The primary effect this has on the estimator in Eq.~\ref{eq:expectation_value_verified} is to re-scale $g(t)\rightarrow p_{\mathrm{ne}}g(t)$ (where $p_{\mathrm{ne}}$ is the probability of no error occurring).
As our circuit is fast-forwarded, under reasonable noise assumptions $p_{\mathrm{ne}}$ is independent of $t$, and this propagates immediately through the fit of Eq.~\ref{eq:phase_function_simple}: $A_0,A_1\rightarrow p_{\mathrm{ne}}A_0, p_{\mathrm{ne}}A_1$.
To recover the noiseless values, we note that the normalization of $|\psi\rangle$ requires $A_0+A_1=1$, which we may enforce by estimating
\begin{equation}
    \langle H_s\rangle = \frac{A_0-A_1}{A_0+A_1}.
\end{equation}
Finally, as expectation values are linear, after repeating this procedure for all $H_s$ required, we may simply sum the result to obtain
\begin{equation}
    \langle H\rangle = \sum_s\langle H_s\rangle.\label{eq:expectation_value_sum}
\end{equation}
Note that each $H_s$ will have different values of $A_0$, $A_1$, and $g(t)$ (we have avoided explicitly labeling the above for simplicity).
In practice, the number of samples for estimation of each $\langle H_s\rangle$ should be varied to minimize the error in the final estimation of $\langle H\rangle$ (i.e. importance sampling).

\section{Schemes for verified phase estimation}\label{sec:theory}

\subsection{Review of single-control quantum phase estimation}

Quantum phase estimation (QPE) refers to a family of protocols to learn eigenphases $e^{i\phi_j}$ of a unitary operator $U$.
Equivalently, quantum phase estimation may be used to learn eigenvalues $E_j$ of a Hermitian operator $H$, as each such operator generates a unitary via exponentiation: $U=e^{iHt}$~\cite{Nielsen00Quantum}.
(Such estimation requires limiting the size of $t$ to prevent aliasing - $e^{iE_jt}=e^{iE'_jt}$ if $E_jt=E'_jt+2n\pi$, which makes estimation ambiguous.)
The eigenvalues of $H$ and the eigenphases of $U$ are related by the same exponentiation and correspond to the same eigenstates $|E_j\rangle$ --- if $H|E_j\rangle=E_j|E_j\rangle$, $U|E_j\rangle=e^{i\phi_j}|E_j\rangle$, and $\phi_j=E_jt$.

In the single-control variant of QPE, the phases $\phi_j$ are learnt by imprinting them on a control qubit --- a process known as phase kickback.
Any unitary $U$ may be implemented as a (perhaps approximate) quantum circuit on a quantum `system' register, but quantum mechanics tells us that $e^{i\phi}|\psi\rangle\equiv|\psi\rangle$ for all pure states $|\psi\rangle$ and numbers $\phi\in\RR$.
This implies that if the system register were prepared in the pure state $|E_j\rangle$ and $U$ applied, we would not be able to infer the phase $\phi_j$ from the resulting state $e^{i\phi_j}|E_j\rangle\equiv |E_j\rangle$.
However, a relative phase $\phi$ between two states, $\frac{1}{\sqrt{2}}(|\psi_1\rangle+e^{i\phi}|\psi_2\rangle)$, is a physical observable that may be detected.
Such detection may be achieved by acting the unitary $U$ conditional on the control qubit being in the state $|1\rangle$ (and doing nothing when the control qubit is in the state $|0\rangle$).
This is commonly written as the `controlled' unitary $\mathrm{C}-U$.
When $\mathrm{C}-U$ acts on a system register prepared in an eigenstate $|E_j\rangle$ and a control qubit prepared in the state $(|0\rangle+|1\rangle)/\sqrt{2}$, the global state evolves to
\begin{equation}
\mathrm{C}-U\frac{1}{\sqrt{2}}\left(|0\rangle+|1\rangle\right)|E_j\rangle = \frac{1}{\sqrt{2}}(|0\rangle + e^{i\phi_j}|1\rangle)|E_j\rangle.
\end{equation}
We see that the eigenphase $e^{i\phi_j}$ from the system register is kicked back onto the control qubit, while the system register itself remains unchanged.
We may estimate this eigenphase $e^{i\phi_j}$ by repeatedly performing the QPE protocol, measuring the control qubit in the $X$ or the $Y$ basis, and recording the number of single-shot readouts of $1$ and $0$.
In the Hamiltonian case, from this estimate one may immediately infer $\frac{1}{it}\mathrm{Arg}(e^{i\phi_j})=E_j \mod 2\pi t$.
The error in the estimation of $E_j$ decreases with $t$; asymptotically optimal protocols need to balance this against the ambiguity modulo $2\pi t$ by repeating the estimation at multiple values of $t$~\cite{Higgins09Demonstrating,Wiebe16Efficient,Svore13Faster}.
In terms of estimating the eigenphases $e^{i\phi_j}$ of a unitary $U$, this optimization requires repeating the above procedure for $\mathrm{C}-U^k$ at varying points $k$.

Often, one does not prepare an eigenstate $|E_j\rangle$, but instead prepares a starting state 
\begin{equation}
|\psis\rangle=\sum_ja_j|E_j\rangle.
\end{equation}
Applying $\mathrm{C}-U^k$ to such a state no longer leaves it unchanged, but instead entangles it with the control qubit.
This produces the combined state (on the system+control register)
\begin{align}
|\Psi(k)\rangle &= \mathrm{C}-U^k\frac{1}{\sqrt{2}}\left(|0\rangle+|1\rangle\right)|\psis\rangle\\&= \sum_j\frac{1}{\sqrt{2}}(|0\rangle + e^{ik\theta_j}|1\rangle)|E_j\rangle.\label{eq:psik_defn}
\end{align}
When one has instead performed controlled time evolution (via the unitary $\mathrm{C}-e^{iHt}$), one may instead write
\begin{align}
|\Psi(t)\rangle &= \mathrm{C}-e^{iHt}\frac{1}{\sqrt{2}}\left(|0\rangle+|1\rangle\right)|\psis\rangle\\&= \sum_j\frac{1}{\sqrt{2}}(|0\rangle + e^{iE_jt}|1\rangle)|E_j\rangle.
\end{align}
The sum over $j$ in the above equation looks problematic, but it turns out that the eigenphases $\theta_j$ (or eigenvalues $E_j$) remain encoded on the control qubit, in a sum weighted by the norm square $A_j:=|a_j|^2$ of the initial amplitudes $a_j$.
To be precise, one may trace over the system register to obtain the reduced density matrix of the control qubit
\begin{align}
\rhoc(t)&=\mathrm{Trace}_{\mathrm{sys}}\big[|\Psi(t)\rangle\langle\Psi(t)|\big]\label{eq:rhodef}\\
&=\frac{1}{2}\left(\begin{array}{cc}
1 & g(t) \\ g^*(t) & 1
\end{array}
\right),
\end{align}
with $g(t)$ the phase function of $|\psis\rangle$ under $H$
\begin{equation}
    g(t)=\sum_jA_je^{iE_jt}.
\end{equation}
Estimates of $g(t)$ may be obtained as an expectation value
\begin{align}
g(t)&=2 \, \mathrm{Trace}_{\mathrm{c}}\big[\rhoc(t)|0\rangle\langle 1|\big]\label{eq:gt_trace}\\
&=\mathrm{Trace}_{\mathrm{c}}\big[\rhoc(t) X\big]+i\mathrm{Trace}_{\mathrm{c}}\big[\rhoc(t) Y\big],
\end{align}
of the Pauli operators $X$ and $Y$.
Measuring these expectation values requires rotating the control qubit into the $x$- or $y$-basis, reading it out, and averaging the output over many repetitions (or shots) of the experiment.

For a unitary operator $U$ one may obtain an equivalent phase function
\begin{equation}
    g(k)=\sum_jA_je^{ik\phi_j},
\end{equation}
by estimating
\begin{align}
    g(k)&=2 \, \mathrm{Trace}_{\mathrm{c}}\big[\rhoc(k)|0\rangle\langle 1|\big]\\
    &=\mathrm{Trace}_{\mathrm{c}}\big[\rhoc(k) X\big]+i\mathrm{Trace}_{\mathrm{c}}\big[\rhoc(k) Y\big],\\
    \rhoc(k)&=\mathrm{Trace}_{\mathrm{sys}}\big[|\Psi(k)\rangle\langle\Psi(k)|\big],
\end{align}
with $|\Psi(k)\rangle$ defined in Eq.~\ref{eq:psik_defn}.
The tomography to extract these expectation values is the same as described in the previous paragraph.

Information about the eigenvalues $E_j$ and amplitudes $A_j=|a_j|^2$ may be inferred classically from estimates of $g(t)$ at multiple values of $t$.
When these are estimated sufficiently well, the expectation value of the Hamiltonian may be calculated
\begin{equation}
    \langle H\rangle = \sum_j A_jE_j\label{eq:expval_from_eval}.
\end{equation}
Inference of the amplitudes $A_j$ from $g(t)$ to error $\epsilon$ takes asymptotically time $\Omega(\epsilon^{-2})$ on a quantum device, even when the eigenvalues $E_j$ are already known~\footnote{This may be calculated via Cramer-Rao bounds as the derivative $\frac{\partial g(t)}{\partial A_j}$ is bounded as a function of $t$, which is not true for the derivative $\frac{\partial g(t)}{\partial E_j}$.}.
By propagating variances, this implies equivalent convergence in the estimation of expectation values via Eq.~\ref{eq:expval_from_eval}.
One need not resolve all $2^N$ eigenvalues of an $N$-qubit operator in order to evaluate Eq.~\ref{eq:expval_from_eval}.
Time-series analysis methods~\cite{Somma19Quantum} or integral methods~\cite{Roggero20Spectral} produce a coarse-grained approximation to the spectrum that may be averaged over to obtain expectation values with similar convergence rates.
Alternatively, for simple operators with highly-degenerate spectrum (e.g. Pauli operators), curve fitting will be sufficient to extract the required data (as described in Sec.~\ref{sec:outline})\footnote{The minimum number of points on the curve that require fitting is determined by the number of eigenvalues and amplitudes that need fitting.}.

\subsection{Verifying a phase estimation experiment}~\label{sec:single_control}
As the data from single-control quantum phase estimation is accumulated entirely on the control qubit, one would be tempted to throw the system register away (or rather, reset the register and begin anew).
In the absence of error correction this temptation grows larger; noise levels in near term devices are high enough that coherent states of more than a few qubits degrade over the course of any reasonably-sized algorithm to within a few percent fidelity to the target state --- if not less~\cite{Google20Hartree}.
However, even when corrupted, the information contained within the system register is valuable, as one can use this information to diagnose potential errors in the data to be read from the control qubit.
For instance, in the presence of global symmetries of the Hamiltonian, one could imagine mitigating errors that do not commute with this symmetry via symmetry verification~\cite{Bonet18Low,Mcardle19Error,Jiang19Majorana}.
In verifying these symmetries, we are in effect projecting the system into a subspace of the global Hilbert space which contains the information we desire.
One could imagine constructing ever-smaller Hilbert spaces, which trades circuit complexity for error-detection power.
It turns out that the limit of this construction is achievable: instead of measuring one or more symmetries on the system register, we can instead verify that it has returned to its initial state $|\psis\rangle$.
(This is similar to the echo-type measurements made in randomized benchmarking~\cite{Magesan11Scalable} or quantum Hamiltonian learning~\cite{Wiebe13Hamiltonian}.)

Assuming $|\psis\rangle$ is prepared from the computational basis state $|0\rangle$ by a preparation unitary $U_{\mathrm{p}}$, this measurement may be achieved by applying $U_{\mathrm{p}}^{\dag}$, and reading out each qubit in the computational basis.
One would expect such a measurement to distort the phase function $g(t)$, but this is not so, as we may expand the trace in Eq.~\ref{eq:rhodef} to show that
\begin{align}
&\mathrm{Trace}_{\mathrm{c}}\big[\rhoc(t)|0\rangle\langle 1|\big] \nonumber\\&\hspace{1cm}= \mathrm{Trace}_{\mathrm{c}}\Big[\langle\psis|\Psi(t)\rangle\langle\Psi(t)|\psis\rangle|0\rangle\langle 1|\Big].\label{eq:trace_equals_verification}
\end{align}
Here, the left-hand side of the equation is the expectation value of $\rhoc(t)$ regardless of the state of the control register, and the right-hand side is the (non-normalized) expectation value of $\rhoc(t)$ on verified experiments only.
The lack of normalization means this is not a post-selection technique; instead one assumes the contribution of un-verified states to the final estimation of $\tilde{g}(t)$ is zero.
(By contrast, verified states either contribute $+1$ or $-1$ to the estimation of $\tilde{g}(t)$.)

We can make a physical argument why Eq.~\ref{eq:trace_equals_verification} holds and verification should not affect the estimation of $g(t)$ in the absence of noise.
Let us decompose 
\begin{equation}
\rhoc=\rhoc^{(\mathrm{v})}+\rhoc^{(\mathrm{f})}\label{eq:rho_decomposition}
\end{equation}
into the ensemble of states $\rhoc^{(\mathrm{v})}$ that have passed verification, and those that have failed.
When the control qubit is in the $|0\rangle$ state, the system register is not evolved, so in the absence of noise the state will pass verification every time.
This implies that a verification failure in the absence of noise projects the control qubit into the $|1\rangle$ state; $\rhoc^{(\mathrm{f})}=|1\rangle\langle 1|$.
As $\mathrm{Trace}[|1\rangle\langle 1|0\rangle\langle 1|]=0$, this fraction of states on average contributes nothing to the estimate of $g(t)$.
In other words,
\begin{equation}
    \mathrm{Trace}[\rhoc|0\rangle\langle 1|]=\mathrm{Trace}[\rhoc^{(\mathrm{v})}|0\rangle\langle 1|]=g(t).
\end{equation}
Note that post-selecting (i.e. keeping only the experimental data where verification was passed) would instead prepare the state $\rhoc^{(\mathrm{v})}/\mathrm{Trace}[\rhoc^{(\mathrm{v})}]$.
As $\mathrm{Trace}[\rhoc^{(\mathrm{v})} (X+iY)] / \mathrm{Trace}[\rhoc^{(\mathrm{v})}]\neq g(t)$, this will not give the desired result.
Instead, an explicit protocol for the measurement of a single $g(t)$ within verified single-control phase estimation takes the following form
\begin{alg}[Single-control VPE]\label{alg:single-control}
\textbf{Input}:
\begin{itemize}
    \item Circuits to implement $U_{\mathrm{p}}, U_{\mathrm{p}}^{\dag}$ and controlled time evolution $e^{iHt}$.
    \item Number of repetitions $M$ of measurement in the $x$ and $y$ basis.
\end{itemize}
\textbf{Output}:
\begin{itemize}
    \item An estimate of $g(t)$ with variance $\frac{1}{M}$ and $\frac{1}{M}$ in the real and imaginary part respectively.
\end{itemize}
\textbf{Algorithm}:
\begin{enumerate}
    \item Prepare classical initial variables $g^x=0$, $g^y=0$.
    \item Prepare the system register in a starting state $|\psis\rangle=U_{\mathrm{p}}|0\rangle$ and the control qubit in the state $\frac{1}{\sqrt{2}}(|0\rangle+|1\rangle)$.
    \item Simulate time evolution $e^{iHt}$ conditional on the control qubit.
    \item Apply the inverse circuit $U^{\dag}_{\mathrm{p}}$ to the system register.
    \item Rotate the control qubit into the $X$ or $Y$ basis and measure it to obtain a number $m\in [0,1]$.
    \item If all qubits in the system register read $0$, increment the relevant variable $g^x$ or $g^y$ by $(-1)^{m}$.
    \item Repeat steps 2-6 $M$ times in the $X$ basis and $M$ times in the $Y$ basis, and estimate $g(t)$ by $\tilde{g}(t)=\frac{g^x}{M}+i\frac{g^y}{M}$.
\end{enumerate}
\end{alg}

\subsection{Why verification mitigates errors}\label{sec:mitigation}

The mitigation power from verification is based on the relative size of the Hilbert spaces in which the verified and unverified states of the entire system, $\rho = \rho^{(\mathrm{v})} + \rho^{(\mathrm{f})}$, live.
If we define the Hilbert spaces in which the two ensembles live $\Hh^{(\mathrm{v})}$ and $\Hh^{(\mathrm{f})}$ respectively, we have $\dim[\Hh^{(\mathrm{v})}]=2$, while $\dim[\Hh^{(\mathrm{f})}]=2^{N+1}-2$.
An error that occurs during the circuit is then likely to scatter the system into the set of rejected states.
As an extreme example, the probability that a completely random error (i.e. an error that scatters all states to a random state) at any point in the circuit will yield a state in $\Hh^{(\mathrm{v})}$ can be immediately calculated to be $2/(2^{N+1}-2)\sim 2^{-N}$.
This includes errors during preparation of $|\psis\rangle$ by the unitary $U_{\mathrm{p}}$ and the inversion of $U_{\mathrm{p}}^{\dag}$ to perform the verification itself.
As we are not post-selecting on the verification output $g(t)$ is still affected by this shift, but the distortion may be accounted for in classical post-processing.
In this simple noise model the effect of noise is then to replace the estimate of $g(t)$ by
\begin{equation}
    g_{\mathrm{noise}}(t) = p_{\mathrm{ne}}(t)g(t) + {\cal O}(2^{-N}p_{\mathrm{err}}(t)),\label{eq:noise_effect}
\end{equation}
where $p_{\mathrm{ne}}(t)$ and $p_{\mathrm{err}}(t)$ are the probabilities of no error or some error occurring, respectively.
The same occurs for any error that scatters the state outside $\Hh^{(\mathrm{v})}$, independently of both the type of error that occurs and when it occurs in the circuit.
(In App.~\ref{app:noise_modelling} we derive the specific requirements for this to be the case.)
Assuming that errors occur at a constant rate as a function of the circuit depth, and all scatter the system outside $\Hh^{(\mathrm{v})}$, for fast-forwardable Hamiltonians $p_{\mathrm{ne}}(t)=p_{\mathrm{ne}}$, and
\begin{equation}
    g_{\mathrm{noise}}(t)= p_{\mathrm{ne}}g(t)=\sum_j(p_{\mathrm{ne}}|a_j|^2)e^{iE_jt}.~\label{constant_err}
\end{equation}
This can be seen as a uniform damping of each squared amplitude $A_j$ to $A'_j=p_{\mathrm{ne}}A_j$.
Such damping may be corrected for classically as we know $|\psis\rangle$ is normalized
\begin{equation}
    \sum_jA_j=1\label{eq:normalization},
\end{equation}
and so we may estimate
\begin{equation}
    A_j=\frac{A'_j}{\sum_jA'_j}.\label{eq:renormalization}
\end{equation}
Depending on the classical signal processing method used, one may not obtain estimates of all $A'_j$ and $E_j$, but may instead directly calculate $\sum_jA'_jE_j$ and $\sum_jA'_j$.
In this case, these numbers may be directly substituted into Eq.~\ref{eq:renormalization_energy} to calculate $\langle H\rangle$.
For example, one could use $g_{\mathrm{noise}}(0)=\sum_jA'_j$ as such a reference point.
For non-fast-forwardable Hamiltonian, assuming again that errors occur at a constant rate throughout the circuit and that all scatter the system outside $\Hh^{(\mathrm{v})}$, we have
\begin{equation}
    g_{\mathrm{noise}}(t)=e^{-t/\tau_{\mathrm{err}}}g(t)=\sum_jA_je^{i(E_j+i/\tau)t}.
\end{equation}
This can be seen to be an imaginary shift to the eigenvalues $E_j\rightarrow E_j+i\tau$.
It can be corrected for in signal processing of the phase function by taking only the real parts of the $E_j$ eigenvalues.

The above analysis is not necessarily true for simulation of an arbitrary Hamiltonian under a realistic noise model.
In particular, if the instantaneous state during simulation is a near-eigenstate of the error model, then the correction in Eq.~\ref{eq:noise_effect} may be as large as ${\cal O}(1)$ instead of ${\cal O}(2^{-N})$.
In App.~\ref{app:noise_modelling} we study this in more detail, and specify the conditions under which errors will distort the results of verified phase estimation.

\subsubsection{Sampling costs}\label{sec:sampling_noise_analytics}

The error mitigation from verification comes at the cost of increasing the number of samples require to estimate $g(t)$.
Assuming all errors fall outside the verified subspace, estimating $g(t)$ to precision $\epsilon$ requires estimating $g_{\mathrm{noise}}(t)$ to precision $p_{\mathrm{ne}}\epsilon$.
To obtain $g^x$ in Alg.~\ref{alg:single-control} (and equivalently for $g^y$) we average over a set of $M$ experimnetal outputs that may take the values $\{-1,0,1\}$.
Let us define the $i$th experimental output $g^x_i$, and we have
\begin{align}
    P(g^x_i=\pm 1) = \frac{1}{2}p_{\mathrm{ne}}(1\pm g^x),\\
    P(g^x_i=0) = 1 - p_{\mathrm{ne}}.
\end{align}
Our estimate of the noisy $g_{\mathrm{noise}}(t)$ is then given by
\begin{equation}
    \mathrm{Re}[g_{\mathrm{noise}}(t)] = P(g^x_i=1) - P(g^x_i=-1).
\end{equation}
As each experiment is IID, the variance on our estimates of these probabilities is
\begin{align}
    \mathrm{Var}[P(g^x_i=\pm 1)] &= \frac{1}{M}\frac{1}{2}p_{\mathrm{ne}}(1\pm g^x)\nonumber\\&\times\left(1-\frac{1}{2}p_{\mathrm{ne}}(1\pm g^x)\right),\\
    \mathrm{Cov}[P(g^x_i=1),&P(g^x_i=-1)] \\&= -\frac{1}{4M}p^2_{\mathrm{ne}}\left(1- [g^x]^2\right).
\end{align}
Propagating variances obtains
\begin{equation}
    \mathrm{Var}\left[\mathrm{Re}\left[g_{\mathrm{noise}}(t)\right]\right]=\frac{1}{M}p_{\mathrm{ne}}-\frac{1}{M}p_{\mathrm{ne}}^2[g^x]^2.
\end{equation}
We may then bound the requirements to estimate $g_{\mathrm{noise}}(t)$ to variance $\epsilon^{-2}p_{\mathrm{ne}}^{-2}$ by
\begin{equation}
    M\geq \epsilon^{-2}p_{\mathrm{ne}}^{-1}.
\end{equation}
This is exactly what one would expect from an actual post-selection technique (i.e. where $Mp_{\mathrm{ne}}$ samples were used to estimate $g(t)$).
We remind the reader that $p_{\mathrm{ne}}$ here is the probability of no error occurring over the entire circuit.
As one should expect for an error mitigation technique, this in turn grows exponentially with the size of the circuit required to implement $e^{iHt}$ or $U_{\mathrm{p}}$.
In a simple model, if the error per qubit per moment is $p$ (i.e. assuming qubit decay is more dominant than gate noise in the model), an $N$-qubit circuit of depth $d$ would have
\begin{equation}
    p_{\mathrm{ne}} = (1-p)^{Nd},
\end{equation}
and thus the number of shots required to estimate (the real or imaginary part) of $g(t)$ would scale as
\begin{equation}
    M\sim (1-p)^{-Nd}\epsilon^{-2}.
\end{equation}
This is not to be ignored; verification requires at least doubling the size of the circuit, which if $p_{\mathrm{ne}}=0.01$ (as has been reported~\cite{Google19Quantum} and mitigated successfully~\cite{Google20Hartree} in previous experiments) will increase the measurement count by a factor of 100.
Some of the methods presented in this work involve increasing the circuit depth by factors of up to 14, which will be impractical for large experiments without further circuit optimization.

\subsubsection{Control noise}\label{sec:control_noise}
An important realistic error to consider in QPE is error on the control qubit.
This keeps the system within the verified subspace, and so is not captured by the above analysis.
However the effect of many common error channels may still be mitigated by verification.
For example, let us assume that the circuit decomposition of $\mathrm{C}-U$ involves the control qubit performing only single-qubit gates and controlled operations on the rest of the circuit (which is typically the case).
In this case, one may show that the effect of a depolarizing channel of strength $\lambda$
\begin{equation}
    R_{\mathrm{depol}}[\rho]=(1-\frac{3\lambda}{4})\rho +\frac{\lambda}{4}(X\rho X + Y\rho Y + Z\rho Z),
\end{equation}
acting on the control qubit at any point in the circuit, sends the final state of the system to
\begin{equation}
    (1-\lambda)\rho_{\mathrm{ne}} + \lambda\rho_{\mathrm{err}},
\end{equation}
where $\rho_{\mathrm{ne}}$ is the state in the absence of error, and
\begin{equation}
    \mathrm{Trace}[\langle\psi_s|\rho_{\mathrm{err}}|\psi_s\rangle |0\rangle\langle 1|]=0.
\end{equation}
In this case, the (noisy) estimate of $g(t)$ is sent to $(1-\lambda)g(t)$, and expectation values and eigenvalues may be recovered via the same analysis as in Sec.~\ref{sec:mitigation}.
However, the above analysis will not hold for a more general noise model, and schemes such as randomized compiling~\cite{Wallman16Noise} may be required to unbias the estimate of $g(t)$.
An example of this biasing effect is if an amplitude-damping channel
\begin{align}
    R_{\mathrm{ampdamp}}[\rho] =& (1-\lambda)\rho +\frac{\lambda}{2}(Z+I)\rho (Z+I)\nonumber\\&+ \frac{\lambda}{2}(X+iY)\rho(X-iY),
\end{align}
 is present on the control qubit between the final measurement pre-rotation and readout in the computational basis.
Left unchecked, this will shift the estimate of $g(t)$ to
\begin{equation}
    g_{\mathrm{err}}(t) = (1-\lambda)g(t) + \lambda.
\end{equation}
In addition to damping the true signal $g(t)$, this additive signal presents as a $0$-energy eigenvalue in the spectrum of $g(t)$.
This will not be accounted for by naive renormalization of $\langle H\rangle$ as outlined in Alg.~\ref{alg:VEVE}; the estimation protocol will instead estimate $(1-\lambda)\langle H\rangle$.
Though this could be corrected in post-processing, we suggest that a more stable mitigation is to flip the $|0\rangle$ and $|1\rangle$ states on the control qubit for $50\%$ of experiments.
This may be compiled into the final pre-rotation, and does not increase the total sampling cost of the experiment (only half as many samples need to be taken at each pre-rotation setting for the same accuracy).
We observe similar biases on bitflip noise channels which tend to decay the real and imaginary parts of $g(t)$ asymmetrically.
This may be compensated for in turn by compiling a $\frac{\pi}{4}$ $Z$-rotation on the intial control qubit state, and uncompiling it in the final prerotation.
(One can see that this commutes with all gates in the circuit).
For the noise models studied numerically in this text we have found either one or both of the above compilation schemes sufficient to mitigate control error.
More complicated noise models may required more complicated compilation schemes; extending the above will be an interesting task for future work.

\subsection{Verified control-free phase-estimation}\label{sec:control_free}
Making time evolution conditional on an control qubit does not increase the asymptotic cost of the circuit, but it does require additional overhead.
One might ask the question whether it is possible to perform the phase estimation without this additional overhead.
To investigate, let us reinsert the trace over the system register in Eq.~\ref{eq:trace_equals_verification} by inserting a resolution of the identity ($\langle\psis|=\sum_j\langle\psis|\psi_j\rangle\langle\psi_j|$) and rearranging
\begin{align}\label{eq:rearranged_gt}
    g(t) &= 2 \,\mathrm{Trace}_{\mathrm{sys}}\bigg[|\Psi(t)\rangle\langle\Psi(t)|\psis\rangle|0\rangle\langle 1|\psis| \bigg]\\
    &=\mathrm{Trace}_{\mathrm{sys}}\bigg[(\mathrm{C}-U)\Big(|0\rangle|\psis\rangle + |1\rangle\psis\rangle\Big)\\&\hspace{0.2cm}\times\Big(\langle\psis|\langle 0|+\langle\psis |\langle 1|\Big)(\mathrm{C}-U)^{\dag}|\psis\rangle|0\rangle\langle 1|\langle \psis|\bigg]\nonumber.
\end{align}
Then, let us define $|\Psis\rangle=|\psis|\rangle|1\rangle$ and $|\Psir\rangle=|\psis\rangle|0\rangle$.
Under this definition, Eq.~\ref{eq:rearranged_gt} is the off-diagonal element $|\Psis\rangle\langle\Psir|$ of a system prepared in the state $(|\Psis\rangle+|\Psir\rangle)/\sqrt{2}$ and evolved under the global Hamiltonian 
\begin{equation}
    \left(\begin{array}{cc}0 & 0 \\ 0 & H\end{array}\right)
\end{equation}
To obtain $g(t)$ we have relied on the fact that the `reference' state $|\Psir\rangle$ is a zero-energy eigenstate of the system.

\begin{figure}
    \centering
    \includegraphics[width=\columnwidth]{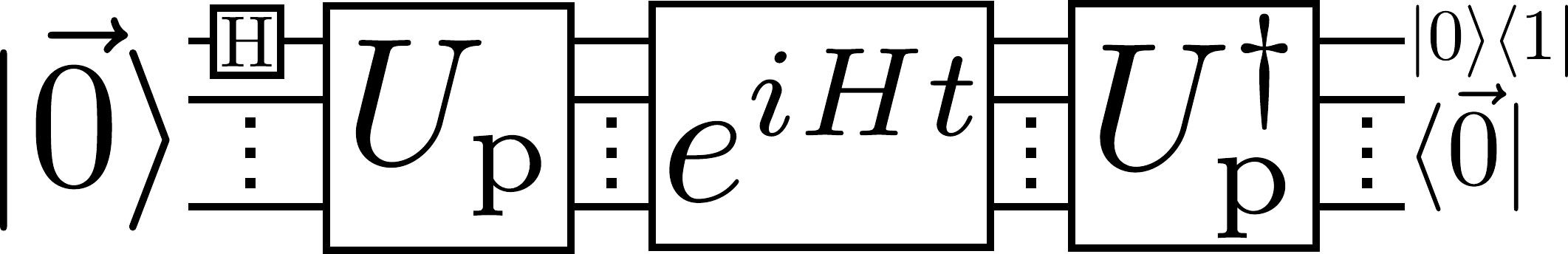}
    \caption{Quantum circuit for control-free verified phase estimation. The preparation unitary $U_{\mathrm{p}}$ is defined in Eq.~\ref{eq:prep_unitary}. The first gate in the circuit is a Hadamard gate (Roman H) on the top-most qubit (labeled the target qubit in text), which should not be confused with the Hamiltonian $H$.}
    \label{fig:control_free_PE}
\end{figure}

As was recently demonstrated in Ref.~\cite{Lu20Algorithms}, the control qubit may then be removed if we have the ability to prepare an alternative reference eigenstate $|\Psir\rangle$ of the Hamiltonian $H$.
The general case, where neither initial state is an eigenstate, was considered in Ref.~\cite{Zintchenko16Randomized} for the purposes of random gap estimation, but estimating single eigenvalues $E_j$ from the random functions generated is somewhat awkward.
This was also recently considered as an extension to the well-known robust QPE scheme~\cite{Kimmel15Robust}, requiring both $|\Psir\rangle$ and $|\Psis\rangle$ to be eigenstates of the system~\cite{Russo20Evaluating}.
For example, in the electronic structure problem in quantum chemistry the number-conserving Hamiltonian has the vacuum as a potential reference state.
Note that $|\Psir\rangle$ need not necessarily be a zero-energy eigenstate of $H$, though the corresponding eigenenergy $E_r$ should be known to high accuracy.
In this case, the control-free phase estimation circuit (Fig.~\ref{fig:control_free_PE}) provides an estimate
\begin{align}
    \mathrm{Trace}\Big[U(|\Psir\rangle+|\Psis\rangle)(\langle\Psir|+\langle\Psis|)U^{\dag}|\Psir\rangle\langle\Psis|\Big]\nonumber&\\
    \hspace{3cm}=e^{-iE_rt}g(t)&,
\end{align}
and the additional phase may be subtracted in post-processing.

The protocol for verified control-free phase estimation does not differ significantly from the single-control case.
Besides the loss of the control qubit and removal of control from the time evolution circuit, we also now require our preparation circuit to prepare the starting state $\frac{1}{\sqrt{2}}(|\Psis\rangle+|\Psir\rangle)$.
We assume that this is achieved by first applying a Hadamard gate to a single target qubit in the system register, placing the system in the state $\frac{1}{\sqrt{2}}(|0\rangle+|\vec{1}_{\mathrm{T}}\rangle)$.
(Here we use the notation $|\vec{1}_{\mathrm{T}}\rangle$ for the basis state where the target qubit is in the $|1\rangle$ state and all other qubits are in $|0\rangle$.)
Then, the desired preparation may be achieved by a preparation unitary $U_{\mathrm{p}}$ which performs the mapping
\begin{equation}
    U_{\mathrm{p}}|0\rangle \mapsto |\Psir\rangle,\qquad U_{\mathrm{p}}|\vec{1}_{\mathrm{T}}\rangle \mapsto |\Psis\rangle.\label{eq:prep_unitary}
\end{equation}
(We use the same notation as for the single-control unitary on purpose, as under the association $|0\rangle|\psis\rangle \leftrightarrow |\Psir\rangle$ and $|1\rangle\leftrightarrow |\Psis\rangle$ one may see the two are equivalent.)
With this definition, estimation of $|\Psir\rangle\langle \Psis|$ may be achieved by inverting $U_{\mathrm{p}}$, as
\begin{equation}
    |\Psir\rangle\langle \Psis| = U_{\mathrm{p}}|0\rangle\langle\vec{1}_{\mathrm{T}}|U_{\mathrm{p}}^{\dag}.
\end{equation}
In particular, after inversion, the reduced density matrix of the target qubit contains the desired phase function $g(t)$, and the verification consists of checking whether all other qubits are measured into $0$.
The full control-free protocol is then
\begin{alg}[Control-free VPE]

\textbf{Input}:
\begin{itemize}
    \item Circuits to prepare a superposition of $|\Psis\rangle$ and $|\Psir\rangle$, invert the preparation, and implement time evolution $e^{iHt}$.
    \item Number of repetitions $M$ and $M$ of measurement in the $x$ and $y$ basis.
\end{itemize}
\textbf{Output}:
\begin{itemize}
    \item An estimate of $g(t)$ with variance $\frac{1}{M}$ and $\frac{1}{M}$ in the real and imaginary part respectively.
\end{itemize}
\textbf{Algorithm}:
\begin{enumerate}
    \item Prepare classical initial variables $g^x=0$, $g^y=0$.
    \item Prepare the system register in a starting state $\frac{1}{\sqrt{2}}(|\Psis\rangle+|\Psir\rangle)=U_{\mathrm{p}}\frac{1}{\sqrt{2}}(|0\rangle+|\vec{1}_{\mathrm{T}}\rangle)$.
    \item Apply the unitary $U^k$, (or equivalently simulate time evolution $e^{iHt}$), conditional on the control qubit.
    \item Apply the inverse circuit $U^{\dag}_{\mathrm{p}}$ to the system register.
    \item Rotate the target qubit into the $X$ or $Y$ basis and measure it to obtain a number $m\in 0,1$.
    \item Measure all other qubits, and if they all read out $0$, increment the relevant variable $g^x$ or $g^y$ by $(-1)^{m}$.
    \item Repeat steps 2-6 $M$ times in the $X$ basis and $M$ times in the $Y$ basis, and estimate $g(t)$ by $\tilde{g}(t)=\frac{g^x}{M}+i\frac{g^y}{M}$.
\end{enumerate}
\end{alg}
The analysis of Sec.~\ref{sec:mitigation} is identical for the control-free case, with the absence of the issue of control noise.
However, we note that at the beginning and the end of any experiment, single-qubit noise on the target qubit behaves similarly to control qubit noise.
This necessitates averaging over multiple initial and final rotations of the target qubit to prevent bias in the estimation of $g(t)$.

The above analysis implies that the algorithms studied in Ref.~\cite{Lu20Algorithms,Zintchenko16Randomized} should be amenable to verification immediately as well.
It also provides some additional explanation for the error-robustness observed in the robust phase estimation of Ref.~\cite{Russo20Evaluating}.

\section{Verified expectation value estimation}\label{sec:VEVE}

In many circumstances, one wishes not to know the eigenvalues of a Hermitian operator $H$, but instead its expectation value $\langle H\rangle$ under a specified state $|\Psi\rangle$.
For instance, in a variational quantum eigensolver~\cite{Mcclean16Theory}, one prepares a state $|\Psi(\vec{\theta})\rangle=U(\vec{\theta})|0\rangle$ dependent on a set of classical input parameters $\vec{\theta}$, then measures the expectation value $E(\vec{\theta})=\langle\Psi(\vec{\theta})|H|\Psi(\vec{\theta})\rangle$.
This is then optimized over $\vec{\theta}$ in a classical outer loop, with the optimized state $|\Psi(\vec{\theta}_{\mathrm{opt}})\rangle$ hopefully a good approximation of the true ground state $|E_0\rangle$.
In quantum variational algorithms it is typical that $\langle\Psi(\vec{\theta})|H|\Psi(\vec{\theta})\rangle$ is estimated by means of partial state tomography~\cite{Bonet20Nearly,Cotler20Quantum,Huggins19Efficient}. 
However, noise in the preparation unitary $U(\vec{\theta})$ causes an errant state $\rho_{\mathrm{err}}(\vec{\theta})\neq|\Psi(\vec{\theta})\rangle\langle\Psi(\vec{\theta})|$ to be prepared and tomographed, propagating the preparation error directly to a final estimation error.
The noise analysis in Sec.~\ref{sec:mitigation} extends to both the preparation and mitigation unitaries, so if verified phase estimation is used to provide estimates of eigenvalues and amplitudes, one may reconstruct
\begin{equation}
    \langle\Psi(\vec{\theta})|H|\Psi(\vec{\theta})\rangle=\sum_j|a_j|^2E_j,\label{eq:single_expectation_value}
\end{equation}
and inherit the mitigation power of the verification protocol.
This has the added advantage that control errors in the preparation circuit (which, being a repeated error, are not mitigated against) are able to be compensated for during the outer optimization loop of the VQE, as is well-known~\cite{Mcclean16Theory,Google20Hartree}.
Quantum phase estimation has previously been suggested as an alternative to partial state tomography for expectation value estimation, both to improve the rate of estimation~\cite{Wang19Accelerated}, and to provide a witness for the presence of eigenstates of the Hamiltonian~\cite{Santagati18Witnessing}.
The verification protocols described in this work should be applicable to these methods as well.
A general algorithm for verified expectation value estimation takes the form
\begin{alg}[Verified expectation value estimation]\label{alg:VEVE}
\textbf{Input}:
\begin{itemize}
    \item (Noisy) circuits to implement $U_{\mathrm{p}}$, $U_{\mathrm{p}}^{\dag}$ and controlled time evolution $e^{iHt}$.
    \item A set of $t$ values.
    \item Number of repetitions $M$ and $M$ of measurement in the $x$ and $y$ basis (that can be $t$-dependent).
    \item A method for classical signal processing (e.g. a curve fitting algorithm).
\end{itemize}
\textbf{Output}:
\begin{itemize}
    \item An estimate of $\langle H\rangle$.
\end{itemize}
\textbf{Algorithm}
\begin{enumerate}
    \item Estimate $g_{\mathrm{err}}(t)$ for all given points $t$ using Alg.~\ref{alg:single-control} to the chosen precision.
    \item Obtain estimates for individual $E_j$ and $A'_j$ values via classical signal processing.
    \item Estimate $\langle H\rangle$ as
    \begin{equation}
        \langle H\rangle = \frac{\sum_j A'_jE_j}{\sum_j A'_j}.\label{eq:renormalization_energy}
    \end{equation}
\end{enumerate}
\end{alg}

One might worry that the sum in Eq.~\ref{eq:single_expectation_value} is over an exponentially large number of eigenstates $|E_j\rangle$.
However one need not resolve all eigenvalues $E_j$ in order to accurately estimate the expectation value $\langle\Psi(\vec{\theta})|H|\Psi(\vec{\theta})\rangle$; if eigenvalues within $\delta$ of each other are binned, the resulting expectation value will be accurate to within $\delta$.
We may formalize this by considering the spectral function $g_{\mathrm{S}}$ of $|\psis\rangle$ under $H$,
\begin{equation}
    g_{\mathrm{S}}(E)=\sum_jA_j\delta(E-E_j).
\end{equation}
This can be seen to be the Fourier transform of the phase function $g(t)$ [strictly, $g(t)$ is the inverse Fourier transform of $g_{\mathrm{S}}(E/2\pi)$], and a coarse-grained approximation may be obtained via time-series methods~\cite{Somma19Quantum} or integral methods~\cite{Roggero20Spectral} with rigorous bounds on each.
Numerically, we find signal processing methods such as Prony's method~\cite{Obrien19Quantum} also perform acceptably (see Sec.~\ref{sec:sampling_noise}).
For fast-forwardable Hamiltonians (such as Pauli operators), one often already knows the target eigenvalues of the problem.
Furthermore, the eigenspectrum of these Hamiltonians is often highly degenerate, making simple curve fitting a practical (and attractive) alternative.

Instead of analysing the phase function at many points as described above, one may expand
\begin{align}
    \mathrm{Im}[g(t)]&=\sum_j|a_j|^2\sin(E_jt)\\&=t\sum_j|a_j|^2E_j+\frac{1}{3}t^3\sum_j|a_j|^2E_j^3+{\cal O}(t^5)\\\frac{1}{t}\mathrm{Im}[g(t)]&=\langle\Psi(\vec{\theta})|H|\Psi(\vec{\theta})\rangle + {\cal O}(t^2),
\end{align}
and simply estimate ${\rm Im}(g(t))$ for short times $t$.
This is similar to the manner in which eigenphases are estimated in the WAVES protocol~\cite{Santagati18Witnessing} (sans verification).
In this case, the normalisation of the resulting amplitudes (Eq.~\ref{eq:renormalization}) must be achieved by the condition that $g(0)=\sum_jA_j$, yielding
\begin{equation}
    \langle H\rangle=\frac{{\rm Im}[g(t)]}{t|g(0)|}+{\cal O}(t^2).\label{eq:noisy_single_point_estimation_single_H}
\end{equation}

\subsection{Fast-forwarded and parallelized Hamiltonian decompositions}\label{sec:parallel}

As expectation values are linear, we may estimate $\langle H\rangle$ by splitting it into multiple terms, estimating the expectation values of each term individually, and re-summing;
\begin{equation}
    H=\sum_sH_s\rightarrow \langle H\rangle=\sum_s\langle H_s\rangle.
\end{equation}
If individual $H_b$ may be simulated at lower circuit depth, this can reduce the accumulation of unmitigated errors, at the cost of requiring more simulation.
This ability becomes especially useful if one chooses the $H_s$ to be fast-forwardable.
Here, we define a fast-forwardable Hamiltonian $H_s$ as one for which a circuit implementation of $e^{iH_st}$ has constant depth in $t$.
The circuit depth required to simulate $e^{iHt}$ for arbitrary $H$ is bounded below as ${\cal O}(t)$~\cite{Berry07Efficient}, but for certain operators this may be improved on~\cite{Atia17Fast}.
For example, as the Pauli operators $\{\One,X,Y,Z\}^{\otimes N}$ are both fast-forwardable and form a basis for the set of $N$-qubit Hermitian operators, a set of $H_s$ terms may be taken from these to decompose an arbitrary Hamiltonian.
As another example, given an instance of the electronic structure problem, one may attempt a low-rank factorization of the interaction operator into a sum of ${\cal O}(N)$ diagonalizable (and thus fast-forwardable) terms \cite{Motta18Low}.

In order to speed up estimation of expectation values of multiple terms $H_s$ in a decomposed Hamiltonian $H=\sum_sH_s$, it may be possible perform the verified phase estimation step of each $H_s$ in parallel.
For example, we can perform time evolution of $L$ multiple summands, each controlled by a different control qubit, in between the preparation and verification steps of a single instance.
In the absence of verification, such parallelization will not affect the outcome of quantum phase estimation of any individual $H_s$, so long as all terms estimated in parallel commute.
This follows immediately from the fact that the time evolution for one such term does not evolve the system between eigenspaces of another.
This is complicated by the addition of verification, as the additional circuitry means that the system may evolve away from $|\psis\rangle$ despite a specific control qubit being in $|0\rangle$.
In App.~\ref{app:parallel}, we show that this gives rise to a set of spurious signals in the estimated phase function $g^{(s)}(t)$:
\begin{equation}
    g_q^{(s)}(t)=\sum_{v,j,j'}B^{(s)}_{j,j'}e^{iF^{(s)}_{v,j,j'}t}.
\end{equation}
Here, the ghost eigenvalues are
\begin{equation}
    F^{(s)}_{v,j,j'}=E_j^{(s)}+\sum_{s'\neq s}v_{s'}\left(E_j^{(s)}-E_{j'}^{(s')}\right),
\end{equation}
where $E_j^{(s')}$ are the true eigenvalues of the Hamiltonians $H_{s'}$, and $v$ is a $L$-bit vector written in binary (i.e. $v_s\in{0,1}$).
The corresponding, $v$-independent amplitudes are
\begin{equation}
    B_{j,j'}=\frac{1}{2^{L}}A_jA_{j'}.
\end{equation}
Although this is a far more complicated signal than the standard phase function $g(t)$, we calculate in App.~\ref{app:parallel} that it yields the same expectation value; i.e.
\begin{equation}
    \sum_{v,j,j'}B_{j,j'}F^{(s)}_{v,j,j'}=\langle H_s\rangle.
\end{equation}
This implies that verified parallel phase estimation may proceed in much the same way as the series protocol.

\begin{figure}
    \centering
    \includegraphics[width=\columnwidth]{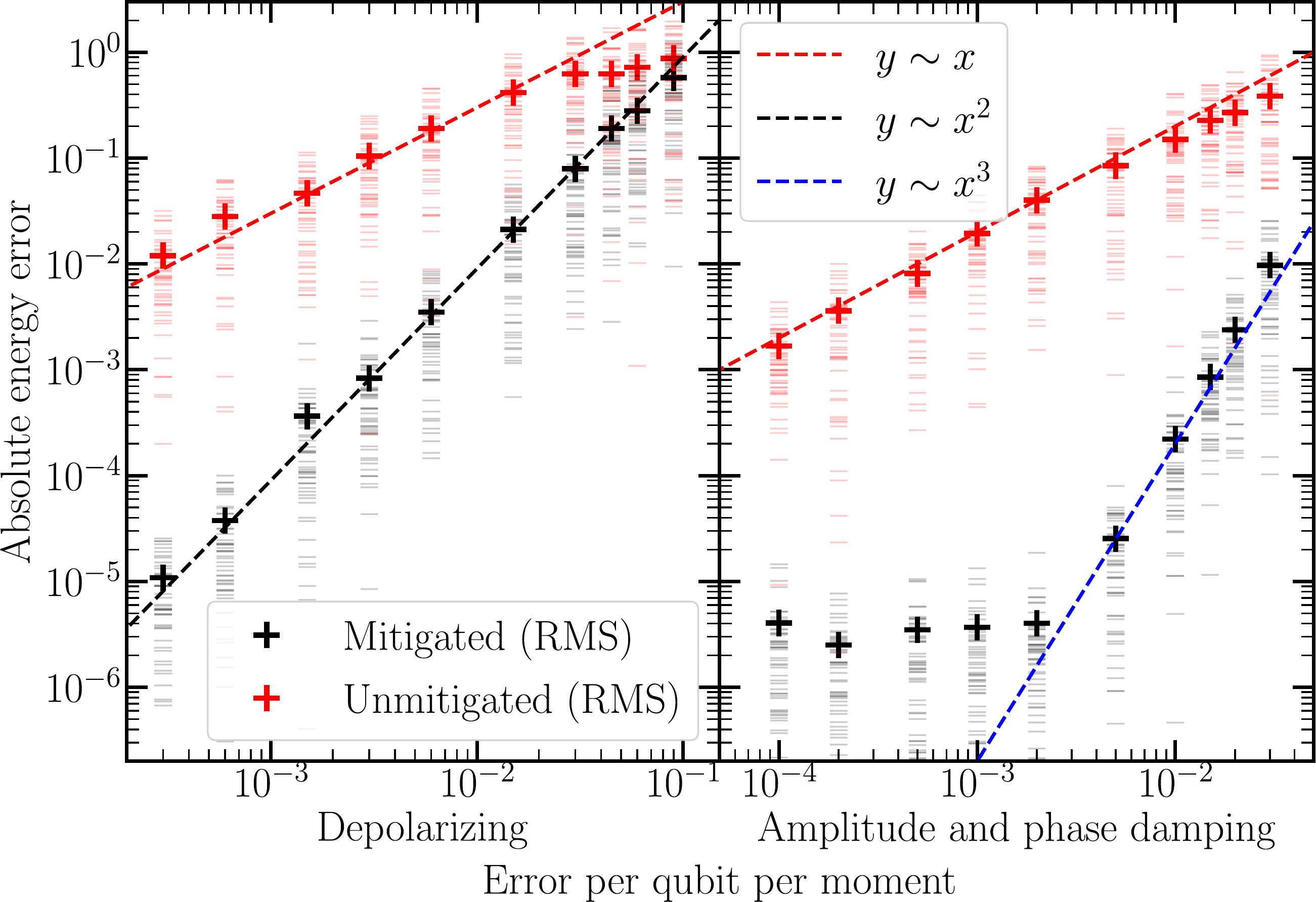}
    \caption{Mitigation of a 4-qubit Givens rotation circuit via verified phase estimation. (left) Error in estimation of random states in a free-fermion system (Eq.~\ref{eq:free_fermion_chain}) under a uniform depolarizing channel. (right) Error in the same estimation, but this time under an amplitude and phase damping model. In both plots, the RMS error (crosses) is calculated over $50$ different estimations for each error rate using either standard partial state tomography (red) or using verified control-free phase estimation. Individual data points (dashes) are additionally shown. For reference, dashed lines showing linear (red), quadratic (black), and cubic (blue) dependence on the gate error rate are plotted.}
    \label{fig:Givens_expt_est}
\end{figure}

\section{Numerical Experiments}\label{sec:results}

To investigate the mitigation capability of verified phase estimation, we first use it for expectation value estimation.
To prepare states, we take different variational ansatze with randomly-drawn parameters.
We compare the performance of verified and unverified circuits across multiple target Hamiltonians, noise strengths and noise models, to attempt to identify trends in the method.
All simulations were executed using the Cirq quantum software development framework ~\cite{cirq} and simulators therein.
Hamiltonians and complex circuits were further generated using code from the OpenFermion~\cite{McClean20Openfermion} libraries.
Except for when mentioned, the Cirq noise models were chosen to be a constant error rate per qubit per moment, where a moment is a period of the circuit where gates occur.
Equivalently, this can be thought of as an error rate per qubit per gate, but including error on idling gates as well.

\subsection{Given's rotation circuits for free-fermion Hamiltonians}

We first test the mitigation ability of the verification protocol on an instance of a ``Givens rotation circuit'' of the form developed for implementing rotations of single-particle fermionic basis functions in \cite{Kivlichan18Quantum}.
This circuit takes the form
\begin{equation}
    U(\vec{\theta})=\mathrm{exp}\left(i\sum_{j,l}\vec{\theta}_{j,l}c^{\dag}_jc_l\right),\label{eq:free_fermion_unitary}
\end{equation}
where $c_j^{\dag}$ and $c_j$ are the creation and annihilation operators for a fermion on site $j$, and $\theta_{j,l}=\theta_{l,j}$.
Such a circuit is classically simulatable, but it is a critical piece of infrastructure in quantum computing applications for quantum chemistry~\cite{Google20Hartree,Huggins19Efficient,Motta18Low,Berry19Qubitization,Burg20Quantum}.
It is also low depth: it may be decomposed exactly by a sequence of matchgates~\cite{Valiant01Quantum}, with optimal compilation in a circuit depth of exactly $N$.
When acting on a $N$-qubit register prepared in the state $\prod_{n=0}^{N_f-1}X_n|0\rangle$, this may prepare an arbitrary ground state of a free-fermion Hamiltonian with $N_f$ particles by an appropriate choice of $\vec{\theta}$.
In this work, we take a simple free-fermion Hamiltonian as an example - namely a one-dimensional chain:
\begin{equation}
    H=-t\sum_jc^{\dag}_{j}c_{j+1}+\mathrm{h.c.}\label{eq:free_fermion_chain}
\end{equation}
Such a Hamiltonian may be diagonalized,
\begin{equation}
    H=V^{\dag}\sum_{\alpha}\epsilon_{\alpha}c^{\dag}_{\alpha}c_{\alpha}V,
\end{equation}
where $V$ here takes the same form as in Eq.~\ref{eq:free_fermion_unitary}.
This decomposition allows immediately for the fast-forwarding of time evolution, as
\begin{align}
    e^{iHt}&=V^{\dag}e^{i\sum_{\alpha}\epsilon_{\alpha}c^{\dag}_{\alpha}c_{\alpha}}V\\
    &=V^{\dag}\prod_{\alpha}e^{i\epsilon_{\alpha}c^{\dag}_{\alpha}c_{\alpha}}.
\end{align}

\begin{figure}
    \centering
    \includegraphics[width=\columnwidth]{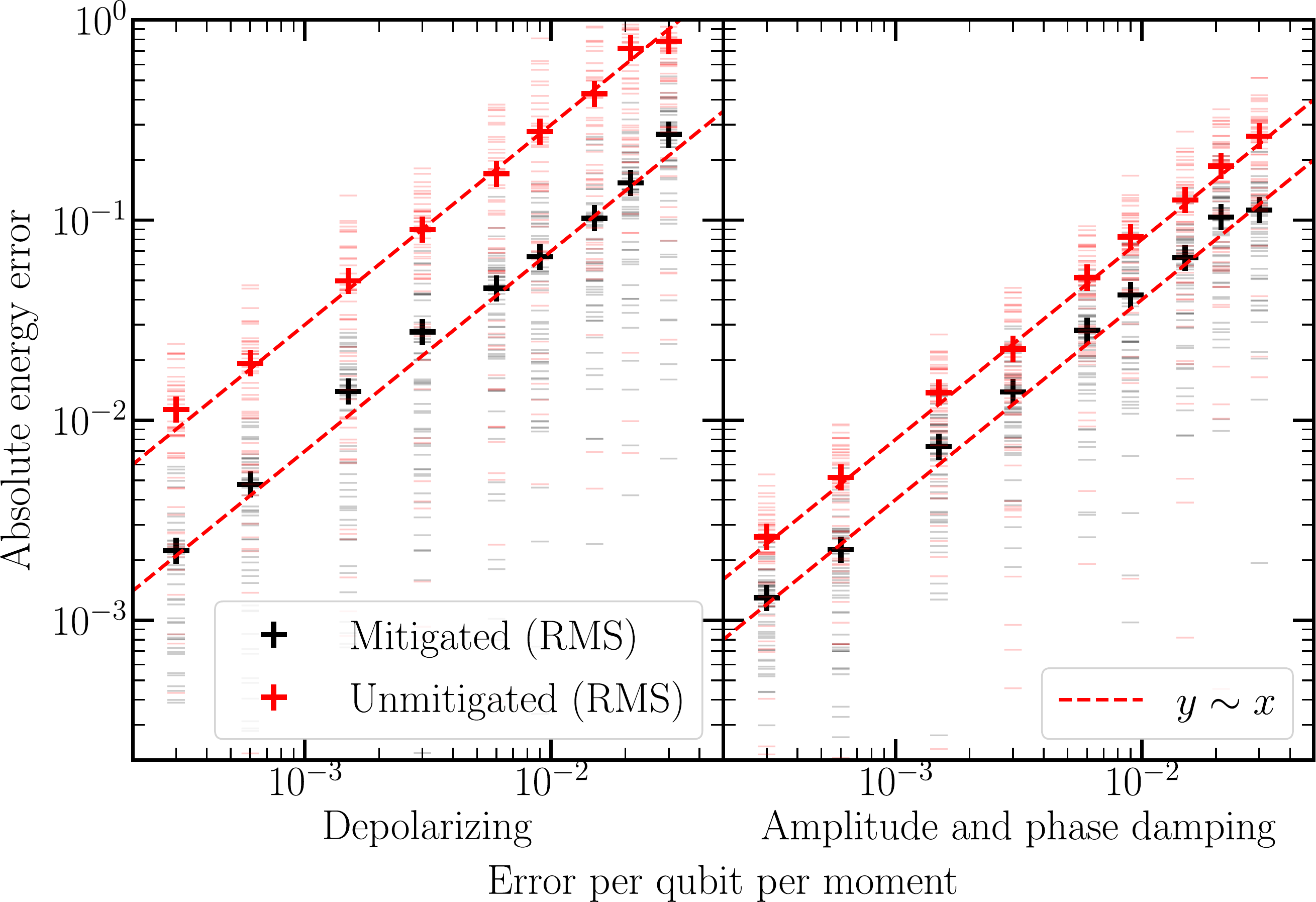}
    \caption{Mitigation of a 4-qubit VHA circuit via verified phase estimation. (left) Error in estimation of the energy of random states generated by the quantum approximate optimization ansatz in the critical phase of the transverse-field Ising model (Eq.~\ref{eq:TFIM}) under a uniform depolarizing channel. (right) Error in the same estimation, but this time under an amplitude and phase damping model. In both plots, the RMS error (crosses) is calculated over $50$ different estimations for each error rate (with randomly-chosen ansatz parameters) using either standard partial state tomography (red) or using verified control-free phase estimation. Individual data points (dashes) are additionally shown. For reference, dashed lines showing linear (red) dependence on the gate error rate are plotted.}
    \label{fig:TFIM_expt_est}
\end{figure}

As the Givens rotation circuits conserve particle number, the vacuum $|0\rangle$ may be used as a reference state for control-free verified estimation.
A superposition of this reference state and starting state $U(\vec{\theta})\prod_{n=1}^{N_f}X_n|0\rangle$ may be prepared by acting the matchgate circuit on the GHZ state 
\begin{equation}
    |\mathrm{GHZ}_{N_f}\rangle = \frac{1}{\sqrt{2}}\left(|0\rangle + \prod_{n=1}^{N_f}X_n|0\rangle\right),
\end{equation}
which may itself be prepared by e.g. a chain of CNOT gates:
\begin{equation}
    |\mathrm{GHZ}_{N_f}\rangle = \prod_{j=N_f-1}^{1}\mathrm{CNOT}_{j-1,j}\mathrm{H}_0|0\rangle.
\end{equation}
Note here the backwards product that runs left-to-right (i.e. the CNOT gate between qubit 1 and qubit 0 is executed first).
Following the definitions in Sec.~\ref{sec:control_free} for verified control-free phase estimation, we can write the complete preparation unitary as
\begin{equation}
    U_{\mathrm{p}}=U(\vec{\theta})\prod_{j=N_f-1}^{1}\mathrm{CNOT}_{j-1,j}.
\end{equation}
Then, as the product of two Givens rotation circuits is itself a Givens rotation circuit \cite{Kivlichan18Quantum}, we may compile $VU(\vec{\theta})=U(\vec{\theta}')$ and implement this in a single Givens rotation circuit.

\begin{figure}
    \centering
    \includegraphics[width=\columnwidth]{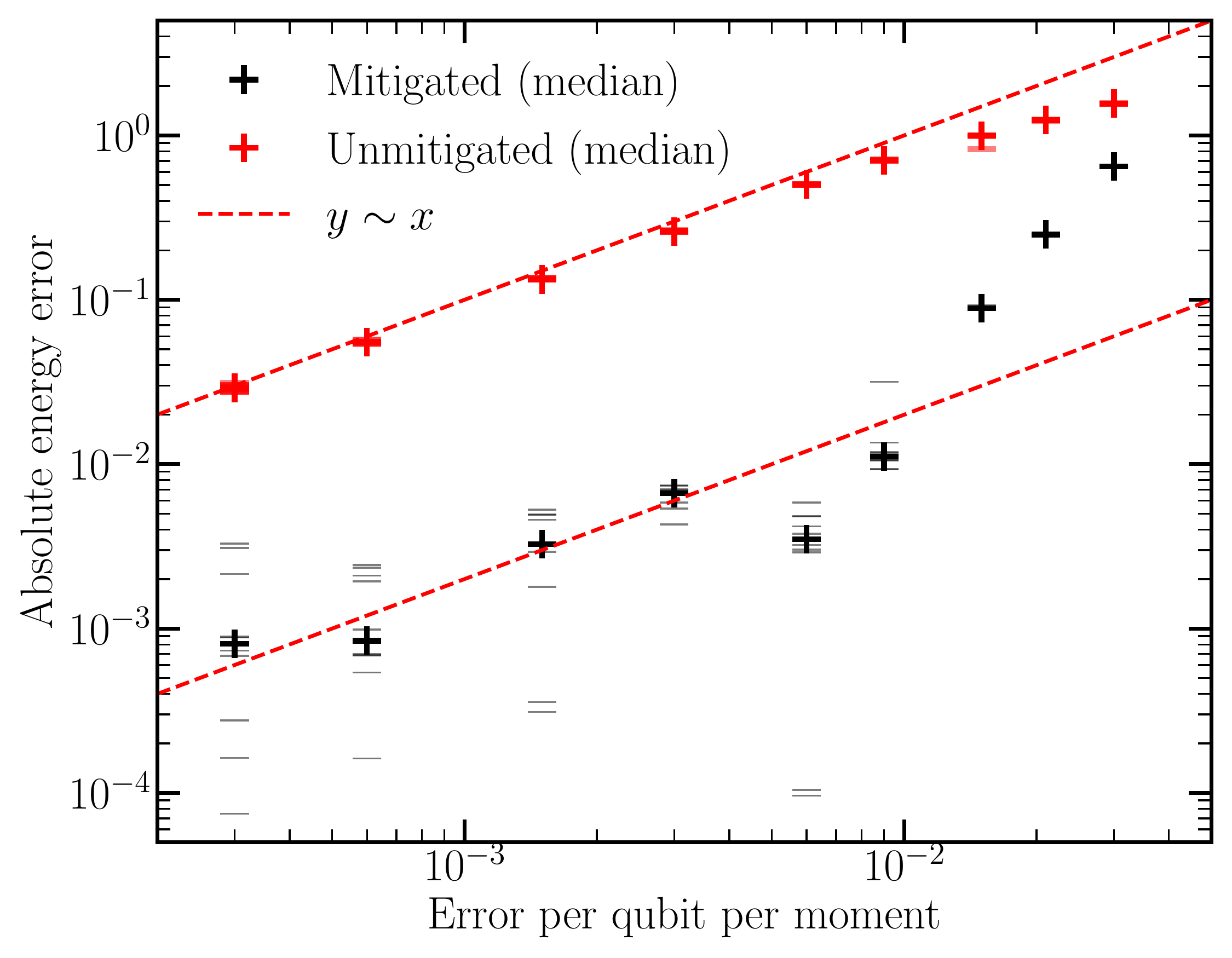}
    \caption{Error in estimating the ground state energy of a 4-site transverse-field Ising model (Eq.~\ref{eq:TFIM}) by variational optimization of a VHA ansatz. The resulting expectation values are measured either by verified single-control phase estimation (black) or taken directly from the simulated state state (red). We plot the median (crosses) of the absolute energy error over 10 optimization attempts, each starting from a different initial point. Individual errors are plotted behind (faint dashes). Guide lines showing a linear dependence are additionally plotted (red dashed lines).}
    \label{fig:TFIM_VQE}
\end{figure}

The complete VPE circuit for this circuit consists of the GHZ preparation, a single Givens' rotation, a set of single-qubit z-rotations, uncomputing the Givens' rotation, uncomputing the GHZ preparation, and measurement in the $X$ or $Y$ basis.
The resulting circuit for verified phase estimation is more than twice the length of the circuit required for the unmitigated VQE.
We assume here that the VQE tomography does not require any additional overhead, and directly estimate the expectation value from the simulated density matrix.
For verified phase estimation, we extract the phase function from the simulated density matrix, and then process it to estimate expectation values using Prony's method.
In order to not bias the final readout (which can lead to significant error in estimation), we average the rotation into the $X$ and $Y$-bases over both $+\pi/2$ and $-\pi/2$ rotations~\ref{sec:control_noise}.
To simplify the analysis here, we do not include additional sampling noise.
In Fig.~\ref{fig:Givens_expt_est}, we plot the RMS error for two error models over a range of noise models and strengths.
For each noise model and at each strength we sample $50$ random choices for the initial parameters $\vec{\theta}$ (and set $t=1$ in Eq.~\ref{eq:free_fermion_chain}).
In the presence of a uniform single-qubit depolarizing channel (Fig.~\ref{fig:Givens_expt_est}, left), we see that the verified error displays a clear $\epsilon\sim p^2$ trend (where $\epsilon$ is the error in the final estimation, and $p$ is the error per qubit per moment).
This implies that the effect of all single errors in this noise model are suppressed by the error mitigation (or fortuitously cancel), but that pairs of errors near to each other in time may affect results.
Under the effect of an amplitude and phase damping channel (Fig.~\ref{fig:Givens_expt_est}, right), the suppression is even starker; we see a clear $\epsilon\sim p^3$ trend till the error drops to below $10^{-5}$, providing up to four orders of magnitude gain in precision.
Below $10^{-5}$ the error plateaus.
This is due to numerical stability issues with Prony's method, and not a fundamental limit of the procedure~\footnote{Proof of this can be found in e.g. Fig.~\ref{fig:single_term_fsw}, where phase fitting obtained a $\sim 100$-fold reduction in this noise floor, which is typical for all simulations performed thus far.}.
This level of estimation error only becomes relevant after $>10^{10}p_{\mathrm{err}}^{-2}$ individual shots have been taken (with $p_{\mathrm{err}}$ the probability of an error over the entire circuit).
As such, we expect this to not be relevant for most experiments.
The lower error rate makes some sense: amplitude damping errors can only ever reduce the number of excitations in the circuit, and so by themselves can never return to a state with non-zero overlap with $|\psis\rangle$.
However, the precise mode for the leading contribution to the error rate is still somewhat unclear.

\subsection{The variational Hamiltonian ansatz for the transverse-field Ising model}\label{sec:TFIM_results}

We next attempt the verification of a completely different model and ansatz.
The transverse-field Ising model (TFIM) is a well-known spin system, with Hamiltonian
\begin{equation}
    H=J_z\sum_jZ_j + J_x\sum_jX_jX_{j+1},\label{eq:TFIM}
\end{equation}
where we take the sum $j+1$ modulo $N$ (i.e. periodic boundary conditions).
In one dimension, this model has a critical phase when $J_z=J_x$, making this a simple model to study interesting quantum phenomena.
Exact ground states of this model may be found by the variational Hamiltonian ansatz (VHA)~\cite{Wecker15Towards} for any values of $J_x$ and $J_z$~\cite{Ho19Efficient}.
The VHA consists of alternating the Ising model and transverse field terms $p$ times, with at each layer $p$ the amount of time to be treated as a free variable:
\begin{equation}
    U(\vec{\theta})=\prod_{p}e^{i\theta_{p,Z}\sum_jZ_j}e^{i\theta_{p,X}\sum_jX_jX_{j+1}}.
\end{equation}
(Note that for this given model the VHA is equivalent to the quantum alternating operator ansatz of Ref.~\cite{Farhi14Quantum}.)
The TFIM does not have any simple eigenstates, and nor does the VHA, so simple methods of control-free verified phase estimation are not available.
Instead, we attempt single-control verified phase estimation.
To lower the error incurred during the circuit, we perform VPE in series for every term in Eq.~\ref{eq:TFIM}.
Unfortunately, verification works significantly less well in this setting, as is shown in Fig.~\ref{fig:TFIM_expt_est}.
For both noise models considered, we see a clear $\epsilon\sim p$ trend with $\epsilon$ the energy error in the final result, and $p$ the error per qubit per moment.
This suggests that errors that map the noiseless state into one with nontrivial overlap with the verified density matrix are dominant in this circuit.
Regardless, we note that verification does provide an $\sim 8$-fold improvement in error rate over the unmitigated circuit, despite the verification circuit requiring one additional qubit and being three times as long.
This result is lessened in the presence of amplitude and phase damping noise, till the point where the mitigation only improves estimation by a factor of 2.

Variational optimization is well-known to mitigate certain types of coherent noise (e.g. coherent parameter drift)~\cite{Mcclean16Theory,Omalley16Scalable}; it also appears to provide some mitigation of incoherent noise when in combination with verified phase estimation.
In Fig.~\ref{fig:TFIM_VQE}, we perform a variational outer loop over the circuit studied in Fig.~\ref{fig:TFIM_expt_est}.
Although the $\epsilon\sim p$ behaviour appears to roughly remain in the latter half of the optimization, the gain from error mitigation improves from $2-8$x to around $50$x, a significant improvement.
We note that the optimization is no longer variationally bound - below about $10^{-2}$ error per qubit per moment, the results are scattered relatively evenly on either side of the true value.
By contrast, in the absence of sampling noise partial state tomography results will always be variationally bound.
We suspect this result may be due to the fact that slightly different circuits need to be run to measure different terms, yielding an 'effective state' that lies slightly outside the positive cone of allowed physical quantum states.
Though this effect does not appear to be particularly severe in this case, further study may be needed to see it does not become an issue in larger experiments.

\subsection{Fermionic swap networks for electronic structure Hamiltonians}\label{sec:fsw_results}

\begin{figure}[h!]
    \centering
    \includegraphics[width=\columnwidth]{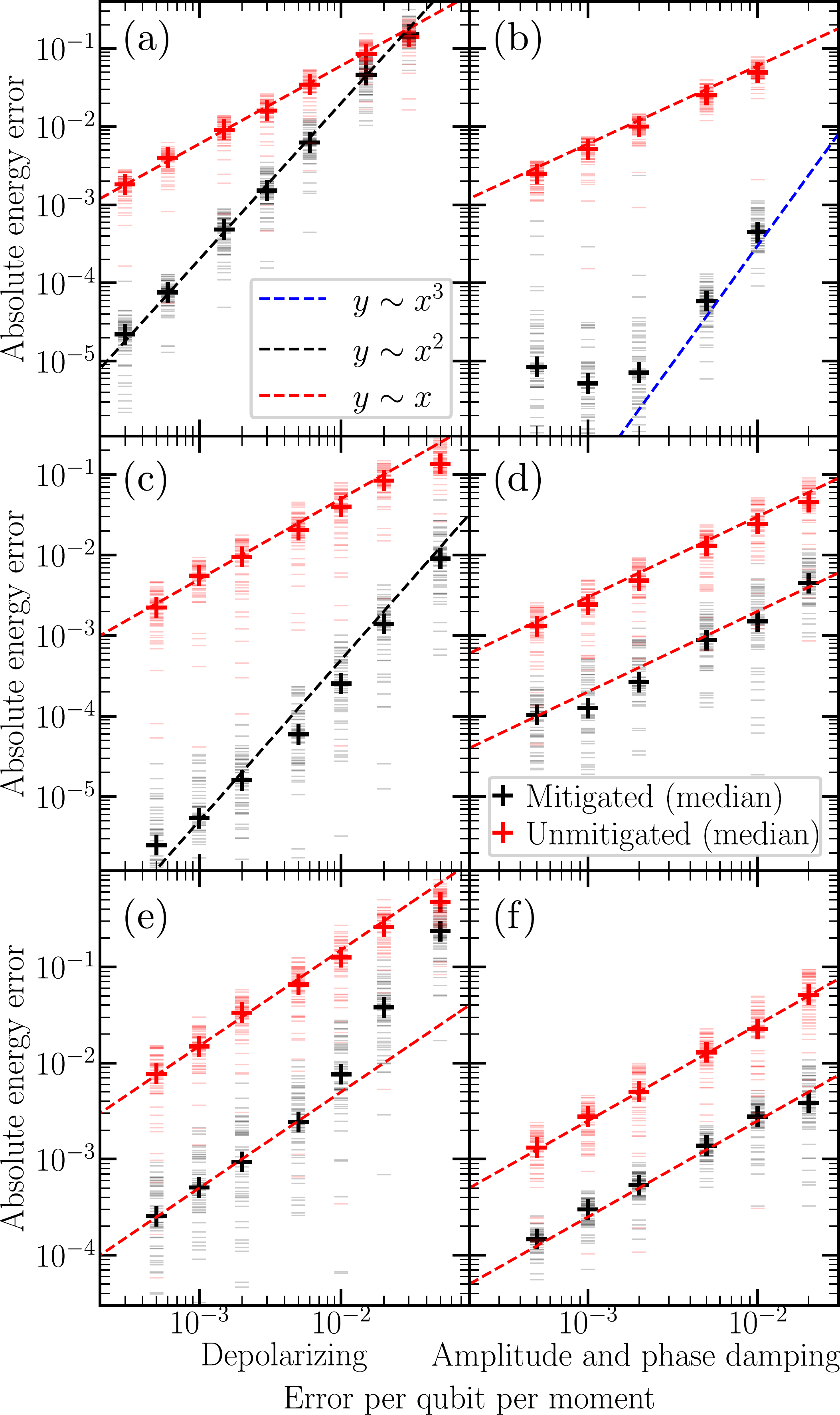}
    \caption{Mitigation of a 4-qubit fermionic swap network via verified phase estimation. Three different VPE protocols are explored --- a low-rank factorization (top row, a-b), a control-free number-conserving Pauli decomposition (middle row, c-d), and a single-control Pauli decomposition (bottom row, e-f). Details of all decompositions are given in the text. The low-rank factorization was studied for the H$_2$ Hamiltonian at the equilibrium bond distance with a swap network of depth $4$, while the other two models were studied at a bond distance of $2$ Angstrom with a swap network of depth $6$. All protocols are tested under depolarizing (left column, a,c,e) and amplitude and phase damping (right column, b,d,f) noise models. In all plots, the median error (crosses) is calculated over $50$ different estimations for each error rate using either standard partial state tomography (red) or using verified control-free phase estimation. Individual data points (dashes) are additionally shown. For reference, dashed lines showing linear (red), quadratic (black) and cubic (blue) dependence on the gate error rate are plotted.}
    \label{fig:fsw_expt_est}
\end{figure}

As a final system for simulation, we move to studying the ability to verify molecular hydrogen on four qubits using a fermionic swap network.
This ansatz was first studied in~\cite{Kivlichan18Quantum}; it consists of a network of two-qubit fermionic simulation gates, which take the form
\begin{equation}
    U_{\mathrm{fsim}}(\theta, \phi)=\left(\begin{array}{cccc}
    1 & 0 & 0 & 0\\
    0 & \cos(\theta) & i\sin(\theta) & 0\\
    0 & i\sin(\theta) & \cos(\theta) & 0 \\
    0 & 0 & 0 & e^{i\phi}
    \end{array}\right).
\end{equation}
The parameters $\theta$ and $\phi$ are then left free to be optimized during the circuit.
Molecular hydrogen is a simple example of the full electronic structure Hamiltonian, which takes the form
\begin{equation}
    H=\sum_{i,j}t_{i,j}c^{\dag}_ic_j + \sum_{i,j,k,l}V_{i,j,k,l}c^{\dag}_ic^{\dag}_jc_kc_l.\label{eq:electronic_structure_hamiltonian}
\end{equation}
Solving this Hamiltonian for mid-to-large system sizes ($\sim60+$ qubits) with strong interactions is a key target application for quantum computers~\cite{Berry19Qubitization,Burg20Quantum,Bauer20Quantum}.

We study three different methods for verified expectation value estimation of the electronic structure Hamiltonian.
Following a transformation from fermionic to qubit operators, Eq.~\ref{eq:electronic_structure_hamiltonian} may be decomposed over Pauli operators for single-control VPE, as was performed for the transverse-field Ising model in Sec.~\ref{sec:TFIM_results}.
However, in order to perform control-free VPE on these terms, we require a reference state.
Individual fermionic terms in Eq.~\ref{eq:electronic_structure_hamiltonian} are number-conserving - so the fermionic vacuum is a good reference state for these, but this is not the case for individual Pauli terms.
To circumvent this problem, we split Eq.~\ref{eq:electronic_structure_hamiltonian} into fermionic terms (summed with their Hermitian conjugate), and decompose these into Pauli operators.
(One can check that the resulting Pauli operators commute, and so their time evolution may be easily fast-forwarded.)
Both of the above decompositions are scalable, in that the number of terms summed and the circuit depth to implement $e^{iHt}$ in all cases is constant in the system size --- assuming that we may adjust our fermion-to-qubit transform for each circuit.
However, for the small example considered the time evolution remains large compared to the state preparation circuit, and the VPE circuits in both cases are $3-4$ times the depth of the original VQE.

Alternatively, by performing a low-rank factorization of the Coulomb operator, we may write $H$ in the form \cite{Motta18Low}
\begin{align}
    H&=\sum_{i,j}t'_{i,j}c^{\dag}_ic_j + \sum_lU^{\dag}_l\left[\sum_{i,j}t^{(l)}_{i,j}c^{\dag}_ic_j\right]^2U_l\\&=H^{(0)}+\sum_lH^{(l)}.
\end{align}
Each such term in this factorization is fast-forwardable.
$H^{(0)}$ is a free-fermion Hamiltonian and may be simulated via the methods discussed earlier in this section.
The interacting factors $H^{(l)}$ may also be diagonalized by diagonalizing the single-particle $t^{(l)}_{i,j}$ matrices.
One finds
\begin{align}
    H^{(l)}&=U^{\dag}_lV^{\dag}_l[\sum_{\alpha}\epsilon_{\alpha}c^{\dag}_{\alpha}c_{\alpha}]^2V_lU_l,\\
    e^{iH^{(l)}t}&= U^{\dag}_lV^{\dag}_l\prod_{\alpha\beta}e^{i\epsilon_{\alpha}\epsilon_{\beta}c^{\dag}_{\alpha}c_{\alpha}c^{\dag}_{\beta}c_{\beta}}V_lU_l,
\end{align}
which may be easily implemented on superconducting hardware, as $e^{i\epsilon_{\alpha}\epsilon_{\beta}c^{\dag}_{\alpha}c_{\alpha}c^{\dag}_{\beta}c_{\beta}}$ is realised by a C-Phase gate.
All of the above Hamiltonians, as well as the fermionic swap network itself, conserve particle number, and so we may again use the vacuum as a reference state for verified control-free quantum phase estimation.
We do not consider the single-control version for comparison in this case.
The resulting circuit is over ten times as long as the VQE itself, as unlike the Givens rotation circuit, we are unable to compile the final basis rotation into the ansatz.

The mitigation power of VPE differs vastly between the different choices of decomposition used, and the different noise models chosen.
In Fig.~\ref{fig:fsw_expt_est}, we plot the effect of mitigating depolarizing, and amplitude and phase damping channels, using the three decompositions described above.
We see that control-free [Fig.~\ref{fig:fsw_expt_est}(a-d)] VPE typically outperforms single-control VPE [Fig.~\ref{fig:fsw_expt_est}(e-f)], despite the single-control VPE circuits being in all cases smaller (due to the lack of coherent state preparation).
Under a depolarizing noise model, both control-free VPE implementations [Fig.~\ref{fig:fsw_expt_est}(a,c)] demonstrate a second-order sensitivity to the physical qubit error rate, consistent with the previous results in Fig.~\ref{fig:Givens_expt_est}.
In this case, the Pauli decomposition clearly outperforms the low-rank factorization, which we attribute to the large reduction ($\sim 2-3\times$) in total circuit depth.
However, although the low-rank factorization repeats the third-order sensitivity to amplitude and phase damping seen in Fig.~\ref{fig:Givens_expt_est} [Fig.~\ref{fig:fsw_expt_est}(b)], this is not observed in the Pauli decomposition case [Fig.~\ref{fig:fsw_expt_est}(d)].
We investigate this further in App.~\ref{app:single_term_decomposition}, and find that this first-order error can be traced back to the verified estimation of a single term --- the two-body interaction term.
We attribute this to the fact that the time evolution circuit for this term breaks number conservation (which is not the case for any other term in the sum), which makes it more susceptible to amplitude damping noise.
Understanding this feature in detail, and determining whether better circuit optimizations exist, are clear targets for future research.
In any case, all three implementations of VPE studied show at least an order of magnitude improvement compared to partial state tomography, and in some cases up to three orders of magnitude improvement, demonstrating the power of this technique.

\subsection{Sampling costs}\label{sec:sampling_noise}

In a realistic experiment, direct estimation of any expectation value requires repeatedly re-preparing the target state and measuring in an appropriate basis to accumulate statistics on the probability of seeing a given $0$ or $1$ measurement.
In verified phase estimation, this repetition must be performed instead on the control qubit (for single-control) or target qubit (for control-free) to accumulate the phase function.
Re-preparation is necessary between subsequent measurements, as such a measurement collapses the global wavefunction, erasing the information about the probability to be estimated.
This implies that each repetition carries substantial cost, and the rate of convergence of error estimation is a critical bottleneck in any variational algorithm.
Although one might expect quantum phase estimation to speed up this estimation (which has been proposed previously~\cite{Wang19Accelerated}), this is only the case when one is estimating eigenvalues of the target Hamiltonian in a specific QPE instance.
We wish to divide up our Hamiltonian for fast-forwarding purposes, and in most cases the resulting terms will not be simultaneously diagonalizable, so no set of mutual eigenstates will exist; instead, the results of Sec.~\ref{sec:sampling_noise_analytics} will hold.
Furthermore, as our expectation value estimation requires to sum over multiple different amplitudes, we should not expect this to improve over the cost of partial state tomography (which requires non-commuting terms to be measured on separate preparations of the state).
The error in expectation value estimation will further depend on the type of classical post-processing used.

\begin{figure}
    \centering
    \includegraphics[width=\columnwidth]{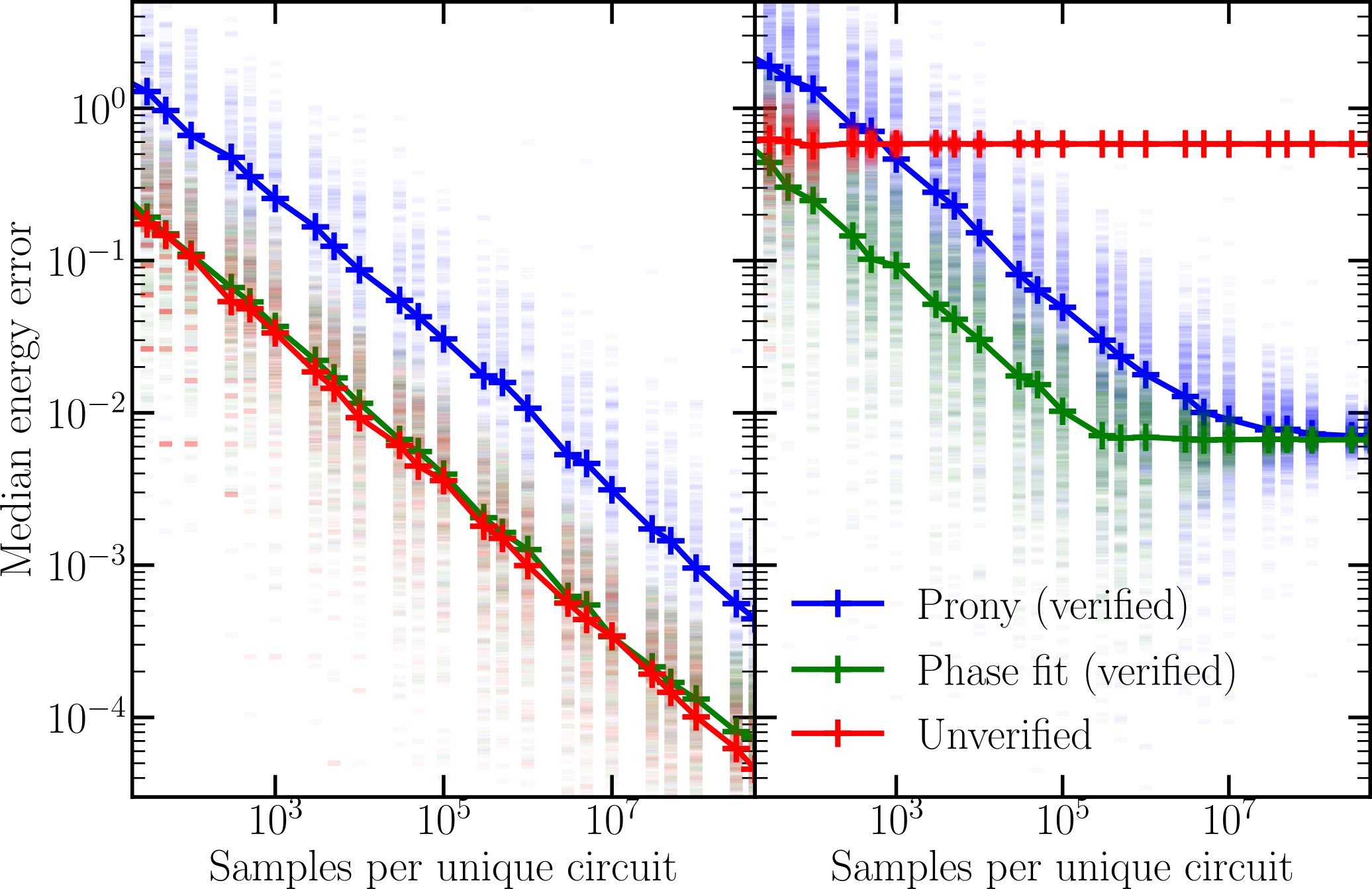}
    \caption{Convergence of the estimation of a single point in a 4-site transverse-field Ising model with the number of samples taken, using verified phase estimation processed either with Prony's method (blue) or by fitting known phases to the phase function (green), or standard partial state tomography (red) on individual Pauli terms. (Left) convergence in the absence of error. (Right) convergence in the presence of $1\%$ depolarizing error per qubit per moment. In each subfigure we plot the median energy error (crosses and lines) over $200$ simulations, which are plotted themselves behind (faint dashes).}
    \label{fig:sampling_noise_no_error}
\end{figure}

In Fig.~\ref{fig:sampling_noise_no_error}, we compare the convergence of two types of classical post-processing to that of standard partial state tomography.
We perform this simulation on the 4-spin VHA-TFIM system studied in Fig.~\ref{fig:TFIM_expt_est} and Fig.~\ref{fig:TFIM_VQE}, on a representative point in the spectrum (the error-free variational minimum).
The first method (green) assumes knowledge about the eigenvalues of the fast-forwarded Hamiltonians, in which case one need only fit the amplitudes, while the second (blue) first estimates the eigenvalues using Prony's method before fitting the amplitudes to the resulting signal.
(We compensate for the presence of spurious phases in Prony's method by a slight adjustment described in App.~\ref{app:prony_compensation}.)
All methods of estimation are seen to converge at a rate $\epsilon\sim M^{-1/2}$, where $\epsilon$ is the estimation error and $M$ is the number of samples taken.

We see that using the prior knowledge of the phases gives a significant advantage in convergence, with the resulting error rate being almost an order of magnitude worse when using Prony's method.
This advantage persists in the presence of a depolarizing channel ($1\%$ error rate), although the convergence of all methods flattens as they approach the sampling-noise-free estimation value.
We note that both classical post-processing methods converge to the same result here, as expected.
It is unclear whether the good overlap between the unverified circuit and the phase fitting method is due to them both achieving a lower bound for convergence or just coincidence.
Further investigation here would be a good target for future work.
The addition of noise makes convergence more costly.
This increase can be bounded below by removing the fraction of experiments where at least one error has occurred (as we are at best effectively removing these results).
Confirming this trend would also be a good target for future work.

\section{Conclusion}

In this work, we presented a new method for error mitigation, based on verification of the system register in a single-control quantum phase estimation routine.
We further extended this method to a scheme for verification of control-free quantum phase estimation.
By writing a complex Hamiltonian as a sum of fast-forwardable parts and using this technique to estimate the expectation value of each part, this becomes a powerful error mitigation tool for near-term experiments such as variational algorithms.
Errors that take the system away from the small verified subspace do not affect the mitigated QPE results (at the cost of requiring additional repetitions of the circuit).
We performed numerical studies of this error mitigation capability of the verification protocol on three different systems, finding the suppression of all single depolarizing errors when a Givens rotation circuit or a fermionic swap network prepare random states of a small fermionic system.
The suppression is further magnified in the presence of amplitude and phase damping, resulting in a gain of up to four orders of magnitude in accuracy.
For a simulation of the transverse-field Ising model the error suppression is less pronounced. However, we find that variational optimization improves the error mitigation to a gain in accuracy of about 50-fold.
We further demonstrated that the combination of variational optimization and verification mitigates against constant control error (which is not naturally mitigated by the verification itself).
However, we found that the choice of post-processing technique in the classical post-processing may affect the estimation error by a factor of $10$ in the presence of sampling noise.

Though verified phase estimation as presented already appears to be one of the most powerful error mitigation techniques available to NISQ-era quantum computing, further avenues for optimization exist.
The wide range of possible options for verification, how to divide the Hamiltonian, and classical post-processing method all provide metaparameters which we have not yet determined how to optimize for any specific problem.
Furthermore, circuits which quickly scramble errors would appear to make verification more reliable.
Whether this observation can be used for meaningful optimization is a clear target for future work.
Similarly, as errors need to have the instantaneous state as a near-eigenstate to not fail verification, the errors that verified phase estimation is most-susceptible to must commute, and could potentially be corrected with a classical error correcting code.
As these codes require much less overhead than full-blown QEC, this may be a practical method to ensure universal suppression of single-qubit errors.
Future work could also investigate whether verified phase estimation may be combined efficiently with other error mitigation techniques.
More generally, it would be timely to benchmark the zoo of error mitigation techniques against one another, and determine which combination of techniques works best in a range of situations.\\

\subsection*{Acknowledgements}

The authors acknowledge helpful discussions with Vedran Dunjko, Gian-Luca Anselmetti, Christian Gogolin, Zhang Jiang, Kostyantyn Kechedzhi, Vadim Smelyanskiy and Nathan Wiebe.

\bibliography{VPE}

\begin{thebibliography}{69}%
\makeatletter
\providecommand \@ifxundefined [1]{%
 \@ifx{#1\undefined}
}%
\providecommand \@ifnum [1]{%
 \ifnum #1\expandafter \@firstoftwo
 \else \expandafter \@secondoftwo
 \fi
}%
\providecommand \@ifx [1]{%
 \ifx #1\expandafter \@firstoftwo
 \else \expandafter \@secondoftwo
 \fi
}%
\providecommand \natexlab [1]{#1}%
\providecommand \enquote  [1]{``#1''}%
\providecommand \bibnamefont  [1]{#1}%
\providecommand \bibfnamefont [1]{#1}%
\providecommand \citenamefont [1]{#1}%
\providecommand \href@noop [0]{\@secondoftwo}%
\providecommand \href [0]{\begingroup \@sanitize@url \@href}%
\providecommand \@href[1]{\@@startlink{#1}\@@href}%
\providecommand \@@href[1]{\endgroup#1\@@endlink}%
\providecommand \@sanitize@url [0]{\catcode `\\12\catcode `\$12\catcode
  `\&12\catcode `\#12\catcode `\^12\catcode `\_12\catcode `\%12\relax}%
\providecommand \@@startlink[1]{}%
\providecommand \@@endlink[0]{}%
\providecommand \url  [0]{\begingroup\@sanitize@url \@url }%
\providecommand \@url [1]{\endgroup\@href {#1}{\urlprefix }}%
\providecommand \urlprefix  [0]{URL }%
\providecommand \Eprint [0]{\href }%
\providecommand \doibase [0]{http://dx.doi.org/}%
\providecommand \selectlanguage [0]{\@gobble}%
\providecommand \bibinfo  [0]{\@secondoftwo}%
\providecommand \bibfield  [0]{\@secondoftwo}%
\providecommand \translation [1]{[#1]}%
\providecommand \BibitemOpen [0]{}%
\providecommand \bibitemStop [0]{}%
\providecommand \bibitemNoStop [0]{.\EOS\space}%
\providecommand \EOS [0]{\spacefactor3000\relax}%
\providecommand \BibitemShut  [1]{\csname bibitem#1\endcsname}%
\let\auto@bib@innerbib\@empty
\bibitem [{\citenamefont {Arute}\ \emph {et~al.}(2019)\citenamefont {Arute},
  \citenamefont {Arya}, \citenamefont {Babbush}, \citenamefont {Bacon},
  \citenamefont {Bardin}, \citenamefont {Barends}, \citenamefont {Biswas},
  \citenamefont {Boixo}, \citenamefont {Brandao}, \citenamefont {Buell},
  \citenamefont {Burkett}, \citenamefont {Chen}, \citenamefont {Chen},
  \citenamefont {Chiaro}, \citenamefont {Collins}, \citenamefont {Courtney},
  \citenamefont {Dunsworth}, \citenamefont {Farhi}, \citenamefont {Foxen},
  \citenamefont {Fowler}, \citenamefont {Gidney}, \citenamefont {Giustina},
  \citenamefont {Graff}, \citenamefont {Guerin}, \citenamefont {Habegger},
  \citenamefont {Harrigan}, \citenamefont {Hartmann}, \citenamefont {Ho},
  \citenamefont {Hoffmann}, \citenamefont {Huang}, \citenamefont {Humble},
  \citenamefont {Isakov}, \citenamefont {Jeffrey}, \citenamefont {Jiang},
  \citenamefont {Kafri}, \citenamefont {Kechedzhi}, \citenamefont {Kelly},
  \citenamefont {Klimov}, \citenamefont {Knysh}, \citenamefont {Korotkov},
  \citenamefont {Kostritsa}, \citenamefont {Landhuis}, \citenamefont
  {Lindmark}, \citenamefont {Lucero}, \citenamefont {Lyakh}, \citenamefont
  {Mandrà}, \citenamefont {McClean}, \citenamefont {McEwen}, \citenamefont
  {Megrant}, \citenamefont {Mi}, \citenamefont {Michielsen}, \citenamefont
  {Mohseni}, \citenamefont {Mutus}, \citenamefont {Naaman}, \citenamefont
  {Neeley}, \citenamefont {Neill}, \citenamefont {Niu}, \citenamefont {Ostby},
  \citenamefont {Petukhov}, \citenamefont {Platt}, \citenamefont {Quintana},
  \citenamefont {Rieffel}, \citenamefont {Roushan}, \citenamefont {Rubin},
  \citenamefont {Sank}, \citenamefont {Satzinger}, \citenamefont {Smelyanskiy},
  \citenamefont {Sung}, \citenamefont {Trevithick}, \citenamefont
  {Vainsencher}, \citenamefont {Villalonga}, \citenamefont {White},
  \citenamefont {Yao}, \citenamefont {Yeh}, \citenamefont {Zalcman},
  \citenamefont {Neven},\ and\ \citenamefont {Martinis}}]{Google19Quantum}%
  \BibitemOpen
  \bibfield  {author} {\bibinfo {author} {\bibfnamefont {Frank}\ \bibnamefont
  {Arute}}, \bibinfo {author} {\bibfnamefont {Kunal}\ \bibnamefont {Arya}},
  \bibinfo {author} {\bibfnamefont {Ryan}\ \bibnamefont {Babbush}}, \bibinfo
  {author} {\bibfnamefont {Dave}\ \bibnamefont {Bacon}}, \bibinfo {author}
  {\bibfnamefont {Joseph~C.}\ \bibnamefont {Bardin}}, \bibinfo {author}
  {\bibfnamefont {Rami}\ \bibnamefont {Barends}}, \bibinfo {author}
  {\bibfnamefont {Rupak}\ \bibnamefont {Biswas}}, \bibinfo {author}
  {\bibfnamefont {Sergio}\ \bibnamefont {Boixo}}, \bibinfo {author}
  {\bibfnamefont {Fernando G. S.~L.}\ \bibnamefont {Brandao}}, \bibinfo
  {author} {\bibfnamefont {David~A.}\ \bibnamefont {Buell}}, \bibinfo {author}
  {\bibfnamefont {Brian}\ \bibnamefont {Burkett}}, \bibinfo {author}
  {\bibfnamefont {Yu}~\bibnamefont {Chen}}, \bibinfo {author} {\bibfnamefont
  {Zijun}\ \bibnamefont {Chen}}, \bibinfo {author} {\bibfnamefont {Ben}\
  \bibnamefont {Chiaro}}, \bibinfo {author} {\bibfnamefont {Roberto}\
  \bibnamefont {Collins}}, \bibinfo {author} {\bibfnamefont {William}\
  \bibnamefont {Courtney}}, \bibinfo {author} {\bibfnamefont {Andrew}\
  \bibnamefont {Dunsworth}}, \bibinfo {author} {\bibfnamefont {Edward}\
  \bibnamefont {Farhi}}, \bibinfo {author} {\bibfnamefont {Brooks}\
  \bibnamefont {Foxen}}, \bibinfo {author} {\bibfnamefont {Austin}\
  \bibnamefont {Fowler}}, \bibinfo {author} {\bibfnamefont {Craig}\
  \bibnamefont {Gidney}}, \bibinfo {author} {\bibfnamefont {Marissa}\
  \bibnamefont {Giustina}}, \bibinfo {author} {\bibfnamefont {Rob}\
  \bibnamefont {Graff}}, \bibinfo {author} {\bibfnamefont {Keith}\ \bibnamefont
  {Guerin}}, \bibinfo {author} {\bibfnamefont {Steve}\ \bibnamefont
  {Habegger}}, \bibinfo {author} {\bibfnamefont {Matthew~P.}\ \bibnamefont
  {Harrigan}}, \bibinfo {author} {\bibfnamefont {Michael~J.}\ \bibnamefont
  {Hartmann}}, \bibinfo {author} {\bibfnamefont {Alan}\ \bibnamefont {Ho}},
  \bibinfo {author} {\bibfnamefont {Markus}\ \bibnamefont {Hoffmann}}, \bibinfo
  {author} {\bibfnamefont {Trent}\ \bibnamefont {Huang}}, \bibinfo {author}
  {\bibfnamefont {Travis~S.}\ \bibnamefont {Humble}}, \bibinfo {author}
  {\bibfnamefont {Sergei~V.}\ \bibnamefont {Isakov}}, \bibinfo {author}
  {\bibfnamefont {Evan}\ \bibnamefont {Jeffrey}}, \bibinfo {author}
  {\bibfnamefont {Zhang}\ \bibnamefont {Jiang}}, \bibinfo {author}
  {\bibfnamefont {Dvir}\ \bibnamefont {Kafri}}, \bibinfo {author}
  {\bibfnamefont {Kostyantyn}\ \bibnamefont {Kechedzhi}}, \bibinfo {author}
  {\bibfnamefont {Julian}\ \bibnamefont {Kelly}}, \bibinfo {author}
  {\bibfnamefont {Paul~V.}\ \bibnamefont {Klimov}}, \bibinfo {author}
  {\bibfnamefont {Sergey}\ \bibnamefont {Knysh}}, \bibinfo {author}
  {\bibfnamefont {Alexander}\ \bibnamefont {Korotkov}}, \bibinfo {author}
  {\bibfnamefont {Fedor}\ \bibnamefont {Kostritsa}}, \bibinfo {author}
  {\bibfnamefont {David}\ \bibnamefont {Landhuis}}, \bibinfo {author}
  {\bibfnamefont {Mike}\ \bibnamefont {Lindmark}}, \bibinfo {author}
  {\bibfnamefont {Erik}\ \bibnamefont {Lucero}}, \bibinfo {author}
  {\bibfnamefont {Dmitry}\ \bibnamefont {Lyakh}}, \bibinfo {author}
  {\bibfnamefont {Salvatore}\ \bibnamefont {Mandrà}}, \bibinfo {author}
  {\bibfnamefont {Jarrod~R.}\ \bibnamefont {McClean}}, \bibinfo {author}
  {\bibfnamefont {Matthew}\ \bibnamefont {McEwen}}, \bibinfo {author}
  {\bibfnamefont {Anthony}\ \bibnamefont {Megrant}}, \bibinfo {author}
  {\bibfnamefont {Xiao}\ \bibnamefont {Mi}}, \bibinfo {author} {\bibfnamefont
  {Kristel}\ \bibnamefont {Michielsen}}, \bibinfo {author} {\bibfnamefont
  {Masoud}\ \bibnamefont {Mohseni}}, \bibinfo {author} {\bibfnamefont {Josh}\
  \bibnamefont {Mutus}}, \bibinfo {author} {\bibfnamefont {Ofer}\ \bibnamefont
  {Naaman}}, \bibinfo {author} {\bibfnamefont {Matthew}\ \bibnamefont
  {Neeley}}, \bibinfo {author} {\bibfnamefont {Charles}\ \bibnamefont {Neill}},
  \bibinfo {author} {\bibfnamefont {Murphy~Yuezhen}\ \bibnamefont {Niu}},
  \bibinfo {author} {\bibfnamefont {Eric}\ \bibnamefont {Ostby}}, \bibinfo
  {author} {\bibfnamefont {Andre}\ \bibnamefont {Petukhov}}, \bibinfo {author}
  {\bibfnamefont {John~C.}\ \bibnamefont {Platt}}, \bibinfo {author}
  {\bibfnamefont {Chris}\ \bibnamefont {Quintana}}, \bibinfo {author}
  {\bibfnamefont {Eleanor~G.}\ \bibnamefont {Rieffel}}, \bibinfo {author}
  {\bibfnamefont {Pedram}\ \bibnamefont {Roushan}}, \bibinfo {author}
  {\bibfnamefont {Nicholas~C.}\ \bibnamefont {Rubin}}, \bibinfo {author}
  {\bibfnamefont {Daniel}\ \bibnamefont {Sank}}, \bibinfo {author}
  {\bibfnamefont {Kevin~J.}\ \bibnamefont {Satzinger}}, \bibinfo {author}
  {\bibfnamefont {Vadim}\ \bibnamefont {Smelyanskiy}}, \bibinfo {author}
  {\bibfnamefont {Kevin~J.}\ \bibnamefont {Sung}}, \bibinfo {author}
  {\bibfnamefont {Matthew~D.}\ \bibnamefont {Trevithick}}, \bibinfo {author}
  {\bibfnamefont {Amit}\ \bibnamefont {Vainsencher}}, \bibinfo {author}
  {\bibfnamefont {Benjamin}\ \bibnamefont {Villalonga}}, \bibinfo {author}
  {\bibfnamefont {Theodore}\ \bibnamefont {White}}, \bibinfo {author}
  {\bibfnamefont {Z.~Jamie}\ \bibnamefont {Yao}}, \bibinfo {author}
  {\bibfnamefont {Ping}\ \bibnamefont {Yeh}}, \bibinfo {author} {\bibfnamefont
  {Adam}\ \bibnamefont {Zalcman}}, \bibinfo {author} {\bibfnamefont {Hartmut}\
  \bibnamefont {Neven}}, \ and\ \bibinfo {author} {\bibfnamefont {John~M.}\
  \bibnamefont {Martinis}},\ }\bibfield  {title} {\enquote {\bibinfo {title}
  {Quantum supremacy using a programmable superconducting processor},}\ }\href
  {https://www.nature.com/articles/s41586-019-1666-5} {\bibfield  {journal}
  {\bibinfo  {journal} {Nature}\ }\textbf {\bibinfo {volume} {574}},\ \bibinfo
  {pages} {505--510} (\bibinfo {year} {2019})}\BibitemShut {NoStop}%
\bibitem [{\citenamefont {Quantum}\ and\ \citenamefont
  {Collaborators}(2020{\natexlab{a}})}]{Google20Quantum}%
  \BibitemOpen
  \bibfield  {author} {\bibinfo {author} {\bibfnamefont {Google~AI}\
  \bibnamefont {Quantum}}\ and\ \bibinfo {author} {\bibnamefont
  {Collaborators}},\ }\bibfield  {title} {\enquote {\bibinfo {title} {Quantum
  approximate optimization of non-planar graph problems on a planar
  superconducting processor},}\ }\href {https://arxiv.org/pdf/2004.04197.pdf}
  {\bibfield  {journal} {\bibinfo  {journal} {ArXiv:2004.04197}\ } (\bibinfo
  {year} {2020}{\natexlab{a}})}\BibitemShut {NoStop}%
\bibitem [{\citenamefont {Pagano}\ \emph {et~al.}(2019)\citenamefont {Pagano},
  \citenamefont {Bapat}, \citenamefont {Becker}, \citenamefont {Collins},
  \citenamefont {De}, \citenamefont {Hess}, \citenamefont {Kaplan},
  \citenamefont {Kyprianidis}, \citenamefont {Tan}, \citenamefont {Baldwin},
  \citenamefont {Brady}, \citenamefont {Deshpande}, \citenamefont {Liu},
  \citenamefont {Jordan}, \citenamefont {Gorshkov},\ and\ \citenamefont
  {Monroe}}]{Pagano19Quantum}%
  \BibitemOpen
  \bibfield  {author} {\bibinfo {author} {\bibfnamefont {G.}~\bibnamefont
  {Pagano}}, \bibinfo {author} {\bibfnamefont {A.}~\bibnamefont {Bapat}},
  \bibinfo {author} {\bibfnamefont {P.}~\bibnamefont {Becker}}, \bibinfo
  {author} {\bibfnamefont {K.~S.}\ \bibnamefont {Collins}}, \bibinfo {author}
  {\bibfnamefont {A.}~\bibnamefont {De}}, \bibinfo {author} {\bibfnamefont
  {P.~W.}\ \bibnamefont {Hess}}, \bibinfo {author} {\bibfnamefont {H.~B.}\
  \bibnamefont {Kaplan}}, \bibinfo {author} {\bibfnamefont {A.}~\bibnamefont
  {Kyprianidis}}, \bibinfo {author} {\bibfnamefont {W.~L.}\ \bibnamefont
  {Tan}}, \bibinfo {author} {\bibfnamefont {C.}~\bibnamefont {Baldwin}},
  \bibinfo {author} {\bibfnamefont {L.~T.}\ \bibnamefont {Brady}}, \bibinfo
  {author} {\bibfnamefont {A.}~\bibnamefont {Deshpande}}, \bibinfo {author}
  {\bibfnamefont {F.}~\bibnamefont {Liu}}, \bibinfo {author} {\bibfnamefont
  {S.}~\bibnamefont {Jordan}}, \bibinfo {author} {\bibfnamefont {A.~V.}\
  \bibnamefont {Gorshkov}}, \ and\ \bibinfo {author} {\bibfnamefont
  {C.}~\bibnamefont {Monroe}},\ }\bibfield  {title} {\enquote {\bibinfo {title}
  {Quantum approximate optimization of the long-range ising model with a
  trapped-ion quantum simulator},}\ }\href {https://arxiv.org/abs/1906.02700}
  {\bibfield  {journal} {\bibinfo  {journal} {ArXiv:1906.02700}\ } (\bibinfo
  {year} {2019})}\BibitemShut {NoStop}%
\bibitem [{\citenamefont {Quantum}\ and\ \citenamefont
  {Collaborators}(2020{\natexlab{b}})}]{Google20Hartree}%
  \BibitemOpen
  \bibfield  {author} {\bibinfo {author} {\bibfnamefont {Google~AI}\
  \bibnamefont {Quantum}}\ and\ \bibinfo {author} {\bibnamefont
  {Collaborators}},\ }\bibfield  {title} {\enquote {\bibinfo {title}
  {Hartree-fock on a superconducting qubit quantum computer},}\ }\href
  {https://arxiv.org/abs/2004.04174} {\bibfield  {journal} {\bibinfo  {journal}
  {Science}\ }\textbf {\bibinfo {volume} {In Press}} (\bibinfo {year}
  {2020}{\natexlab{b}})}\BibitemShut {NoStop}%
\bibitem [{\citenamefont {Kandala}\ \emph {et~al.}(2017)\citenamefont
  {Kandala}, \citenamefont {Mezzacapo}, \citenamefont {Temme}, \citenamefont
  {Takita}, \citenamefont {Brink}, \citenamefont {Chow},\ and\ \citenamefont
  {Gambetta}}]{Kandala17Hardware}%
  \BibitemOpen
  \bibfield  {author} {\bibinfo {author} {\bibfnamefont {Abhinav}\ \bibnamefont
  {Kandala}}, \bibinfo {author} {\bibfnamefont {Antonio}\ \bibnamefont
  {Mezzacapo}}, \bibinfo {author} {\bibfnamefont {Kristan}\ \bibnamefont
  {Temme}}, \bibinfo {author} {\bibfnamefont {Maika}\ \bibnamefont {Takita}},
  \bibinfo {author} {\bibfnamefont {Markus}\ \bibnamefont {Brink}}, \bibinfo
  {author} {\bibfnamefont {Jerry~M.}\ \bibnamefont {Chow}}, \ and\ \bibinfo
  {author} {\bibfnamefont {Jay~M.}\ \bibnamefont {Gambetta}},\ }\bibfield
  {title} {\enquote {\bibinfo {title} {Hardware-efficient variational quantum
  eigensolver for small molecules and quantum magnets},}\ }\href
  {https://www.nature.com/articles/nature23879} {\bibfield  {journal} {\bibinfo
   {journal} {Nature}\ }\textbf {\bibinfo {volume} {549}},\ \bibinfo {pages}
  {242--246} (\bibinfo {year} {2017})}\BibitemShut {NoStop}%
\bibitem [{\citenamefont {Hempel}\ \emph {et~al.}(2018)\citenamefont {Hempel},
  \citenamefont {Maier}, \citenamefont {Romero}, \citenamefont {McClean},
  \citenamefont {Monz}, \citenamefont {Shen}, \citenamefont {Jurcevic},
  \citenamefont {Lanyon}, \citenamefont {Love}, \citenamefont {Babbush},
  \citenamefont {Aspuru-Guzik}, \citenamefont {Blatt},\ and\ \citenamefont
  {Roos}}]{Hempel18Quantum}%
  \BibitemOpen
  \bibfield  {author} {\bibinfo {author} {\bibfnamefont {Cornelius}\
  \bibnamefont {Hempel}}, \bibinfo {author} {\bibfnamefont {Christine}\
  \bibnamefont {Maier}}, \bibinfo {author} {\bibfnamefont {Jonathan}\
  \bibnamefont {Romero}}, \bibinfo {author} {\bibfnamefont {Jarrod}\
  \bibnamefont {McClean}}, \bibinfo {author} {\bibfnamefont {Thomas}\
  \bibnamefont {Monz}}, \bibinfo {author} {\bibfnamefont {Heng}\ \bibnamefont
  {Shen}}, \bibinfo {author} {\bibfnamefont {Petar}\ \bibnamefont {Jurcevic}},
  \bibinfo {author} {\bibfnamefont {Ben}\ \bibnamefont {Lanyon}}, \bibinfo
  {author} {\bibfnamefont {Peter}\ \bibnamefont {Love}}, \bibinfo {author}
  {\bibfnamefont {Ryan}\ \bibnamefont {Babbush}}, \bibinfo {author}
  {\bibfnamefont {Alan}\ \bibnamefont {Aspuru-Guzik}}, \bibinfo {author}
  {\bibfnamefont {Rainer}\ \bibnamefont {Blatt}}, \ and\ \bibinfo {author}
  {\bibfnamefont {Christian}\ \bibnamefont {Roos}},\ }\bibfield  {title}
  {\enquote {\bibinfo {title} {Quantum chemistry calculations on a trapped-ion
  quantum simulator},}\ }\href
  {https://journals.aps.org/prx/abstract/10.1103/PhysRevX.8.031022} {\bibfield
  {journal} {\bibinfo  {journal} {Phys. Rev. X}\ }\textbf {\bibinfo {volume}
  {8}},\ \bibinfo {pages} {031022} (\bibinfo {year} {2018})}\BibitemShut
  {NoStop}%
\bibitem [{\citenamefont {Otterbach}\ \emph {et~al.}(2017)\citenamefont
  {Otterbach}, \citenamefont {Manenti}, \citenamefont {Alidoust}, \citenamefont
  {Bestwick}, \citenamefont {Block}, \citenamefont {Bloom}, \citenamefont
  {Caldwell}, \citenamefont {Didier}, \citenamefont {Fried}, \citenamefont
  {Hong}, \citenamefont {Karalekas}, \citenamefont {Osborn}, \citenamefont
  {Papageorge}, \citenamefont {Peterson}, \citenamefont {Prawiroatmodjo},
  \citenamefont {Rubin}, \citenamefont {Ryan}, \citenamefont {Scarabelli},
  \citenamefont {Scheer}, \citenamefont {Sete}, \citenamefont {Sivarajah},
  \citenamefont {Smith}, \citenamefont {Staley}, \citenamefont {Tezak},
  \citenamefont {Zeng}, \citenamefont {Hudson}, \citenamefont {Johnson},
  \citenamefont {Reagor}, \citenamefont {da~Silva},\ and\ \citenamefont
  {Rigetti}}]{Otterbach17Unsupervised}%
  \BibitemOpen
  \bibfield  {author} {\bibinfo {author} {\bibfnamefont {J.~S.}\ \bibnamefont
  {Otterbach}}, \bibinfo {author} {\bibfnamefont {R.}~\bibnamefont {Manenti}},
  \bibinfo {author} {\bibfnamefont {N.}~\bibnamefont {Alidoust}}, \bibinfo
  {author} {\bibfnamefont {A.}~\bibnamefont {Bestwick}}, \bibinfo {author}
  {\bibfnamefont {M.}~\bibnamefont {Block}}, \bibinfo {author} {\bibfnamefont
  {B.}~\bibnamefont {Bloom}}, \bibinfo {author} {\bibfnamefont
  {S.}~\bibnamefont {Caldwell}}, \bibinfo {author} {\bibfnamefont
  {N.}~\bibnamefont {Didier}}, \bibinfo {author} {\bibfnamefont {E.~Schuyler}\
  \bibnamefont {Fried}}, \bibinfo {author} {\bibfnamefont {S.}~\bibnamefont
  {Hong}}, \bibinfo {author} {\bibfnamefont {P.}~\bibnamefont {Karalekas}},
  \bibinfo {author} {\bibfnamefont {C.~B.}\ \bibnamefont {Osborn}}, \bibinfo
  {author} {\bibfnamefont {A.}~\bibnamefont {Papageorge}}, \bibinfo {author}
  {\bibfnamefont {E.~C.}\ \bibnamefont {Peterson}}, \bibinfo {author}
  {\bibfnamefont {G.}~\bibnamefont {Prawiroatmodjo}}, \bibinfo {author}
  {\bibfnamefont {N.}~\bibnamefont {Rubin}}, \bibinfo {author} {\bibfnamefont
  {Colm~A.}\ \bibnamefont {Ryan}}, \bibinfo {author} {\bibfnamefont
  {D.}~\bibnamefont {Scarabelli}}, \bibinfo {author} {\bibfnamefont
  {M.}~\bibnamefont {Scheer}}, \bibinfo {author} {\bibfnamefont {E.~A.}\
  \bibnamefont {Sete}}, \bibinfo {author} {\bibfnamefont {P.}~\bibnamefont
  {Sivarajah}}, \bibinfo {author} {\bibfnamefont {Robert~S.}\ \bibnamefont
  {Smith}}, \bibinfo {author} {\bibfnamefont {A.}~\bibnamefont {Staley}},
  \bibinfo {author} {\bibfnamefont {N.}~\bibnamefont {Tezak}}, \bibinfo
  {author} {\bibfnamefont {W.~J.}\ \bibnamefont {Zeng}}, \bibinfo {author}
  {\bibfnamefont {A.}~\bibnamefont {Hudson}}, \bibinfo {author} {\bibfnamefont
  {Blake~R.}\ \bibnamefont {Johnson}}, \bibinfo {author} {\bibfnamefont
  {M.}~\bibnamefont {Reagor}}, \bibinfo {author} {\bibfnamefont {M.~P.}\
  \bibnamefont {da~Silva}}, \ and\ \bibinfo {author} {\bibfnamefont
  {C.}~\bibnamefont {Rigetti}},\ }\bibfield  {title} {\enquote {\bibinfo
  {title} {Unsupervised machine learning on a hybrid quantum computer},}\
  }\href {https://arxiv.org/abs/1712.05771} {\bibfield  {journal} {\bibinfo
  {journal} {ArXiv:1712.05771}\ } (\bibinfo {year} {2017})}\BibitemShut
  {NoStop}%
\bibitem [{\citenamefont {Gottesman}(1998)}]{Gottesman97Theory}%
  \BibitemOpen
  \bibfield  {author} {\bibinfo {author} {\bibfnamefont {Daniel}\ \bibnamefont
  {Gottesman}},\ }\bibfield  {title} {\enquote {\bibinfo {title} {Theory of
  fault-tolerant quantum computation},}\ }\href
  {https://arxiv.org/abs/quant-ph/9702029} {\bibfield  {journal} {\bibinfo
  {journal} {Phys. Rev. A}\ }\textbf {\bibinfo {volume} {57}} (\bibinfo {year}
  {1998})}\BibitemShut {NoStop}%
\bibitem [{\citenamefont {Kitaev}(2003)}]{Kitaev97Fault}%
  \BibitemOpen
  \bibfield  {author} {\bibinfo {author} {\bibfnamefont {A.~Yu.}\ \bibnamefont
  {Kitaev}},\ }\bibfield  {title} {\enquote {\bibinfo {title} {Fault-tolerant
  quantum computation by anyons},}\ }\href@noop {} {\bibfield  {journal}
  {\bibinfo  {journal} {Ann. Phys.}\ }\textbf {\bibinfo {volume} {303}},\
  \bibinfo {pages} {2--30} (\bibinfo {year} {2003})}\BibitemShut {NoStop}%
\bibitem [{\citenamefont {Fowler}\ \emph {et~al.}(2012)\citenamefont {Fowler},
  \citenamefont {Mariantoni}, \citenamefont {Martinis},\ and\ \citenamefont
  {Cleland}}]{Fowler12Surface}%
  \BibitemOpen
  \bibfield  {author} {\bibinfo {author} {\bibfnamefont {Austin~G.}\
  \bibnamefont {Fowler}}, \bibinfo {author} {\bibfnamefont {Matteo}\
  \bibnamefont {Mariantoni}}, \bibinfo {author} {\bibfnamefont {John~M.}\
  \bibnamefont {Martinis}}, \ and\ \bibinfo {author} {\bibfnamefont
  {Andrew~N.}\ \bibnamefont {Cleland}},\ }\bibfield  {title} {\enquote
  {\bibinfo {title} {Surface codes: Towards practical large-scale quantum
  computation},}\ }\href {https://arxiv.org/abs/1208.0928} {\bibfield
  {journal} {\bibinfo  {journal} {Phys. Rev. A}\ }\textbf {\bibinfo {volume}
  {86}},\ \bibinfo {pages} {032324} (\bibinfo {year} {2012})}\BibitemShut
  {NoStop}%
\bibitem [{\citenamefont {Berry}\ \emph {et~al.}(2019)\citenamefont {Berry},
  \citenamefont {Gidney}, \citenamefont {Motta}, \citenamefont {McClean},\ and\
  \citenamefont {Babbush}}]{Berry19Qubitization}%
  \BibitemOpen
  \bibfield  {author} {\bibinfo {author} {\bibfnamefont {Dominic~W.}\
  \bibnamefont {Berry}}, \bibinfo {author} {\bibfnamefont {Craig}\ \bibnamefont
  {Gidney}}, \bibinfo {author} {\bibfnamefont {Mario}\ \bibnamefont {Motta}},
  \bibinfo {author} {\bibfnamefont {Jarrod~R.}\ \bibnamefont {McClean}}, \ and\
  \bibinfo {author} {\bibfnamefont {Ryan}\ \bibnamefont {Babbush}},\ }\bibfield
   {title} {\enquote {\bibinfo {title} {Qubitization of arbitrary basis quantum
  chemistry leveraging sparsity and low rank factorization},}\ }\href
  {https://quantum-journal.org/papers/q-2019-12-02-208/} {\bibfield  {journal}
  {\bibinfo  {journal} {Quantum}\ }\textbf {\bibinfo {volume} {3}} (\bibinfo
  {year} {2019})}\BibitemShut {NoStop}%
\bibitem [{\citenamefont {Gidney}\ and\ \citenamefont
  {Ekerå}(2019)}]{Gidney19How}%
  \BibitemOpen
  \bibfield  {author} {\bibinfo {author} {\bibfnamefont {Craig}\ \bibnamefont
  {Gidney}}\ and\ \bibinfo {author} {\bibfnamefont {Martin}\ \bibnamefont
  {Ekerå}},\ }\bibfield  {title} {\enquote {\bibinfo {title} {How to factor
  2048 bit rsa integers in 8 hours using 20 million noisy qubits},}\ }\href
  {https://arxiv.org/abs/1905.09749} {\bibfield  {journal} {\bibinfo  {journal}
  {ArXiv:1905.09749}\ } (\bibinfo {year} {2019})}\BibitemShut {NoStop}%
\bibitem [{\citenamefont {von Burg}\ \emph {et~al.}(2020)\citenamefont {von
  Burg}, \citenamefont {Low}, \citenamefont {Häner}, \citenamefont {Steiger},
  \citenamefont {Reiher}, \citenamefont {Roetteler},\ and\ \citenamefont
  {Troyer}}]{Burg20Quantum}%
  \BibitemOpen
  \bibfield  {author} {\bibinfo {author} {\bibfnamefont {Vera}\ \bibnamefont
  {von Burg}}, \bibinfo {author} {\bibfnamefont {Guang~Hao}\ \bibnamefont
  {Low}}, \bibinfo {author} {\bibfnamefont {Thomas}\ \bibnamefont {Häner}},
  \bibinfo {author} {\bibfnamefont {Damian~S.}\ \bibnamefont {Steiger}},
  \bibinfo {author} {\bibfnamefont {Markus}\ \bibnamefont {Reiher}}, \bibinfo
  {author} {\bibfnamefont {Martin}\ \bibnamefont {Roetteler}}, \ and\ \bibinfo
  {author} {\bibfnamefont {Matthias}\ \bibnamefont {Troyer}},\ }\bibfield
  {title} {\enquote {\bibinfo {title} {Quantum computing enhanced computational
  catalysis},}\ }\href {https://arxiv.org/abs/2007.14460} {\bibfield  {journal}
  {\bibinfo  {journal} {ArXiv:2007.14460}\ } (\bibinfo {year}
  {2020})}\BibitemShut {NoStop}%
\bibitem [{\citenamefont {Sanders}\ \emph {et~al.}(2020)\citenamefont
  {Sanders}, \citenamefont {Berry}, \citenamefont {Costa}, \citenamefont
  {Tessler}, \citenamefont {Wiebe}, \citenamefont {Gidney}, \citenamefont
  {Neven},\ and\ \citenamefont {Babbush}}]{Sanders20Compilation}%
  \BibitemOpen
  \bibfield  {author} {\bibinfo {author} {\bibfnamefont {Yuval~R.}\
  \bibnamefont {Sanders}}, \bibinfo {author} {\bibfnamefont {Dominic~W.}\
  \bibnamefont {Berry}}, \bibinfo {author} {\bibfnamefont {Pedro C.~S.}\
  \bibnamefont {Costa}}, \bibinfo {author} {\bibfnamefont {Louis~W.}\
  \bibnamefont {Tessler}}, \bibinfo {author} {\bibfnamefont {Nathan}\
  \bibnamefont {Wiebe}}, \bibinfo {author} {\bibfnamefont {Craig}\ \bibnamefont
  {Gidney}}, \bibinfo {author} {\bibfnamefont {Hartmut}\ \bibnamefont {Neven}},
  \ and\ \bibinfo {author} {\bibfnamefont {Ryan}\ \bibnamefont {Babbush}},\
  }\bibfield  {title} {\enquote {\bibinfo {title} {Compilation of
  fault-tolerant quantum heuristics for combinatorial optimization},}\ }\href
  {https://arxiv.org/abs/2007.07391} {\bibfield  {journal} {\bibinfo  {journal}
  {ArXiv:2007.07391}\ } (\bibinfo {year} {2020})}\BibitemShut {NoStop}%
\bibitem [{\citenamefont {Campbell}\ \emph {et~al.}(2019)\citenamefont
  {Campbell}, \citenamefont {Khurana},\ and\ \citenamefont
  {Montanaro}}]{Campbell18Applying}%
  \BibitemOpen
  \bibfield  {author} {\bibinfo {author} {\bibfnamefont {Earl}\ \bibnamefont
  {Campbell}}, \bibinfo {author} {\bibfnamefont {Ankur}\ \bibnamefont
  {Khurana}}, \ and\ \bibinfo {author} {\bibfnamefont {Ashley}\ \bibnamefont
  {Montanaro}},\ }\bibfield  {title} {\enquote {\bibinfo {title} {Applying
  quantum algorithms to constraint satisfaction problems},}\ }\href
  {https://quantum-journal.org/papers/q-2019-07-18-167/} {\bibfield  {journal}
  {\bibinfo  {journal} {Quantum}\ }\textbf {\bibinfo {volume} {3}} (\bibinfo
  {year} {2019})}\BibitemShut {NoStop}%
\bibitem [{\citenamefont {Preskill}(2018)}]{Preskill18Quantum}%
  \BibitemOpen
  \bibfield  {author} {\bibinfo {author} {\bibfnamefont {John}\ \bibnamefont
  {Preskill}},\ }\bibfield  {title} {\enquote {\bibinfo {title} {Quantum
  computing in the nisq era and beyond},}\ }\href
  {https://quantum-journal.org/papers/q-2018-08-06-79/} {\bibfield  {journal}
  {\bibinfo  {journal} {Quantum}\ }\textbf {\bibinfo {volume} {2}} (\bibinfo
  {year} {2018})}\BibitemShut {NoStop}%
\bibitem [{\citenamefont {Farhi}\ \emph {et~al.}(2014)\citenamefont {Farhi},
  \citenamefont {Goldstone},\ and\ \citenamefont {Gutmann}}]{Farhi14Quantum}%
  \BibitemOpen
  \bibfield  {author} {\bibinfo {author} {\bibfnamefont {Edward}\ \bibnamefont
  {Farhi}}, \bibinfo {author} {\bibfnamefont {Jeffrey}\ \bibnamefont
  {Goldstone}}, \ and\ \bibinfo {author} {\bibfnamefont {Sam}\ \bibnamefont
  {Gutmann}},\ }\bibfield  {title} {\enquote {\bibinfo {title} {A quantum
  approximate optimization algorithm},}\ }\href
  {https://arxiv.org/abs/1411.4028} {\bibfield  {journal} {\bibinfo  {journal}
  {ArXiv:1411.4028}\ } (\bibinfo {year} {2014})}\BibitemShut {NoStop}%
\bibitem [{\citenamefont {McClean}\ \emph {et~al.}(2016)\citenamefont
  {McClean}, \citenamefont {Romero}, \citenamefont {Babbush},\ and\
  \citenamefont {Aspuru-Guzik}}]{Mcclean16Theory}%
  \BibitemOpen
  \bibfield  {author} {\bibinfo {author} {\bibfnamefont {Jarrod~R.}\
  \bibnamefont {McClean}}, \bibinfo {author} {\bibfnamefont {Jonathan}\
  \bibnamefont {Romero}}, \bibinfo {author} {\bibfnamefont {Ryan}\ \bibnamefont
  {Babbush}}, \ and\ \bibinfo {author} {\bibfnamefont {Alán}\ \bibnamefont
  {Aspuru-Guzik}},\ }\bibfield  {title} {\enquote {\bibinfo {title} {The theory
  of variational hybrid quantum-classical algorithms},}\ }\href
  {https://iopscience.iop.org/article/10.1088/1367-2630/18/2/023023} {\bibfield
   {journal} {\bibinfo  {journal} {New J. Phys.}\ }\textbf {\bibinfo {volume}
  {18}} (\bibinfo {year} {2016})}\BibitemShut {NoStop}%
\bibitem [{\citenamefont {da~Silva}\ \emph {et~al.}(2016)\citenamefont
  {da~Silva}, \citenamefont {Ludermir},\ and\ \citenamefont
  {de~Oliveira}}]{Silva16Quantum}%
  \BibitemOpen
  \bibfield  {author} {\bibinfo {author} {\bibfnamefont {Adenilton~J.}\
  \bibnamefont {da~Silva}}, \bibinfo {author} {\bibfnamefont {Teresa~B.}\
  \bibnamefont {Ludermir}}, \ and\ \bibinfo {author} {\bibfnamefont
  {Wilson~R.}\ \bibnamefont {de~Oliveira}},\ }\bibfield  {title} {\enquote
  {\bibinfo {title} {Quantum perceptron over a field and neural network
  architecture selection in a quantum computer},}\ }\href
  {https://arxiv.org/abs/1602.00709} {\bibfield  {journal} {\bibinfo  {journal}
  {Neu. Net.}\ }\textbf {\bibinfo {volume} {76}},\ \bibinfo {pages} {55--64}
  (\bibinfo {year} {2016})}\BibitemShut {NoStop}%
\bibitem [{\citenamefont {Chen}\ \emph {et~al.}(2018)\citenamefont {Chen},
  \citenamefont {Wossnig}, \citenamefont {Severini}, \citenamefont {Neven},\
  and\ \citenamefont {Mohseni}}]{Chen18Universal}%
  \BibitemOpen
  \bibfield  {author} {\bibinfo {author} {\bibfnamefont {Hongxiang}\
  \bibnamefont {Chen}}, \bibinfo {author} {\bibfnamefont {Leonard}\
  \bibnamefont {Wossnig}}, \bibinfo {author} {\bibfnamefont {Simone}\
  \bibnamefont {Severini}}, \bibinfo {author} {\bibfnamefont {Hartmut}\
  \bibnamefont {Neven}}, \ and\ \bibinfo {author} {\bibfnamefont {Masoud}\
  \bibnamefont {Mohseni}},\ }\bibfield  {title} {\enquote {\bibinfo {title}
  {Universal discriminative quantum neural networks},}\ }\href
  {https://arxiv.org/abs/1805.08654} {\bibfield  {journal} {\bibinfo  {journal}
  {ArXiv:1805.08654}\ } (\bibinfo {year} {2018})}\BibitemShut {NoStop}%
\bibitem [{\citenamefont {Bravyi}\ \emph {et~al.}(2019)\citenamefont {Bravyi},
  \citenamefont {Gosset},\ and\ \citenamefont {Movassagh}}]{Bravyi19Classical}%
  \BibitemOpen
  \bibfield  {author} {\bibinfo {author} {\bibfnamefont {Sergey}\ \bibnamefont
  {Bravyi}}, \bibinfo {author} {\bibfnamefont {David}\ \bibnamefont {Gosset}},
  \ and\ \bibinfo {author} {\bibfnamefont {Ramis}\ \bibnamefont {Movassagh}},\
  }\bibfield  {title} {\enquote {\bibinfo {title} {Classical algorithms for
  quantum mean values},}\ }\href {https://arxiv.org/abs/1909.11485} {\bibfield
  {journal} {\bibinfo  {journal} {ArXiv:1909.11485}\ } (\bibinfo {year}
  {2019})}\BibitemShut {NoStop}%
\bibitem [{\citenamefont {Sagastizabal}\ \emph {et~al.}(2019)\citenamefont
  {Sagastizabal}, \citenamefont {Bonet-Monroig}, \citenamefont {Singh},
  \citenamefont {Rol}, \citenamefont {Bultink}, \citenamefont {Fu},
  \citenamefont {Price}, \citenamefont {Ostroukh}, \citenamefont
  {Muthusubramanian}, \citenamefont {Bruno}, \citenamefont {Beekman},
  \citenamefont {Haider}, \citenamefont {O'Brien},\ and\ \citenamefont
  {DiCarlo}}]{Sagastizabal19Error}%
  \BibitemOpen
  \bibfield  {author} {\bibinfo {author} {\bibfnamefont {R.}~\bibnamefont
  {Sagastizabal}}, \bibinfo {author} {\bibfnamefont {X.}~\bibnamefont
  {Bonet-Monroig}}, \bibinfo {author} {\bibfnamefont {M.}~\bibnamefont
  {Singh}}, \bibinfo {author} {\bibfnamefont {M.A.}\ \bibnamefont {Rol}},
  \bibinfo {author} {\bibfnamefont {C.C.}\ \bibnamefont {Bultink}}, \bibinfo
  {author} {\bibfnamefont {X.}~\bibnamefont {Fu}}, \bibinfo {author}
  {\bibfnamefont {C.H.}\ \bibnamefont {Price}}, \bibinfo {author}
  {\bibfnamefont {V.P.}\ \bibnamefont {Ostroukh}}, \bibinfo {author}
  {\bibfnamefont {N.}~\bibnamefont {Muthusubramanian}}, \bibinfo {author}
  {\bibfnamefont {A.}~\bibnamefont {Bruno}}, \bibinfo {author} {\bibfnamefont
  {M.}~\bibnamefont {Beekman}}, \bibinfo {author} {\bibfnamefont
  {N.}~\bibnamefont {Haider}}, \bibinfo {author} {\bibfnamefont {T.E.}\
  \bibnamefont {O'Brien}}, \ and\ \bibinfo {author} {\bibfnamefont
  {L.}~\bibnamefont {DiCarlo}},\ }\bibfield  {title} {\enquote {\bibinfo
  {title} {Error mitigation by symmetry verification on a variational quantum
  eigensolver},}\ }\href {https://arxiv.org/abs/1902.11258} {\bibfield
  {journal} {\bibinfo  {journal} {Phys. Rev. A}\ }\textbf {\bibinfo {volume}
  {100}},\ \bibinfo {pages} {010302} (\bibinfo {year} {2019})}\BibitemShut
  {NoStop}%
\bibitem [{\citenamefont {McClean}\ \emph {et~al.}(2017)\citenamefont
  {McClean}, \citenamefont {Schwartz}, \citenamefont {Carter},\ and\
  \citenamefont {de~Jong}}]{Mcclean17Hybrid}%
  \BibitemOpen
  \bibfield  {author} {\bibinfo {author} {\bibfnamefont {Jarrod~R.}\
  \bibnamefont {McClean}}, \bibinfo {author} {\bibfnamefont {Mollie~E.}\
  \bibnamefont {Schwartz}}, \bibinfo {author} {\bibfnamefont {Jonathan}\
  \bibnamefont {Carter}}, \ and\ \bibinfo {author} {\bibfnamefont {Wibe~A.}\
  \bibnamefont {de~Jong}},\ }\bibfield  {title} {\enquote {\bibinfo {title}
  {Hybrid quantum-classical hierarchy for mitigation of decoherence and
  determination of excited states},}\ }\href {https://arxiv.org/abs/1603.05681}
  {\bibfield  {journal} {\bibinfo  {journal} {Phys. Rev. A}\ }\textbf {\bibinfo
  {volume} {95}},\ \bibinfo {pages} {042308} (\bibinfo {year}
  {2017})}\BibitemShut {NoStop}%
\bibitem [{\citenamefont {McClean}\ \emph
  {et~al.}(2020{\natexlab{a}})\citenamefont {McClean}, \citenamefont {Jiang},
  \citenamefont {Rubin}, \citenamefont {Babbush},\ and\ \citenamefont
  {Neven}}]{mcclean2020decoding}%
  \BibitemOpen
  \bibfield  {author} {\bibinfo {author} {\bibfnamefont {Jarrod~R}\
  \bibnamefont {McClean}}, \bibinfo {author} {\bibfnamefont {Zhang}\
  \bibnamefont {Jiang}}, \bibinfo {author} {\bibfnamefont {Nicholas~C}\
  \bibnamefont {Rubin}}, \bibinfo {author} {\bibfnamefont {Ryan}\ \bibnamefont
  {Babbush}}, \ and\ \bibinfo {author} {\bibfnamefont {Hartmut}\ \bibnamefont
  {Neven}},\ }\bibfield  {title} {\enquote {\bibinfo {title} {Decoding quantum
  errors with subspace expansions},}\ }\href@noop {} {\bibfield  {journal}
  {\bibinfo  {journal} {Nature Communications}\ }\textbf {\bibinfo {volume}
  {11}},\ \bibinfo {pages} {1--9} (\bibinfo {year}
  {2020}{\natexlab{a}})}\BibitemShut {NoStop}%
\bibitem [{\citenamefont {Temme}\ \emph {et~al.}(2017)\citenamefont {Temme},
  \citenamefont {Bravyi},\ and\ \citenamefont {Gambetta}}]{Temme16Error}%
  \BibitemOpen
  \bibfield  {author} {\bibinfo {author} {\bibfnamefont {Kristan}\ \bibnamefont
  {Temme}}, \bibinfo {author} {\bibfnamefont {Sergey}\ \bibnamefont {Bravyi}},
  \ and\ \bibinfo {author} {\bibfnamefont {Jay~M.}\ \bibnamefont {Gambetta}},\
  }\bibfield  {title} {\enquote {\bibinfo {title} {Error mitigation for
  short-depth quantum circuits},}\ }\href {https://arxiv.org/abs/1612.02058}
  {\bibfield  {journal} {\bibinfo  {journal} {Phys. Rev. Lett.}\ }\textbf
  {\bibinfo {volume} {119}},\ \bibinfo {pages} {180509} (\bibinfo {year}
  {2017})}\BibitemShut {NoStop}%
\bibitem [{\citenamefont {Kandala}\ \emph {et~al.}(2019)\citenamefont
  {Kandala}, \citenamefont {Temme}, \citenamefont {Córcoles}, \citenamefont
  {Mezzacapo}, \citenamefont {Chow},\ and\ \citenamefont
  {Gambetta}}]{Kandala19Error}%
  \BibitemOpen
  \bibfield  {author} {\bibinfo {author} {\bibfnamefont {Abhinav}\ \bibnamefont
  {Kandala}}, \bibinfo {author} {\bibfnamefont {Kristan}\ \bibnamefont
  {Temme}}, \bibinfo {author} {\bibfnamefont {Antonio~D.}\ \bibnamefont
  {Córcoles}}, \bibinfo {author} {\bibfnamefont {Antonio}\ \bibnamefont
  {Mezzacapo}}, \bibinfo {author} {\bibfnamefont {Jerry~M.}\ \bibnamefont
  {Chow}}, \ and\ \bibinfo {author} {\bibfnamefont {Jay~M.}\ \bibnamefont
  {Gambetta}},\ }\bibfield  {title} {\enquote {\bibinfo {title} {Error
  mitigation extends the computational reach of a noisy quantum processor},}\
  }\href {https://www.nature.com/articles/s41586-019-1040-7} {\bibfield
  {journal} {\bibinfo  {journal} {Nature}\ }\textbf {\bibinfo {volume} {567}},\
  \bibinfo {pages} {491--495} (\bibinfo {year} {2019})}\BibitemShut {NoStop}%
\bibitem [{\citenamefont {Endo}\ \emph {et~al.}(2018)\citenamefont {Endo},
  \citenamefont {Benjamin},\ and\ \citenamefont {Li}}]{Endo17Practical}%
  \BibitemOpen
  \bibfield  {author} {\bibinfo {author} {\bibfnamefont {Suguru}\ \bibnamefont
  {Endo}}, \bibinfo {author} {\bibfnamefont {Simon~C.}\ \bibnamefont
  {Benjamin}}, \ and\ \bibinfo {author} {\bibfnamefont {Ying}\ \bibnamefont
  {Li}},\ }\bibfield  {title} {\enquote {\bibinfo {title} {Practical quantum
  error mitigation for near-future applications},}\ }\href
  {https://journals.aps.org/prx/abstract/10.1103/PhysRevX.8.031027} {\bibfield
  {journal} {\bibinfo  {journal} {Phys. Rev. X}\ }\textbf {\bibinfo {volume}
  {8}},\ \bibinfo {pages} {031027} (\bibinfo {year} {2018})}\BibitemShut
  {NoStop}%
\bibitem [{\citenamefont {Rubin}\ \emph {et~al.}(2018)\citenamefont {Rubin},
  \citenamefont {Babbush},\ and\ \citenamefont {McClean}}]{Rubin18Application}%
  \BibitemOpen
  \bibfield  {author} {\bibinfo {author} {\bibfnamefont {Nicholas~C.}\
  \bibnamefont {Rubin}}, \bibinfo {author} {\bibfnamefont {Ryan}\ \bibnamefont
  {Babbush}}, \ and\ \bibinfo {author} {\bibfnamefont {Jarrod}\ \bibnamefont
  {McClean}},\ }\bibfield  {title} {\enquote {\bibinfo {title} {Application of
  fermionic marginal constraints to hybrid quantum algorithms},}\ }\href
  {https://iopscience.iop.org/article/10.1088/1367-2630/aab919} {\bibfield
  {journal} {\bibinfo  {journal} {New J. Phys.}\ }\textbf {\bibinfo {volume}
  {20}},\ \bibinfo {pages} {053020} (\bibinfo {year} {2018})}\BibitemShut
  {NoStop}%
\bibitem [{\citenamefont {McArdle}\ \emph {et~al.}(2019)\citenamefont
  {McArdle}, \citenamefont {Yuan},\ and\ \citenamefont
  {Benjamin}}]{Mcardle19Error}%
  \BibitemOpen
  \bibfield  {author} {\bibinfo {author} {\bibfnamefont {Sam}\ \bibnamefont
  {McArdle}}, \bibinfo {author} {\bibfnamefont {Xiao}\ \bibnamefont {Yuan}}, \
  and\ \bibinfo {author} {\bibfnamefont {Simon}\ \bibnamefont {Benjamin}},\
  }\bibfield  {title} {\enquote {\bibinfo {title} {Error-mitigated digital
  quantum simulation},}\ }\href {https://arxiv.org/abs/1807.02467} {\bibfield
  {journal} {\bibinfo  {journal} {Phys. Rev. Lett.}\ }\textbf {\bibinfo
  {volume} {122}},\ \bibinfo {pages} {180501} (\bibinfo {year}
  {2019})}\BibitemShut {NoStop}%
\bibitem [{\citenamefont {Bonet-Monroig}\ \emph {et~al.}(2018)\citenamefont
  {Bonet-Monroig}, \citenamefont {Sagastizabal}, \citenamefont {Singh},\ and\
  \citenamefont {O'Brien}}]{Bonet18Low}%
  \BibitemOpen
  \bibfield  {author} {\bibinfo {author} {\bibfnamefont {X.}~\bibnamefont
  {Bonet-Monroig}}, \bibinfo {author} {\bibfnamefont {R.}~\bibnamefont
  {Sagastizabal}}, \bibinfo {author} {\bibfnamefont {M.}~\bibnamefont {Singh}},
  \ and\ \bibinfo {author} {\bibfnamefont {T.E.}\ \bibnamefont {O'Brien}},\
  }\bibfield  {title} {\enquote {\bibinfo {title} {Low-cost error mitigation by
  symmetry verification},}\ }\href {https://arxiv.org/abs/1807.10050}
  {\bibfield  {journal} {\bibinfo  {journal} {Phys. Rev. A}\ }\textbf {\bibinfo
  {volume} {98}},\ \bibinfo {pages} {062339} (\bibinfo {year}
  {2018})}\BibitemShut {NoStop}%
\bibitem [{\citenamefont {Huggins}\ \emph {et~al.}(2019)\citenamefont
  {Huggins}, \citenamefont {McClean}, \citenamefont {Rubin}, \citenamefont
  {Jiang}, \citenamefont {Wiebe}, \citenamefont {Whaley},\ and\ \citenamefont
  {Babbush}}]{Huggins19Efficient}%
  \BibitemOpen
  \bibfield  {author} {\bibinfo {author} {\bibfnamefont {William~J.}\
  \bibnamefont {Huggins}}, \bibinfo {author} {\bibfnamefont {Jarrod}\
  \bibnamefont {McClean}}, \bibinfo {author} {\bibfnamefont {Nicholas}\
  \bibnamefont {Rubin}}, \bibinfo {author} {\bibfnamefont {Zhang}\ \bibnamefont
  {Jiang}}, \bibinfo {author} {\bibfnamefont {Nathan}\ \bibnamefont {Wiebe}},
  \bibinfo {author} {\bibfnamefont {K.~Birgitta}\ \bibnamefont {Whaley}}, \
  and\ \bibinfo {author} {\bibfnamefont {Ryan}\ \bibnamefont {Babbush}},\
  }\bibfield  {title} {\enquote {\bibinfo {title} {Efficient and noise
  resilient measurements for quantum chemistry on near-term quantum
  computers},}\ }\href {https://arxiv.org/abs/1907.13117} {\bibfield  {journal}
  {\bibinfo  {journal} {ArXiv:1907.13117}\ } (\bibinfo {year}
  {2019})}\BibitemShut {NoStop}%
\bibitem [{\citenamefont {Jiang}\ \emph {et~al.}(2019)\citenamefont {Jiang},
  \citenamefont {McClean}, \citenamefont {Babbush},\ and\ \citenamefont
  {Neven}}]{Jiang19Majorana}%
  \BibitemOpen
  \bibfield  {author} {\bibinfo {author} {\bibfnamefont {Zhang}\ \bibnamefont
  {Jiang}}, \bibinfo {author} {\bibfnamefont {Jarrod}\ \bibnamefont {McClean}},
  \bibinfo {author} {\bibfnamefont {Ryan}\ \bibnamefont {Babbush}}, \ and\
  \bibinfo {author} {\bibfnamefont {Hartmut}\ \bibnamefont {Neven}},\
  }\bibfield  {title} {\enquote {\bibinfo {title} {Majorana loop stabilizer
  codes for error correction of fermionic quantum simulations},}\ }\href
  {https://arxiv.org/abs/1812.08190} {\bibfield  {journal} {\bibinfo  {journal}
  {Phys. Rev. Appl.}\ }\textbf {\bibinfo {volume} {12}},\ \bibinfo {pages}
  {064041} (\bibinfo {year} {2019})}\BibitemShut {NoStop}%
\bibitem [{\citenamefont {O'Brien}\ \emph {et~al.}(2019)\citenamefont
  {O'Brien}, \citenamefont {Tarasinski},\ and\ \citenamefont
  {Terhal}}]{Obrien19Quantum}%
  \BibitemOpen
  \bibfield  {author} {\bibinfo {author} {\bibfnamefont {Thomas~E}\
  \bibnamefont {O'Brien}}, \bibinfo {author} {\bibfnamefont {Brian}\
  \bibnamefont {Tarasinski}}, \ and\ \bibinfo {author} {\bibfnamefont
  {Barbara~M}\ \bibnamefont {Terhal}},\ }\bibfield  {title} {\enquote {\bibinfo
  {title} {Quantum phase estimation of multiple eigenvalues for small-scale
  (noisy) experiments},}\ }\href
  {https://iopscience.iop.org/article/10.1088/1367-2630/aafb8e} {\bibfield
  {journal} {\bibinfo  {journal} {New J. Phys.}\ }\textbf {\bibinfo {volume}
  {21}},\ \bibinfo {pages} {023022} (\bibinfo {year} {2019})}\BibitemShut
  {NoStop}%
\bibitem [{\citenamefont {Somma}(2019)}]{Somma19Quantum}%
  \BibitemOpen
  \bibfield  {author} {\bibinfo {author} {\bibfnamefont {Rolando~D.}\
  \bibnamefont {Somma}},\ }\bibfield  {title} {\enquote {\bibinfo {title}
  {Quantum eigenvalue estimation via time series analysis},}\ }\href
  {https://iopscience.iop.org/article/10.1088/1367-2630/ab5c60/pdf} {\bibfield
  {journal} {\bibinfo  {journal} {New J. Phys.}\ }\textbf {\bibinfo {volume}
  {21}},\ \bibinfo {pages} {123025} (\bibinfo {year} {2019})}\BibitemShut
  {NoStop}%
\bibitem [{\citenamefont {Lu}\ \emph {et~al.}(2020)\citenamefont {Lu},
  \citenamefont {Bañuls},\ and\ \citenamefont {Cirac}}]{Lu20Algorithms}%
  \BibitemOpen
  \bibfield  {author} {\bibinfo {author} {\bibfnamefont {Sirui}\ \bibnamefont
  {Lu}}, \bibinfo {author} {\bibfnamefont {Mari~Carmen}\ \bibnamefont
  {Bañuls}}, \ and\ \bibinfo {author} {\bibfnamefont {J.~Ignacio}\
  \bibnamefont {Cirac}},\ }\bibfield  {title} {\enquote {\bibinfo {title}
  {Algorithms for quantum simulation at finite energies},}\ }\href@noop {}
  {\bibfield  {journal} {\bibinfo  {journal} {ArXiv:2006.03032}\ } (\bibinfo
  {year} {2020})}\BibitemShut {NoStop}%
\bibitem [{\citenamefont {Russo}\ \emph {et~al.}(2020)\citenamefont {Russo},
  \citenamefont {Rudinger}, \citenamefont {Morrison},\ and\ \citenamefont
  {Baczewski}}]{Russo20Evaluating}%
  \BibitemOpen
  \bibfield  {author} {\bibinfo {author} {\bibfnamefont {A.E.}\ \bibnamefont
  {Russo}}, \bibinfo {author} {\bibfnamefont {K.M.}\ \bibnamefont {Rudinger}},
  \bibinfo {author} {\bibfnamefont {B.C.A.}\ \bibnamefont {Morrison}}, \ and\
  \bibinfo {author} {\bibfnamefont {A.D.}\ \bibnamefont {Baczewski}},\
  }\bibfield  {title} {\enquote {\bibinfo {title} {Evaluating energy
  differences on a quantum computer with robust phase estimation},}\
  }\href@noop {} {\bibfield  {journal} {\bibinfo  {journal} {ArXiv:2007.08697}\
  } (\bibinfo {year} {2020})}\BibitemShut {NoStop}%
\bibitem [{\citenamefont {Berry}\ \emph {et~al.}(2007)\citenamefont {Berry},
  \citenamefont {Ahokas}, \citenamefont {Cleve},\ and\ \citenamefont
  {Sanders}}]{Berry07Efficient}%
  \BibitemOpen
  \bibfield  {author} {\bibinfo {author} {\bibfnamefont {Dominic~W.}\
  \bibnamefont {Berry}}, \bibinfo {author} {\bibfnamefont {Graeme}\
  \bibnamefont {Ahokas}}, \bibinfo {author} {\bibfnamefont {Richard}\
  \bibnamefont {Cleve}}, \ and\ \bibinfo {author} {\bibfnamefont {Barry~C.}\
  \bibnamefont {Sanders}},\ }\bibfield  {title} {\enquote {\bibinfo {title}
  {Efficient quantum algorithms for simulating sparse hamiltonians},}\ }\href
  {https://arxiv.org/abs/quant-ph/0508139} {\bibfield  {journal} {\bibinfo
  {journal} {Comm. Math. Phys.}\ }\textbf {\bibinfo {volume} {270}} (\bibinfo
  {year} {2007})}\BibitemShut {NoStop}%
\bibitem [{\citenamefont {Aharonov}\ and\ \citenamefont
  {Ta-Shma}(2003)}]{Aharonov03Adiabatic}%
  \BibitemOpen
  \bibfield  {author} {\bibinfo {author} {\bibfnamefont {Dorit}\ \bibnamefont
  {Aharonov}}\ and\ \bibinfo {author} {\bibfnamefont {Amnon}\ \bibnamefont
  {Ta-Shma}},\ }\bibfield  {title} {\enquote {\bibinfo {title} {Adiabatic
  quantum state generation and statistical zero knowledge},}\ }\href
  {https://arxiv.org/abs/quant-ph/0301023} {\bibfield  {journal} {\bibinfo
  {journal} {ArXiv:quant-ph/0301023}\ } (\bibinfo {year} {2003})}\BibitemShut
  {NoStop}%
\bibitem [{\citenamefont {Whitfield}\ \emph {et~al.}(2011)\citenamefont
  {Whitfield}, \citenamefont {Biamonte},\ and\ \citenamefont
  {Aspuru-Guzik}}]{Whitfield11Simulation}%
  \BibitemOpen
  \bibfield  {author} {\bibinfo {author} {\bibfnamefont {James~D.}\
  \bibnamefont {Whitfield}}, \bibinfo {author} {\bibfnamefont {Jacob}\
  \bibnamefont {Biamonte}}, \ and\ \bibinfo {author} {\bibfnamefont {Alán}\
  \bibnamefont {Aspuru-Guzik}},\ }\bibfield  {title} {\enquote {\bibinfo
  {title} {Simulation of electronic structure hamiltonians using quantum
  computers},}\ }\href {https://arxiv.org/abs/1001.3855} {\bibfield  {journal}
  {\bibinfo  {journal} {Mol. Phys.}\ }\textbf {\bibinfo {volume} {109}},\
  \bibinfo {pages} {735--750} (\bibinfo {year} {2011})}\BibitemShut {NoStop}%
\bibitem [{\citenamefont {Nielsen}\ and\ \citenamefont
  {Chuang}(2000)}]{Nielsen00Quantum}%
  \BibitemOpen
  \bibfield  {author} {\bibinfo {author} {\bibfnamefont {M.A.}\ \bibnamefont
  {Nielsen}}\ and\ \bibinfo {author} {\bibfnamefont {I.L.}\ \bibnamefont
  {Chuang}},\ }\href {https://books.google.de/books?id=65FqEKQOfP8C} {\emph
  {\bibinfo {title} {Quantum Computation and Quantum Information}}},\ Cambridge
  Series on Information and the Natural Sciences\ (\bibinfo  {publisher}
  {Cambridge University Press},\ \bibinfo {year} {2000})\BibitemShut {NoStop}%
\bibitem [{\citenamefont {Higgins}\ \emph {et~al.}(2009)\citenamefont
  {Higgins}, \citenamefont {Berry}, \citenamefont {Bartlett}, \citenamefont
  {Mitchell}, \citenamefont {Wiseman},\ and\ \citenamefont
  {Pryde.}}]{Higgins09Demonstrating}%
  \BibitemOpen
  \bibfield  {author} {\bibinfo {author} {\bibfnamefont {B.~L.}\ \bibnamefont
  {Higgins}}, \bibinfo {author} {\bibfnamefont {D.~W.}\ \bibnamefont {Berry}},
  \bibinfo {author} {\bibfnamefont {S.~D.}\ \bibnamefont {Bartlett}}, \bibinfo
  {author} {\bibfnamefont {M.~W.}\ \bibnamefont {Mitchell}}, \bibinfo {author}
  {\bibfnamefont {H.~M.}\ \bibnamefont {Wiseman}}, \ and\ \bibinfo {author}
  {\bibfnamefont {G.~J.}\ \bibnamefont {Pryde.}},\ }\bibfield  {title}
  {\enquote {\bibinfo {title} {Demonstrating heisenberg-limited unambiguous
  phase estimation without adaptive measurements.}}\ }\href
  {https://arxiv.org/abs/0809.3308} {\bibfield  {journal} {\bibinfo  {journal}
  {New J. Phys.}\ }\textbf {\bibinfo {volume} {11}},\ \bibinfo {pages} {073023}
  (\bibinfo {year} {2009})}\BibitemShut {NoStop}%
\bibitem [{\citenamefont {Wiebe}\ and\ \citenamefont
  {Granade}(2016)}]{Wiebe16Efficient}%
  \BibitemOpen
  \bibfield  {author} {\bibinfo {author} {\bibfnamefont {Nathan}\ \bibnamefont
  {Wiebe}}\ and\ \bibinfo {author} {\bibfnamefont {Chris}\ \bibnamefont
  {Granade}},\ }\bibfield  {title} {\enquote {\bibinfo {title} {Efficient
  bayesian phase estimation},}\ }\href {https://arxiv.org/abs/1508.00869}
  {\bibfield  {journal} {\bibinfo  {journal} {Phys. Rev. Lett.}\ }\textbf
  {\bibinfo {volume} {117}},\ \bibinfo {pages} {010503} (\bibinfo {year}
  {2016})}\BibitemShut {NoStop}%
\bibitem [{\citenamefont {Svore}\ \emph {et~al.}(2013)\citenamefont {Svore},
  \citenamefont {Hastings},\ and\ \citenamefont {Freedman}}]{Svore13Faster}%
  \BibitemOpen
  \bibfield  {author} {\bibinfo {author} {\bibfnamefont {Krysta~M.}\
  \bibnamefont {Svore}}, \bibinfo {author} {\bibfnamefont {Matthew~B.}\
  \bibnamefont {Hastings}}, \ and\ \bibinfo {author} {\bibfnamefont {Michael}\
  \bibnamefont {Freedman}},\ }\bibfield  {title} {\enquote {\bibinfo {title}
  {Faster phase estimation},}\ }\href {https://arxiv.org/abs/1304.0741}
  {\bibfield  {journal} {\bibinfo  {journal} {Quant. Inf. Comp.}\ }\textbf
  {\bibinfo {volume} {14}},\ \bibinfo {pages} {306--328} (\bibinfo {year}
  {2013})}\BibitemShut {NoStop}%
\bibitem [{Note1()}]{Note1}%
  \BibitemOpen
  \bibinfo {note} {This may be calculated via Cramer-Rao bounds as the
  derivative $\protect \frac {\partial g(t)}{\partial A_j}$ is bounded as a
  function of $t$, which is not true for the derivative $\protect \frac
  {\partial g(t)}{\partial E_j}$.}\BibitemShut {Stop}%
\bibitem [{\citenamefont {Roggero}(2020)}]{Roggero20Spectral}%
  \BibitemOpen
  \bibfield  {author} {\bibinfo {author} {\bibfnamefont {Alessandro}\
  \bibnamefont {Roggero}},\ }\bibfield  {title} {\enquote {\bibinfo {title}
  {Spectral density estimation with the gaussian integral transform},}\ }\href
  {https://arxiv.org/abs/2004.04889} {\bibfield  {journal} {\bibinfo  {journal}
  {ArXiv:2004.04889}\ } (\bibinfo {year} {2020})}\BibitemShut {NoStop}%
\bibitem [{Note2()}]{Note2}%
  \BibitemOpen
  \bibinfo {note} {The minimum number of points on the curve that require
  fitting is determined by the number of eigenvalues and amplitudes that need
  fitting.}\BibitemShut {Stop}%
\bibitem [{\citenamefont {Magesan}\ \emph {et~al.}(2011)\citenamefont
  {Magesan}, \citenamefont {Gambetta}, ,\ and\ \citenamefont
  {Emerson}}]{Magesan11Scalable}%
  \BibitemOpen
  \bibfield  {author} {\bibinfo {author} {\bibfnamefont {Easwar}\ \bibnamefont
  {Magesan}}, \bibinfo {author} {\bibfnamefont {J.~M.}\ \bibnamefont
  {Gambetta}}, , \ and\ \bibinfo {author} {\bibfnamefont {Joseph}\ \bibnamefont
  {Emerson}},\ }\bibfield  {title} {\enquote {\bibinfo {title} {Scalable and
  robust randomized benchmarking of quantum processes},}\ }\href
  {https://arxiv.org/abs/1009.3639} {\bibfield  {journal} {\bibinfo  {journal}
  {Phys. Rev. Lett.}\ }\textbf {\bibinfo {volume} {106}},\ \bibinfo {pages}
  {180504} (\bibinfo {year} {2011})}\BibitemShut {NoStop}%
\bibitem [{\citenamefont {Wiebe}\ \emph {et~al.}(2014)\citenamefont {Wiebe},
  \citenamefont {Granade}, \citenamefont {Ferrie},\ and\ \citenamefont
  {Cory}}]{Wiebe13Hamiltonian}%
  \BibitemOpen
  \bibfield  {author} {\bibinfo {author} {\bibfnamefont {Nathan}\ \bibnamefont
  {Wiebe}}, \bibinfo {author} {\bibfnamefont {Christopher}\ \bibnamefont
  {Granade}}, \bibinfo {author} {\bibfnamefont {Christopher}\ \bibnamefont
  {Ferrie}}, \ and\ \bibinfo {author} {\bibfnamefont {D.~G.}\ \bibnamefont
  {Cory}},\ }\bibfield  {title} {\enquote {\bibinfo {title} {Hamiltonian
  learning and certification using quantum resources},}\ }\href
  {https://arxiv.org/abs/1309.0876} {\bibfield  {journal} {\bibinfo  {journal}
  {Phys. Rev. Lett.}\ }\textbf {\bibinfo {volume} {112}},\ \bibinfo {pages}
  {190501} (\bibinfo {year} {2014})}\BibitemShut {NoStop}%
\bibitem [{\citenamefont {Wallman}\ and\ \citenamefont
  {Emerson}(2016)}]{Wallman16Noise}%
  \BibitemOpen
  \bibfield  {author} {\bibinfo {author} {\bibfnamefont {Joel~J}\ \bibnamefont
  {Wallman}}\ and\ \bibinfo {author} {\bibfnamefont {Joseph}\ \bibnamefont
  {Emerson}},\ }\bibfield  {title} {\enquote {\bibinfo {title} {Noise tailoring
  for scalable quantum computation via randomized compiling},}\ }\href@noop {}
  {\bibfield  {journal} {\bibinfo  {journal} {Physical Review A}\ }\textbf
  {\bibinfo {volume} {94}},\ \bibinfo {pages} {052325} (\bibinfo {year}
  {2016})}\BibitemShut {NoStop}%
\bibitem [{\citenamefont {Zintchenko}\ and\ \citenamefont
  {Wiebe}(2016)}]{Zintchenko16Randomized}%
  \BibitemOpen
  \bibfield  {author} {\bibinfo {author} {\bibfnamefont {Ilia}\ \bibnamefont
  {Zintchenko}}\ and\ \bibinfo {author} {\bibfnamefont {Nathan}\ \bibnamefont
  {Wiebe}},\ }\bibfield  {title} {\enquote {\bibinfo {title} {Randomized gap
  and amplitude estimation},}\ }\href@noop {} {\bibfield  {journal} {\bibinfo
  {journal} {Phys. Rev. A}\ }\textbf {\bibinfo {volume} {93}},\ \bibinfo
  {pages} {062306} (\bibinfo {year} {2016})}\BibitemShut {NoStop}%
\bibitem [{\citenamefont {Kimmel}\ \emph {et~al.}(2015)\citenamefont {Kimmel},
  \citenamefont {Low},\ and\ \citenamefont {Yoder}}]{Kimmel15Robust}%
  \BibitemOpen
  \bibfield  {author} {\bibinfo {author} {\bibfnamefont {Shelby}\ \bibnamefont
  {Kimmel}}, \bibinfo {author} {\bibfnamefont {Guang~Hao}\ \bibnamefont {Low}},
  \ and\ \bibinfo {author} {\bibfnamefont {Theodore~J.}\ \bibnamefont
  {Yoder}},\ }\bibfield  {title} {\enquote {\bibinfo {title} {Robust
  calibration of a universal single-qubit gate-set via robust phase
  estimation},}\ }\href {https://arxiv.org/abs/1502.02677} {\bibfield
  {journal} {\bibinfo  {journal} {Phys. Rev. A}\ }\textbf {\bibinfo {volume}
  {92}},\ \bibinfo {pages} {062315} (\bibinfo {year} {2015})}\BibitemShut
  {NoStop}%
\bibitem [{\citenamefont {Bonet-Monroig}\ \emph {et~al.}(2020)\citenamefont
  {Bonet-Monroig}, \citenamefont {Babbush},\ and\ \citenamefont
  {O'Brien}}]{Bonet20Nearly}%
  \BibitemOpen
  \bibfield  {author} {\bibinfo {author} {\bibfnamefont {Xavier}\ \bibnamefont
  {Bonet-Monroig}}, \bibinfo {author} {\bibfnamefont {Ryan}\ \bibnamefont
  {Babbush}}, \ and\ \bibinfo {author} {\bibfnamefont {Thomas~E}\ \bibnamefont
  {O'Brien}},\ }\bibfield  {title} {\enquote {\bibinfo {title} {Nearly optimal
  measurement scheduling for partial tomography of quantum states},}\ }\href
  {https://arxiv.org/abs/1908.05628} {\bibfield  {journal} {\bibinfo  {journal}
  {Phys. Rev. X}\ }\textbf {\bibinfo {volume} {In Press}} (\bibinfo {year}
  {2020})}\BibitemShut {NoStop}%
\bibitem [{\citenamefont {Cotler}\ and\ \citenamefont
  {Wilczek}(2020)}]{Cotler20Quantum}%
  \BibitemOpen
  \bibfield  {author} {\bibinfo {author} {\bibfnamefont {Jordan}\ \bibnamefont
  {Cotler}}\ and\ \bibinfo {author} {\bibfnamefont {Frank}\ \bibnamefont
  {Wilczek}},\ }\bibfield  {title} {\enquote {\bibinfo {title} {Quantum
  overlapping tomography},}\ }\href {https://arxiv.org/abs/1908.02754}
  {\bibfield  {journal} {\bibinfo  {journal} {Phys. Rev. Lett.}\ }\textbf
  {\bibinfo {volume} {124}} (\bibinfo {year} {2020})}\BibitemShut {NoStop}%
\bibitem [{\citenamefont {Wang}\ \emph {et~al.}(2019)\citenamefont {Wang},
  \citenamefont {Higgott},\ and\ \citenamefont {Brierley}}]{Wang19Accelerated}%
  \BibitemOpen
  \bibfield  {author} {\bibinfo {author} {\bibfnamefont {Daochen}\ \bibnamefont
  {Wang}}, \bibinfo {author} {\bibfnamefont {Oscar}\ \bibnamefont {Higgott}}, \
  and\ \bibinfo {author} {\bibfnamefont {Stephen}\ \bibnamefont {Brierley}},\
  }\bibfield  {title} {\enquote {\bibinfo {title} {Accelerated variational
  quantum eigensolver},}\ }\href {https://arxiv.org/abs/1802.00171} {\bibfield
  {journal} {\bibinfo  {journal} {Phys. Rev. Lett.}\ }\textbf {\bibinfo
  {volume} {122}},\ \bibinfo {pages} {140504} (\bibinfo {year}
  {2019})}\BibitemShut {NoStop}%
\bibitem [{\citenamefont {Santagati}\ \emph {et~al.}(2018)\citenamefont
  {Santagati}, \citenamefont {Wang}, \citenamefont {Gentile}, \citenamefont
  {Paesani}, \citenamefont {Wiebe}, \citenamefont {McClean}, \citenamefont
  {Morley-Short}, \citenamefont {Shadbolt}, \citenamefont {Bonneau},
  \citenamefont {Silverstone}, \citenamefont {Tew}, \citenamefont {Zhou},
  \citenamefont {O’Brien},\ and\ \citenamefont
  {Thompson}}]{Santagati18Witnessing}%
  \BibitemOpen
  \bibfield  {author} {\bibinfo {author} {\bibfnamefont {Raffaele}\
  \bibnamefont {Santagati}}, \bibinfo {author} {\bibfnamefont {Jianwei}\
  \bibnamefont {Wang}}, \bibinfo {author} {\bibfnamefont {Antonio~A.}\
  \bibnamefont {Gentile}}, \bibinfo {author} {\bibfnamefont {Stefano}\
  \bibnamefont {Paesani}}, \bibinfo {author} {\bibfnamefont {Nathan}\
  \bibnamefont {Wiebe}}, \bibinfo {author} {\bibfnamefont {Jarrod~R.}\
  \bibnamefont {McClean}}, \bibinfo {author} {\bibfnamefont {Sam}\ \bibnamefont
  {Morley-Short}}, \bibinfo {author} {\bibfnamefont {Peter~J.}\ \bibnamefont
  {Shadbolt}}, \bibinfo {author} {\bibfnamefont {Damien}\ \bibnamefont
  {Bonneau}}, \bibinfo {author} {\bibfnamefont {Joshua~W.}\ \bibnamefont
  {Silverstone}}, \bibinfo {author} {\bibfnamefont {David~P.}\ \bibnamefont
  {Tew}}, \bibinfo {author} {\bibfnamefont {Xiaoqi}\ \bibnamefont {Zhou}},
  \bibinfo {author} {\bibfnamefont {Jeremy~L.}\ \bibnamefont {O’Brien}}, \
  and\ \bibinfo {author} {\bibfnamefont {Mark~G.}\ \bibnamefont {Thompson}},\
  }\bibfield  {title} {\enquote {\bibinfo {title} {Witnessing eigenstates for
  quantum simulation of hamiltonian spectra},}\ }\href
  {https://advances.sciencemag.org/content/4/1/eaap9646} {\bibfield  {journal}
  {\bibinfo  {journal} {Sci. Adv.}\ }\textbf {\bibinfo {volume} {4}} (\bibinfo
  {year} {2018})}\BibitemShut {NoStop}%
\bibitem [{\citenamefont {Atia}\ and\ \citenamefont
  {Aharonov}(2017)}]{Atia17Fast}%
  \BibitemOpen
  \bibfield  {author} {\bibinfo {author} {\bibfnamefont {Yosi}\ \bibnamefont
  {Atia}}\ and\ \bibinfo {author} {\bibfnamefont {Dorit}\ \bibnamefont
  {Aharonov}},\ }\bibfield  {title} {\enquote {\bibinfo {title}
  {Fast-forwarding of hamiltonians and exponentially precise measurements},}\
  }\href {https://www.nature.com/articles/s41467-017-01637-7} {\bibfield
  {journal} {\bibinfo  {journal} {Nat. Comm.}\ }\textbf {\bibinfo {volume} {8}}
  (\bibinfo {year} {2017})}\BibitemShut {NoStop}%
\bibitem [{\citenamefont {Motta}\ \emph {et~al.}(2018)\citenamefont {Motta},
  \citenamefont {Ye}, \citenamefont {McClean}, \citenamefont {Li},
  \citenamefont {Minnich}, \citenamefont {Babbush},\ and\ \citenamefont
  {Chan}}]{Motta18Low}%
  \BibitemOpen
  \bibfield  {author} {\bibinfo {author} {\bibfnamefont {Mario}\ \bibnamefont
  {Motta}}, \bibinfo {author} {\bibfnamefont {Erika}\ \bibnamefont {Ye}},
  \bibinfo {author} {\bibfnamefont {Jarrod~R.}\ \bibnamefont {McClean}},
  \bibinfo {author} {\bibfnamefont {Zhendong}\ \bibnamefont {Li}}, \bibinfo
  {author} {\bibfnamefont {Austin~J.}\ \bibnamefont {Minnich}}, \bibinfo
  {author} {\bibfnamefont {Ryan}\ \bibnamefont {Babbush}}, \ and\ \bibinfo
  {author} {\bibfnamefont {Garnet Kin-Lic}\ \bibnamefont {Chan}},\ }\bibfield
  {title} {\enquote {\bibinfo {title} {Low rank representations for quantum
  simulation of electronic structure},}\ }\href
  {https://arxiv.org/abs/1808.02625} {\bibfield  {journal} {\bibinfo  {journal}
  {ArXiv:1808.02625}\ } (\bibinfo {year} {2018})}\BibitemShut {NoStop}%
\bibitem [{cir()}]{cirq}%
  \BibitemOpen
  \bibfield  {title} {\enquote {\bibinfo {title} {Cirq, a python framework for
  creating, editing, and invoking noisy intermediate scale quantum (nisq)
  circuits},}\ }\href@noop {} {\ }\bibinfo {note}
  {\url{https://github.com/quantumlib/Cirq}}\BibitemShut {NoStop}%
\bibitem [{\citenamefont {McClean}\ \emph
  {et~al.}(2020{\natexlab{b}})\citenamefont {McClean}, \citenamefont {Rubin},
  \citenamefont {Sung}, \citenamefont {Kivlichan}, \citenamefont
  {Bonet-Monroig}, \citenamefont {Cao}, \citenamefont {Dai}, \citenamefont
  {Fried}, \citenamefont {Gidney}, \citenamefont {Gimby}, \citenamefont
  {Gokhale}, \citenamefont {Häner}, \citenamefont {Hardikar}, \citenamefont
  {Havlíček}, \citenamefont {Higgott}, \citenamefont {Huang}, \citenamefont
  {Izaac}, \citenamefont {Jiang}, \citenamefont {Liu}, \citenamefont {McArdle},
  \citenamefont {Neeley}, \citenamefont {Thomas O'Brien~and}, \citenamefont
  {Ozfidan}, \citenamefont {Radin}, \citenamefont {Romero}, \citenamefont
  {Sawaya}, \citenamefont {Senjean}, \citenamefont {Setia}, \citenamefont
  {Sim}, \citenamefont {Steiger}, \citenamefont {Steudtner}, \citenamefont
  {Sun}, \citenamefont {Sun}, \citenamefont {Wang}, \citenamefont {Zhang}, ,\
  and\ \citenamefont {Babbush}}]{McClean20Openfermion}%
  \BibitemOpen
  \bibfield  {author} {\bibinfo {author} {\bibfnamefont {Jarrod~R}\
  \bibnamefont {McClean}}, \bibinfo {author} {\bibfnamefont {Nicholas~C}\
  \bibnamefont {Rubin}}, \bibinfo {author} {\bibfnamefont {Kevin~J}\
  \bibnamefont {Sung}}, \bibinfo {author} {\bibfnamefont {Ian~D}\ \bibnamefont
  {Kivlichan}}, \bibinfo {author} {\bibfnamefont {Xavier}\ \bibnamefont
  {Bonet-Monroig}}, \bibinfo {author} {\bibfnamefont {Yudong}\ \bibnamefont
  {Cao}}, \bibinfo {author} {\bibfnamefont {Chengyu}\ \bibnamefont {Dai}},
  \bibinfo {author} {\bibfnamefont {E~Schuyler}\ \bibnamefont {Fried}},
  \bibinfo {author} {\bibfnamefont {Craig}\ \bibnamefont {Gidney}}, \bibinfo
  {author} {\bibfnamefont {Brendan}\ \bibnamefont {Gimby}}, \bibinfo {author}
  {\bibfnamefont {Pranav}\ \bibnamefont {Gokhale}}, \bibinfo {author}
  {\bibfnamefont {Thomas}\ \bibnamefont {Häner}}, \bibinfo {author}
  {\bibfnamefont {Tarini}\ \bibnamefont {Hardikar}}, \bibinfo {author}
  {\bibfnamefont {Vojtěch}\ \bibnamefont {Havlíček}}, \bibinfo {author}
  {\bibfnamefont {Oscar}\ \bibnamefont {Higgott}}, \bibinfo {author}
  {\bibfnamefont {Cupjin}\ \bibnamefont {Huang}}, \bibinfo {author}
  {\bibfnamefont {Josh}\ \bibnamefont {Izaac}}, \bibinfo {author}
  {\bibfnamefont {Zhang}\ \bibnamefont {Jiang}}, \bibinfo {author}
  {\bibfnamefont {Xinle}\ \bibnamefont {Liu}}, \bibinfo {author} {\bibfnamefont
  {Sam}\ \bibnamefont {McArdle}}, \bibinfo {author} {\bibfnamefont {Matthew}\
  \bibnamefont {Neeley}}, \bibinfo {author} {\bibfnamefont {Bryan~O'Gorman}\
  \bibnamefont {Thomas O'Brien~and}}, \bibinfo {author} {\bibfnamefont {Isil}\
  \bibnamefont {Ozfidan}}, \bibinfo {author} {\bibfnamefont {Maxwell~D}\
  \bibnamefont {Radin}}, \bibinfo {author} {\bibfnamefont {Jhonathan}\
  \bibnamefont {Romero}}, \bibinfo {author} {\bibfnamefont {Nicolas P~D}\
  \bibnamefont {Sawaya}}, \bibinfo {author} {\bibfnamefont {Bruno}\
  \bibnamefont {Senjean}}, \bibinfo {author} {\bibfnamefont {Kanav}\
  \bibnamefont {Setia}}, \bibinfo {author} {\bibfnamefont {Sukin}\ \bibnamefont
  {Sim}}, \bibinfo {author} {\bibfnamefont {Damian~S}\ \bibnamefont {Steiger}},
  \bibinfo {author} {\bibfnamefont {Mark}\ \bibnamefont {Steudtner}}, \bibinfo
  {author} {\bibfnamefont {Qiming}\ \bibnamefont {Sun}}, \bibinfo {author}
  {\bibfnamefont {Wei}\ \bibnamefont {Sun}}, \bibinfo {author} {\bibfnamefont
  {Daochen}\ \bibnamefont {Wang}}, \bibinfo {author} {\bibfnamefont {Fang}\
  \bibnamefont {Zhang}}, , \ and\ \bibinfo {author} {\bibfnamefont {Ryan}\
  \bibnamefont {Babbush}},\ }\bibfield  {title} {\enquote {\bibinfo {title}
  {Openfermion: the electronic structure package for quantum computers},}\
  }\href {https://iopscience.iop.org/article/10.1088/2058-9565/ab8ebc}
  {\bibfield  {journal} {\bibinfo  {journal} {Quant. Sci. Tech.}\ }\textbf
  {\bibinfo {volume} {5}} (\bibinfo {year} {2020}{\natexlab{b}})}\BibitemShut
  {NoStop}%
\bibitem [{\citenamefont {Kivlichan}\ \emph {et~al.}(2018)\citenamefont
  {Kivlichan}, \citenamefont {McClean}, \citenamefont {Wiebe}, \citenamefont
  {Craig~Gidney}, \citenamefont {Chan},\ and\ \citenamefont
  {Babbush}}]{Kivlichan18Quantum}%
  \BibitemOpen
  \bibfield  {author} {\bibinfo {author} {\bibfnamefont {Ian~D.}\ \bibnamefont
  {Kivlichan}}, \bibinfo {author} {\bibfnamefont {Jarrod}\ \bibnamefont
  {McClean}}, \bibinfo {author} {\bibfnamefont {Nathan}\ \bibnamefont {Wiebe}},
  \bibinfo {author} {\bibfnamefont {Alán Aspuru-Guzik}\ \bibnamefont
  {Craig~Gidney}}, \bibinfo {author} {\bibfnamefont {Garnet Kin-Lic}\
  \bibnamefont {Chan}}, \ and\ \bibinfo {author} {\bibfnamefont {Ryan}\
  \bibnamefont {Babbush}},\ }\bibfield  {title} {\enquote {\bibinfo {title}
  {Quantum simulation of electronic structure with linear depth and
  connectivity},}\ }\href {https://arxiv.org/abs/1711.04789} {\bibfield
  {journal} {\bibinfo  {journal} {Phys. Rev. Lett.}\ }\textbf {\bibinfo
  {volume} {120}},\ \bibinfo {pages} {110501} (\bibinfo {year}
  {2018})}\BibitemShut {NoStop}%
\bibitem [{\citenamefont {Valiant}(2001)}]{Valiant01Quantum}%
  \BibitemOpen
  \bibfield  {author} {\bibinfo {author} {\bibfnamefont {Leslie~G.}\
  \bibnamefont {Valiant}},\ }\bibfield  {title} {\enquote {\bibinfo {title}
  {Quantum computers that can be simulated classically in polynomial time},}\
  }\href {http://people.seas.harvard.edu/~valiant/stoc01.pdf} {\bibfield
  {journal} {\bibinfo  {journal} {Sym. Theo. Comp.}\ }\textbf {\bibinfo
  {volume} {33}},\ \bibinfo {pages} {114--123} (\bibinfo {year}
  {2001})}\BibitemShut {NoStop}%
\bibitem [{Note3()}]{Note3}%
  \BibitemOpen
  \bibinfo {note} {Proof of this can be found in e.g. Fig.~\ref
  {fig:single_term_fsw}, where phase fitting obtained a $\sim 100$-fold
  reduction in this noise floor, which is typical for all simulations performed
  thus far.}\BibitemShut {Stop}%
\bibitem [{\citenamefont {Wecker}\ \emph {et~al.}(2015)\citenamefont {Wecker},
  \citenamefont {Hastings},\ and\ \citenamefont {Troyer}}]{Wecker15Towards}%
  \BibitemOpen
  \bibfield  {author} {\bibinfo {author} {\bibfnamefont {D.}~\bibnamefont
  {Wecker}}, \bibinfo {author} {\bibfnamefont {M.~B.}\ \bibnamefont
  {Hastings}}, \ and\ \bibinfo {author} {\bibfnamefont {M.}~\bibnamefont
  {Troyer}},\ }\bibfield  {title} {\enquote {\bibinfo {title} {Towards
  practical quantum variational algorithms},}\ }\href
  {https://arxiv.org/abs/1507.08969} {\bibfield  {journal} {\bibinfo  {journal}
  {Phys. Rev. A}\ }\textbf {\bibinfo {volume} {92}},\ \bibinfo {pages} {042303}
  (\bibinfo {year} {2015})}\BibitemShut {NoStop}%
\bibitem [{\citenamefont {Ho}\ and\ \citenamefont
  {Hsieh}(2019)}]{Ho19Efficient}%
  \BibitemOpen
  \bibfield  {author} {\bibinfo {author} {\bibfnamefont {Wen~Wei}\ \bibnamefont
  {Ho}}\ and\ \bibinfo {author} {\bibfnamefont {Timothy~H.}\ \bibnamefont
  {Hsieh}},\ }\bibfield  {title} {\enquote {\bibinfo {title} {Efficient
  variational simulation of non-trivial quantum states},}\ }\href
  {https://scipost.org/10.21468/SciPostPhys.6.3.029} {\bibfield  {journal}
  {\bibinfo  {journal} {SciPost Phys.}\ }\textbf {\bibinfo {volume} {6}}
  (\bibinfo {year} {2019})}\BibitemShut {NoStop}%
\bibitem [{\citenamefont {O’Malley}\ \emph {et~al.}(2016)\citenamefont
  {O’Malley}, \citenamefont {Babbush}, \citenamefont {Kivlichan},
  \citenamefont {Romero}, \citenamefont {McClean}, \citenamefont {Barends},
  \citenamefont {Kelly}, \citenamefont {Roushan}, \citenamefont {Tranter},
  \citenamefont {Ding}, \citenamefont {Campbell}, \citenamefont {Chen},
  \citenamefont {Chen}, \citenamefont {Chiaro}, \citenamefont {Dunsworth},
  \citenamefont {Fowler}, \citenamefont {Jeffrey}, \citenamefont {Lucero},
  \citenamefont {Megrant}, \citenamefont {Mutus}, \citenamefont {Neeley},
  \citenamefont {Neill}, \citenamefont {Quintana}, \citenamefont {Sank},
  \citenamefont {Vainsencher}, \citenamefont {Wenner}, \citenamefont {White},
  \citenamefont {Coveney}, \citenamefont {Love}, \citenamefont {Neven},
  \citenamefont {Aspuru-Guzik},\ and\ \citenamefont
  {Martinis}}]{Omalley16Scalable}%
  \BibitemOpen
  \bibfield  {author} {\bibinfo {author} {\bibfnamefont {P.J.J.}\ \bibnamefont
  {O’Malley}}, \bibinfo {author} {\bibfnamefont {R.}~\bibnamefont {Babbush}},
  \bibinfo {author} {\bibfnamefont {I.D.}\ \bibnamefont {Kivlichan}}, \bibinfo
  {author} {\bibfnamefont {J.}~\bibnamefont {Romero}}, \bibinfo {author}
  {\bibfnamefont {J.R.}\ \bibnamefont {McClean}}, \bibinfo {author}
  {\bibfnamefont {R.}~\bibnamefont {Barends}}, \bibinfo {author} {\bibfnamefont
  {J.}~\bibnamefont {Kelly}}, \bibinfo {author} {\bibfnamefont
  {P.}~\bibnamefont {Roushan}}, \bibinfo {author} {\bibfnamefont
  {A.}~\bibnamefont {Tranter}}, \bibinfo {author} {\bibfnamefont
  {N.}~\bibnamefont {Ding}}, \bibinfo {author} {\bibfnamefont {B.}~\bibnamefont
  {Campbell}}, \bibinfo {author} {\bibfnamefont {Y.}~\bibnamefont {Chen}},
  \bibinfo {author} {\bibfnamefont {Z.}~\bibnamefont {Chen}}, \bibinfo {author}
  {\bibfnamefont {B.}~\bibnamefont {Chiaro}}, \bibinfo {author} {\bibfnamefont
  {A.}~\bibnamefont {Dunsworth}}, \bibinfo {author} {\bibfnamefont {A.G.}\
  \bibnamefont {Fowler}}, \bibinfo {author} {\bibfnamefont {E.}~\bibnamefont
  {Jeffrey}}, \bibinfo {author} {\bibfnamefont {E.}~\bibnamefont {Lucero}},
  \bibinfo {author} {\bibfnamefont {A.}~\bibnamefont {Megrant}}, \bibinfo
  {author} {\bibfnamefont {J.Y.}\ \bibnamefont {Mutus}}, \bibinfo {author}
  {\bibfnamefont {M.}~\bibnamefont {Neeley}}, \bibinfo {author} {\bibfnamefont
  {C.}~\bibnamefont {Neill}}, \bibinfo {author} {\bibfnamefont
  {C.}~\bibnamefont {Quintana}}, \bibinfo {author} {\bibfnamefont
  {D.}~\bibnamefont {Sank}}, \bibinfo {author} {\bibfnamefont {A.}~\bibnamefont
  {Vainsencher}}, \bibinfo {author} {\bibfnamefont {J.}~\bibnamefont {Wenner}},
  \bibinfo {author} {\bibfnamefont {T.C.}\ \bibnamefont {White}}, \bibinfo
  {author} {\bibfnamefont {P.V.}\ \bibnamefont {Coveney}}, \bibinfo {author}
  {\bibfnamefont {P.J.}\ \bibnamefont {Love}}, \bibinfo {author} {\bibfnamefont
  {H.}~\bibnamefont {Neven}}, \bibinfo {author} {\bibfnamefont
  {A.}~\bibnamefont {Aspuru-Guzik}}, \ and\ \bibinfo {author} {\bibfnamefont
  {J.M.}\ \bibnamefont {Martinis}},\ }\bibfield  {title} {\enquote {\bibinfo
  {title} {Scalable quantum simulation of molecular energies},}\ }\href
  {https://journals.aps.org/prx/abstract/10.1103/PhysRevX.6.031007} {\bibfield
  {journal} {\bibinfo  {journal} {Phys. Rev. X}\ }\textbf {\bibinfo {volume}
  {6}},\ \bibinfo {pages} {031007} (\bibinfo {year} {2016})}\BibitemShut
  {NoStop}%
\bibitem [{\citenamefont {Bauer}\ \emph {et~al.}(2020)\citenamefont {Bauer},
  \citenamefont {Bravyi}, \citenamefont {Motta},\ and\ \citenamefont
  {Chan}}]{Bauer20Quantum}%
  \BibitemOpen
  \bibfield  {author} {\bibinfo {author} {\bibfnamefont {Bela}\ \bibnamefont
  {Bauer}}, \bibinfo {author} {\bibfnamefont {Sergey}\ \bibnamefont {Bravyi}},
  \bibinfo {author} {\bibfnamefont {Mario}\ \bibnamefont {Motta}}, \ and\
  \bibinfo {author} {\bibfnamefont {Garnet Kin-Lic}\ \bibnamefont {Chan}},\
  }\bibfield  {title} {\enquote {\bibinfo {title} {Quantum algorithms for
  quantum chemistry and quantum materials science},}\ }\href
  {https://arxiv.org/abs/2001.03685} {\bibfield  {journal} {\bibinfo  {journal}
  {ArXiv:2001.03685}\ } (\bibinfo {year} {2020})}\BibitemShut {NoStop}%
\bibitem [{Note4()}]{Note4}%
  \BibitemOpen
  \bibinfo {note} {We remind the reader here that $\rho ^{(\protect \mathrm
  {v})}$ and $\rho ^{(\protect \mathrm {f})}$ are not normalized, hence our use
  of the word 'ensemble' rather than 'state'.}\BibitemShut {Stop}%
\bibitem [{Note5()}]{Note5}%
  \BibitemOpen
  \bibinfo {note} {We have observed this for instance due to T$_1$ decay on the
  control qubit between final rotation and readout. However, we can correct for
  this easily by measuring in the opposite basis $50\%$ of the
  time}\BibitemShut {NoStop}%
\bibitem [{\citenamefont {{Virtanen}}\ \emph {et~al.}(2020)\citenamefont
  {{Virtanen}}, \citenamefont {{Gommers}}, \citenamefont {{Oliphant}},
  \citenamefont {{Haberland}}, \citenamefont {{Reddy}}, \citenamefont
  {{Cournapeau}}, \citenamefont {{Burovski}}, \citenamefont {{Peterson}},
  \citenamefont {{Weckesser}}, \citenamefont {{Bright}}, \citenamefont {{van
  der Walt}}, \citenamefont {{Brett}}, \citenamefont {{Wilson}}, \citenamefont
  {{Jarrod Millman}}, \citenamefont {{Mayorov}}, \citenamefont {{Nelson}},
  \citenamefont {{Jones}}, \citenamefont {{Kern}}, \citenamefont {{Larson}},
  \citenamefont {{Carey}}, \citenamefont {{Polat}}, \citenamefont {{Feng}},
  \citenamefont {{Moore}}, \citenamefont {{Vand erPlas}}, \citenamefont
  {{Laxalde}}, \citenamefont {{Perktold}}, \citenamefont {{Cimrman}},
  \citenamefont {{Henriksen}}, \citenamefont {{Quintero}}, \citenamefont
  {{Harris}}, \citenamefont {{Archibald}}, \citenamefont {{Ribeiro}},
  \citenamefont {{Pedregosa}}, \citenamefont {{van Mulbregt}},\ and\
  \citenamefont {{Contributors}}}]{scipy}%
  \BibitemOpen
  \bibfield  {author} {\bibinfo {author} {\bibfnamefont {Pauli}\ \bibnamefont
  {{Virtanen}}}, \bibinfo {author} {\bibfnamefont {Ralf}\ \bibnamefont
  {{Gommers}}}, \bibinfo {author} {\bibfnamefont {Travis~E.}\ \bibnamefont
  {{Oliphant}}}, \bibinfo {author} {\bibfnamefont {Matt}\ \bibnamefont
  {{Haberland}}}, \bibinfo {author} {\bibfnamefont {Tyler}\ \bibnamefont
  {{Reddy}}}, \bibinfo {author} {\bibfnamefont {David}\ \bibnamefont
  {{Cournapeau}}}, \bibinfo {author} {\bibfnamefont {Evgeni}\ \bibnamefont
  {{Burovski}}}, \bibinfo {author} {\bibfnamefont {Pearu}\ \bibnamefont
  {{Peterson}}}, \bibinfo {author} {\bibfnamefont {Warren}\ \bibnamefont
  {{Weckesser}}}, \bibinfo {author} {\bibfnamefont {Jonathan}\ \bibnamefont
  {{Bright}}}, \bibinfo {author} {\bibfnamefont {St{\'e}fan~J.}\ \bibnamefont
  {{van der Walt}}}, \bibinfo {author} {\bibfnamefont {Matthew}\ \bibnamefont
  {{Brett}}}, \bibinfo {author} {\bibfnamefont {Joshua}\ \bibnamefont
  {{Wilson}}}, \bibinfo {author} {\bibfnamefont {K.}~\bibnamefont {{Jarrod
  Millman}}}, \bibinfo {author} {\bibfnamefont {Nikolay}\ \bibnamefont
  {{Mayorov}}}, \bibinfo {author} {\bibfnamefont {Andrew R.~J.}\ \bibnamefont
  {{Nelson}}}, \bibinfo {author} {\bibfnamefont {Eric}\ \bibnamefont
  {{Jones}}}, \bibinfo {author} {\bibfnamefont {Robert}\ \bibnamefont
  {{Kern}}}, \bibinfo {author} {\bibfnamefont {Eric}\ \bibnamefont {{Larson}}},
  \bibinfo {author} {\bibfnamefont {CJ}~\bibnamefont {{Carey}}}, \bibinfo
  {author} {\bibfnamefont {{\.I}lhan}\ \bibnamefont {{Polat}}}, \bibinfo
  {author} {\bibfnamefont {Yu}~\bibnamefont {{Feng}}}, \bibinfo {author}
  {\bibfnamefont {Eric~W.}\ \bibnamefont {{Moore}}}, \bibinfo {author}
  {\bibfnamefont {Jake}\ \bibnamefont {{Vand erPlas}}}, \bibinfo {author}
  {\bibfnamefont {Denis}\ \bibnamefont {{Laxalde}}}, \bibinfo {author}
  {\bibfnamefont {Josef}\ \bibnamefont {{Perktold}}}, \bibinfo {author}
  {\bibfnamefont {Robert}\ \bibnamefont {{Cimrman}}}, \bibinfo {author}
  {\bibfnamefont {Ian}\ \bibnamefont {{Henriksen}}}, \bibinfo {author}
  {\bibfnamefont {E.~A.}\ \bibnamefont {{Quintero}}}, \bibinfo {author}
  {\bibfnamefont {Charles~R}\ \bibnamefont {{Harris}}}, \bibinfo {author}
  {\bibfnamefont {Anne~M.}\ \bibnamefont {{Archibald}}}, \bibinfo {author}
  {\bibfnamefont {Ant{\^o}nio~H.}\ \bibnamefont {{Ribeiro}}}, \bibinfo {author}
  {\bibfnamefont {Fabian}\ \bibnamefont {{Pedregosa}}}, \bibinfo {author}
  {\bibfnamefont {Paul}\ \bibnamefont {{van Mulbregt}}}, \ and\ \bibinfo
  {author} {\bibfnamefont {SciPy 1.~0}\ \bibnamefont {{Contributors}}},\
  }\bibfield  {title} {\enquote {\bibinfo {title} {{SciPy 1.0: Fundamental
  Algorithms for Scientific Computing in Python}},}\ }\href {\doibase
  https://doi.org/10.1038/s41592-019-0686-2} {\bibfield  {journal} {\bibinfo
  {journal} {Nature Methods}\ }\textbf {\bibinfo {volume} {17}},\ \bibinfo
  {pages} {261--272} (\bibinfo {year} {2020})}\BibitemShut {NoStop}%
\end{thebibliography}%

\appendix

\section{Error analysis}\label{app:noise_modelling}
Let us formalize the ideas outlined in Sec.~\ref{sec:mitigation} by considering how the verified and unverified Hilbert spaces $\Hh^{(\mathrm{v})}$ and $\Hh^{(\mathrm{f})}$, and the verified and unverified ensembles~\footnote{We remind the reader here that $\rho^{(\mathrm{v})}$ and $\rho^{(\mathrm{f})}$ are not normalized, hence our use of the word 'ensemble' rather than 'state'.} $\rho^{(\mathrm{v})}$ and $\rho^{(\mathrm{f})}$ within them, evolve over the course of a noisy quantum circuit.
We will then attempt to provide some mechanisms for the observed scaling laws in Sec.~\ref{sec:results}.
At the end of the VPE circuit, the verified Hilbert space $\Hh^{(\mathrm{f})}$ is spanned by the two verified basis states.
In single-control VPE, these are $|0\rangle|0\rangle$ and $|1\rangle|0\rangle$, while in control-free VPE these are $|0\rangle$ and $|\vec{1}_{\mathrm{T}}\rangle$.
Let us label these $|0_{\mathrm{v}}\rangle$ and $|1_{\mathrm{v}}\rangle$ respectively, and then we may define the verified Hilbert space as
\begin{equation}
    \Hh^{\mathrm{(\mathrm{v})}}=\mathrm{Span}\{|0_{\mathrm{v}}\rangle,|1_{\mathrm{v}}\rangle\},
\end{equation}
and the verified ensemble as
\begin{equation}
    \rho^{(\mathrm{v})}(t)=P_{\mathrm{v}}\rho(t) P_{\mathrm{v}},\;\; P_{\mathrm{v}} = |0_{\mathrm{v}}\rangle\langle 0_{\mathrm{v}}| + |1_{\mathrm{v}}\rangle\langle 1_{\mathrm{v}}|.\label{eq:verified_projection}
\end{equation}

The system state $\rho$ here is the state at the end of the VPE circuit, let us now consider how the system evolves to get here.
This evolution is not a function of the simulated time $t$, as we may use entirely different circuits to estimate the phase function $g(t)$ and $g(t')$.
Instead, we must frame the evolution of the state on the quantum device over the course of the VPE circuit in terms of the device time $\tau$.
That is, let us fix $t$, and assume that the circuit that implements $U=e^{iHt}$ is split into a set of discrete moments $U(\tau)$ (with the last moment occuring at time $\tau_{\max}$),
\begin{equation}
    U=\prod_{\tau=0}^{\tau_{\max}}U(\tau),
\end{equation}
where each moment consists of a set of gates acting in parallel
\begin{equation}
    U(\tau)=\prod_iU(\tau)_i.
\end{equation}
This is how circuits are represented in the cirq quantum programming framework~\cite{cirq}, and is a good way of approximating the behaviour of a real quantum circuit.

To best understand how noise and verification work together, we must move to the interaction picture, or rather a rotating reference frame.
In the Schr\:{o}dinger picture, the system begins entirely within $\Hh^{(\mathrm{v})}$, as in all cases it is initialized in $|0_{\mathrm{v}}\rangle = |0\rangle$ and immediately rotated to $\frac{1}{\sqrt{2}}(|0_{\mathrm{v}}\rangle + |1_{\mathrm{v}}\rangle)$.
It then evolves out of $\Hh^{(\mathrm{v})}$ as we prepare, evolve, and un-prepare the system, even in the absence of error.
However, for us it is more helpful to consider the states that will be rotated into $\Hh^{(\mathrm{v})}$ at the end of the circuit.
This may be achieved by re-defining the verified basis states in the reference frame 
\begin{align}
    |0_{\mathrm{v}}\rangle \rightarrow \left(\prod_{\tau'>\tau}U(\tau')\right)^{-1}|0_{\mathrm{v}}\rangle\label{eq:rrf_0state}\\
    |1_{\mathrm{v}}\rangle \rightarrow \left(\prod_{\tau'>\tau}U(\tau')\right)^{-1}|1_{\mathrm{v}}\rangle.\label{eq:rrf_1state}
\end{align}
(This a slightly non-standard choice of reference frame, as we are shifting backwards in time from the final state, rather than forwards in time from the initial state, but it makes our error analysis far easier.)
In the absence of error, this is the Heisenberg picture: our system remains in the state
\begin{align}
    \rho &= |\rho\rangle\langle\rho|\\
    |\rho\rangle &= \frac{1}{\sqrt{2}}|0_{\mathrm{v}}\rangle + \frac{g(t)}{\sqrt{2}}|1_{\mathrm{v}}\rangle + \sqrt{\frac{1-|g(t)|^2}{2}}|\rho^{(\mathrm{f})}\rangle,
\end{align}
throughout the entire circuit.
Here $|\rho^{(\mathrm{f})}\rangle$ is the fraction of the state that will eventually fail verification --- 
\begin{equation}
\rho^{(\mathrm{f})}=|\rho^{(\mathrm{f})}\rangle\langle\rho^{(\mathrm{f})}|.
\end{equation}
We may project our system at any device time $\tau$ into the verified Hilbert space via Eq.~\ref{eq:verified_projection}, but with the basis states in their rotating reference frame (Eq.~\ref{eq:rrf_0state} and Eq.~\ref{eq:rrf_1state}).

Noise may be added to the above by treating it as a perturbation and switching to the interaction picture.
Without loss of generality, we may say the effect of noise is to shift the unitary $U(\tau)$ at each moment
\begin{equation}
    U(\tau) \rightarrow R_{U(\tau)}(1-p_{\tau} + p_{\tau}E_{\tau}),
\end{equation}
where $R_{U}$ is the process map associated with a unitary $U$
\begin{equation}
    R_{U}[\rho] = U^{\dag}\rho U,
\end{equation}
$E_{\tau}$ is the process map associated with errors during the moment $\tau$, and $p_{\tau}$ is the probability of any such errors occurring.
In the interaction picture, the action of the circuit has been shifted into our basis states, and $U(\tau) = 1$.
So, we may write our final state in the presence of error as
\begin{align}
    \rho^{(\mathrm{err})} =& \frac{1}{\Nn}\left\{\rho + \sum_{\tau_0 < \tau}p'_{\tau_0}E_{\tau_0}[\rho]\right.\nonumber\\ &\left.+ \sum_{\tau_1 < \tau_0 < \tau}p'_{\tau_0}p'_{\tau_1}E_{\tau_0}\left[E_{\tau_1}[\rho]\right] + \ldots\right\},
\end{align}
where $\Nn$ is a normalization coefficient
\begin{equation}
    \Nn=\prod_{\tau}\frac{1}{1-p_{\tau}},
\end{equation}
and $p'_{\tau}$ are the rescaled probabilities
\begin{equation}
    p'_{\tau}=\frac{p_{\tau}}{1-p_{\tau}}.
\end{equation}
If desired, one can recognise this also as a discrete form of the well-known time-ordered integrals in quantum mechanics --- a time-ordered sum
\begin{equation}
    \rho^{(\mathrm{err})} =\frac{1}{\Nn} \Tt \mathrm{exp}\left(\sum_{\tau} p'_{\tau}E_{\tau}\right)[\rho],
\end{equation}
where $\Tt$ is the time-ordering operator.
Our projection onto the verified subspace is linear, so we may consider it on each of the individual terms in the sum.
Assuming $p'(\tau)$ is small for all $\tau$, the first-order corrections to $\rho^{(v)}$ occur from errors $E_{\tau}$ during a single timestep.
These corrections take the form
\begin{equation}
    P_{\mathrm{v}}p'_{\tau}E_{\tau}[\rho]P_{\mathrm{v}}= p'_{\tau}\left(\begin{array}{cc}p_{0,\tau} & \frac{1}{2}g^{(\mathrm{err})}_{\tau}(t) \\ \frac{1}{2}g^{(\mathrm{err})\dag}_{\tau}(t) & p_{1,\tau}\end{array}\right),\label{eq:error_dm}
\end{equation}
where
\begin{align}
    p_{0,\tau}&=\langle 0_{\mathrm{v}}|E_{\tau}[\rho]|0_{\mathrm{v}}\rangle\\
    p_{1,\tau}&=\langle 1_{\mathrm{v}}|E_{\tau}[\rho]|1_{\mathrm{v}}\rangle\\
    g^{(\mathrm{err})}_{\tau}(t)&=\langle 1_{\mathrm{v}}|E_{\tau}[\rho]|0_{\mathrm{v}}\rangle\label{eq:gk_error_first_order}
\end{align}
The off-diagonal element in this matrix gives the contribution to the phase function $g(t)$
\begin{equation}
    g(t)\rightarrow \frac{1}{\Nn} g(t) + \frac{1}{\Nn}\sum_{\tau}p'_{\tau}g_{\tau}^{(\mathrm{err})}(t).
\end{equation}
One may generalize this to higher-order terms.
For example, the second-order contribution to the error takes the form
\begin{equation}
    \frac{1}{\Nn}\sum_{\tau_0 < \tau_1}p'_{\tau_0}p'_{\tau_1}\langle 1_{\mathrm{v}}|E_{\tau_1}[E_{\tau_0}[\rho]]|0_{\mathrm{v}}\rangle\label{eq:gk_error_second_order}
\end{equation}

The mitigation power from verification requires two conditions: that the dependence of the normalization $\Nn$ on the simulated time $t$ is simple, and that the off-diagonal error contributions (Eq.~\ref{eq:gk_error_first_order}) are small.
We expect both conditions to often be the case.
The positivity of $E_{\tau}[\rho]$ implies that
\begin{equation}
    g^{(\mathrm{err})}_{\tau} + p_{0,\tau} + \sum_{n=1}^{2^N-2}\langle\rho^{(\mathrm{f})}_n|E_{\tau}[\rho]|0_{\mathrm{v}}\rangle \leq 1,\label{eq:exp_sum}
\end{equation}
where $|\rho^{(\mathrm{f})}\rangle$ is an appropriately chosen basis for $\Hh^{(\mathrm{f})}$.
On average all terms are equally-weighted, so $g^{(\mathrm{err})}_{\tau}\sim 2^{-N}$.
As such, negligible $g^{(\mathrm{err})}_{\tau}$ should be the norm rather than the exception; we need reason to expect that the error channel $E_{\tau}$ will not scatter us out of the verified subspace.
If $g^{(\mathrm{err})}_{\tau}=0$, the effect of $E_{\tau}$ on $g(t)$ occurs via the damping by $\Nn$, which itself may depend on $t$.
However, $\Nn$ depends only on the rate at which errors occur, and is oblivious to their specific form.
If a Hamiltonian is fast-forwardable, $e^{iHt}$ may be implemented in time constant in $t$.
Then assuming a constant error rate per moment of the circuit, our phase function is dampened by a constant amount,
\begin{equation}
    g^{\mathrm{err}}(t)= \frac{1}{\Nn}g(t),\label{eq:g_const}
\end{equation}
which may be corrected for by renormalization (Eq.\ref{eq:renormalization}).
If a Hamiltonian is not fast-fowardable, $e^{iHt}$ must take real time $\tau_{\max}(t)={\cal O}(t)$ to simulate to constant error.
Assuming this is the case, and that we have a constant error rate per moment of the circuit, the damping from each possible error $E_{\tau}$ is multiplicative, and our estimation takes the form
\begin{equation}
    g^{\mathrm{err}}(t)= e^{-t/T_1}g(t).\label{eq:g_decay}
\end{equation}
Here, $T_1$ is defined as the (simulated) time $t$ over which enough errors $E$ have accumulated that
\begin{equation}
    \frac{1}{\Nn(\tau)}(E)=e^{-1}.
\end{equation}
This constant damping may be considered an imaginary shift to the eigenvalues $E_j$; $E_j\rightarrow E_j+\frac{1}{T_1}$.
It may be removed by classical postprocessing techniques~\cite{Kimmel15Robust,Obrien19Quantum,Wiebe16Efficient}.
However, the shrinking of the signal increases the sampling requirements to estimate $g(t)$ exponentially in $t$.

Although random error channels are exponentially suppressed by verification (following Eq.~\ref{eq:exp_sum}), realistic error models are biased, and may apply undesired phases to $g^{\mathrm{err}}_{\tau}(t)$ instead of setting it to $0$.
The density matrix in Eq.~\ref{eq:error_dm} is not normalized, but it must be positive, which implies
\begin{align}
    |g^{(\mathrm{err})}_{\tau}|^2 &< \frac{p_{0,\tau}p_{1,\tau}}{p_{0,\tau}+p_{1,\tau}}.
\end{align}
This means that errors must either fail to scatter \emph{both} $|0_{\mathrm{v}}\rangle$ and $|1_{\mathrm{v}}\rangle$, or rotate between these states and the failed state $|\rho^{(f)}\rangle$.
When control-free methods are used, $|0_{\mathrm{v}}\rangle$ is separated from $|1_{\mathrm{v}}\rangle$ and $|\rho^{(f)}\rangle$ by highly non-local excitations, which are non-physical error channels.
However, when single-control methods are used, $|0_{\mathrm{v}}\rangle$ is coupled to $|1_{\mathrm{v}}\rangle$ and $|\rho^{(f)}\rangle$ by control qubit errors.
These control qubit errors deform the Bloch sphere defined by $|0_{\mathrm{v}}\rangle = |0\rangle |0\rangle$ and
\begin{equation}
    g(t)|1_{\mathrm{v}}\rangle + \sqrt{1-|g(t)|^2}|\rho^{(f)}\rangle = |0\rangle|\vec{1}\rangle.
\end{equation}
When this deformation is asymmetric around the z-axis, or a rotation, $g(t)$ may be quickly corrupted\footnote{We have observed this for instance due to T$_1$ decay on the control qubit between final rotation and readout. However, we can correct for this easily by measuring in the opposite basis $50\%$ of the time}.
However, symmetric noise (such as a depolarizing channel, or T$_1$ or T$_2$ channels during the bulk of the circuit) can be seen to simply dampen $g(t)$ in an identical manner to $\Nn$.
That is, the dampening will depend only on the rate at which these errors occur.
Such dampening will be cancelled by renormalization, as observed in Fig.~\ref{fig:TFIM_split_noise}.

Errors that do not rotate between $|0_{\mathrm{v}}\rangle$ and $|1_{\mathrm{v}}\rangle$, but still contribute non-trivially to $g^{(\mathrm{err})}_{\tau}(t)$ to first order must have both $|0_{\mathrm{v}}\rangle$ and $|1_{\mathrm{v}}\rangle$ as approximate eigenstates of the error channel.
This suggests a reason why control-free VPE is more noise-robust to noise than single-control VPE: the starting and reference states are very different when looked at locally, which makes it less likely that a single local error will have both states as near-eigenstates.
It also suggests a reason why we might expect the suppression of errors to only second-order: if the same error occurs in subsequent moments (in a local frame), and the basis states $|0_{\mathrm{v}}\rangle = |0_{\mathrm{v}}(\tau)\rangle$ have not evolved significantly between these moments, the second error will almost (but not completely) cancel out the first, driving the system back into the verified subspace in an uncorrectable manner.
This implies that a circuit which more quickly scrambles the basis states $|0_{\mathrm{v}}\rangle$ and $|1_{\mathrm{v}}\rangle$ between moments should be less susceptible to error than one where the states evolve slowly.
Understanding the dynamics of these noisy circuits in more detail is a clear target for future work.

\section{Effect of parallelizing QPE}\label{app:parallel}
In this appendix we investigate the phase function obtained during the parallel estimation of multiple commuting Hamiltonians, and demonstrate that the resulting expectation values from this estimation are not affected by the parallelization and verification process combined.
Let us consider the case where we have two commuting Hamiltonians $H_0$, $H_1$.
In this case, we may write a simultaneous eigenbasis $|E_j\rangle$ for both Hamiltonians --- $H_b|E_j\rangle=E_j^{(b)}|E_j\rangle$.
Let $|\psis\rangle=\sum_ja_j|E_j\rangle$, and we may calculate the controlled-time-evolved global state $|\Psi(t)\rangle$ to be
\begin{equation}
    \sum_ja_j(|0\rangle+e^{iE^{(0)}_jt}|1\rangle)(|0\rangle +e^{iE^{(1)}_jt}|1\rangle)|E_j\rangle.
\end{equation}
Tracing out control qubit $1$ obtains the following reduced density matrix for the system + control qubit $0$
\begin{align}
    &\sum_{j,j'}a_ja_{j'}^*\left[1+e^{i\left(E_j^{(1)}-E_{j'}^{(1)}\right)}\right]\nonumber\\
    &\times\left[\left(|0\rangle+e^{iE_j^{(0)}t}|1\rangle\right)|E_j\rangle\langle E_{j'}|\left(\langle 0|+e^{-iE_{j'}^{(0)}t}\langle 1\right)|\right].
\end{align}
The issue here then comes from this additional factor $\left[1+e^{i\left(E_j^{(1)}-E_{j'}^{(1)}\right)}\right]$ at the front.
Note that (as we should expect) this goes away upon tracing out the system register, as the trace over $|E_j\rangle\langle E_j'|$ yields (dropping all additional terms in the above expression).
\begin{equation}
    \sum_{l}\langle E_l|E_j\rangle\langle E_{j'}|E_l\rangle = \delta_{j,j'}
\end{equation}
However, post-selection implies that we take the expectation value with regards to $|\psis\rangle$, obtaining
\begin{equation}
    \langle\psis|E_j\rangle\langle E_{j'}|\psis\rangle = a_{j}^*a_{j'}.
\end{equation}
The off-diagonal element of the control qubit density matrix can then be found to be
\begin{align}
    &\frac{1}{2}\sum_{j,j'}|a_j|^2|a_{j'}|^2\left(1+e^{i(E_j^{(1)}-E_{j'}^{(1)})t}\right)e^{iE_j^{(0)}t}\\
    &=\frac{1}{2}\sum_j|a_j|^2e^{iE_j^{(0)}t} \nonumber\\&+ \frac{1}{2}\sum_j|a_j|^2e^{i(E_j^{(1)}+E_j^{(0)})t}\sum_{j'}|a_{j'}|^2e^{-iE_{j'}^{(1)}t}\label{eq:two_control_parallel}.
\end{align}
One can see that this is a linear combination of products of the phase functions of $H_0$, $H_1$, and $H_0+H_1$.
In theory the eigenvalues $E_j^{(0)}$ and amplitudes squared $|a_j|^2$ are still present in this function, and could be extracted via classical postprocessing.
However, the $\frac{1}{2}$ coefficient implies we need $4$ times as many single-shot experiments for the estimation of $|a_j|^2$ to the same error (compared to a standard QPE experiment for $H_0$).
Extending this to $L>2$ summands, the off-diagonal for the $s$th control qubit can be written:
\begin{equation}
    \frac{1}{2^{L}}\sum_{j,j'}|a_j|^2|a_{j'}|^2e^{iE_j^{(s)}t}\prod_{s'\neq s}\left(1+e^{i(E_j^{(s')}-E_{j'}^{(s')})t}\right),\label{eq:manyphase}
\end{equation}
and we see that the signal corresponding to `just' $g(t)$ is exponentially small.
However, all is not lost.
Inspecting the form of Eq.~\ref{eq:manyphase}, we see that we may expand this as a sum of $2^{L}J^2$ separate (possibly degenerate) spurious energies $F^{(s)}_{v,j,j'}$, indexed by a $L$-bit binary integer $v$ and the original $j$ and $j'$ indices
\begin{equation}
    F^{(s)}_{v,j,j'} = E_j^{(s)} + \sum_{s'\neq s}v_{s'}(E_j^{(s')}-E_{j'}^{(s')}),
\end{equation}
with corresponding ($v$-independent) spurious amplitudes
\begin{equation}
    B_{j,j'}=\frac{1}{2^{L}}|a_j|^2|a_{j'}|^2.
\end{equation}
(Note that as stated these energies are automatically at least doubly-degenerate as $v_s$ does not appear in the equation for $F^{(s)}_{v,j,j'}$.)
If we then calculate the weighted average of the $F^{(s)}_{v,j,j'}$ (which is what we would do if we processed the signal as if the parallelization had not occurred), we find
\begin{align}
    \sum_{v,j,j'}B_{j,j'}F^{(s)}_{v,j,j'}=&\frac{1}{2^{L}}\sum_{v,j,j'}|a_j|^2|a_{j'}|^2E_{j}^{(s)}\nonumber\\&+\frac{1}{2^{L}}\sum_{v,j,j'}B_{j,j'}\sum_{s'\neq s}v_{s'}E_j^{(s')}\nonumber\\&-\frac{1}{2^{L}}\sum_{v,j,j'}B_{j,j'}\sum_{s'\neq s}v_{s'}E_{j'}^{(s')}.
\end{align}
As $j$ and $j'$ are just dummy indices, and as $B_{j,j'}=B_{j',j}$, the last two terms cancel, and as $\sum_v=2^{L}$ and $\sum_{j'}|a_{j'}|^2=1$, we have
\begin{equation}
    \sum_{v,j,j'}B_{j,j'}F^{(s)}_{v,j,j'}=\sum_{j}|a_j|^2E_{j}^{(s)}=\langle H_s\rangle.
\end{equation}
This implies that expectation values may be extracted via parallel verified phase estimation, even though the signal itself may be significantly more complex.
For the case of Pauli $H_s$ operators, the spectrum $F_{v,j,j'}^{(s)}$ is highly degenerate --- it is the set of odd integers $\{-2L+1,-2L+3,\ldots,2L-3,2L-1\}$.
(This parallels the spectrum of a spin-$\frac{2L-1}{2}$ operator, which one might not expect following Hund's rules for the combination of spin-$\frac{1}{2}$ systems, which is curious.)
This must be taken into account when signal processing by amplitude-fitting, as one would otherwise miss components of the energy.
However, the overhead for this is only linear in the number of simultaneously-estimated terms.

\section{Compensation for spurious eigenvalues due to sampling noise}\label{app:prony_compensation}

\begin{figure}
    \centering
    \includegraphics[width=\columnwidth]{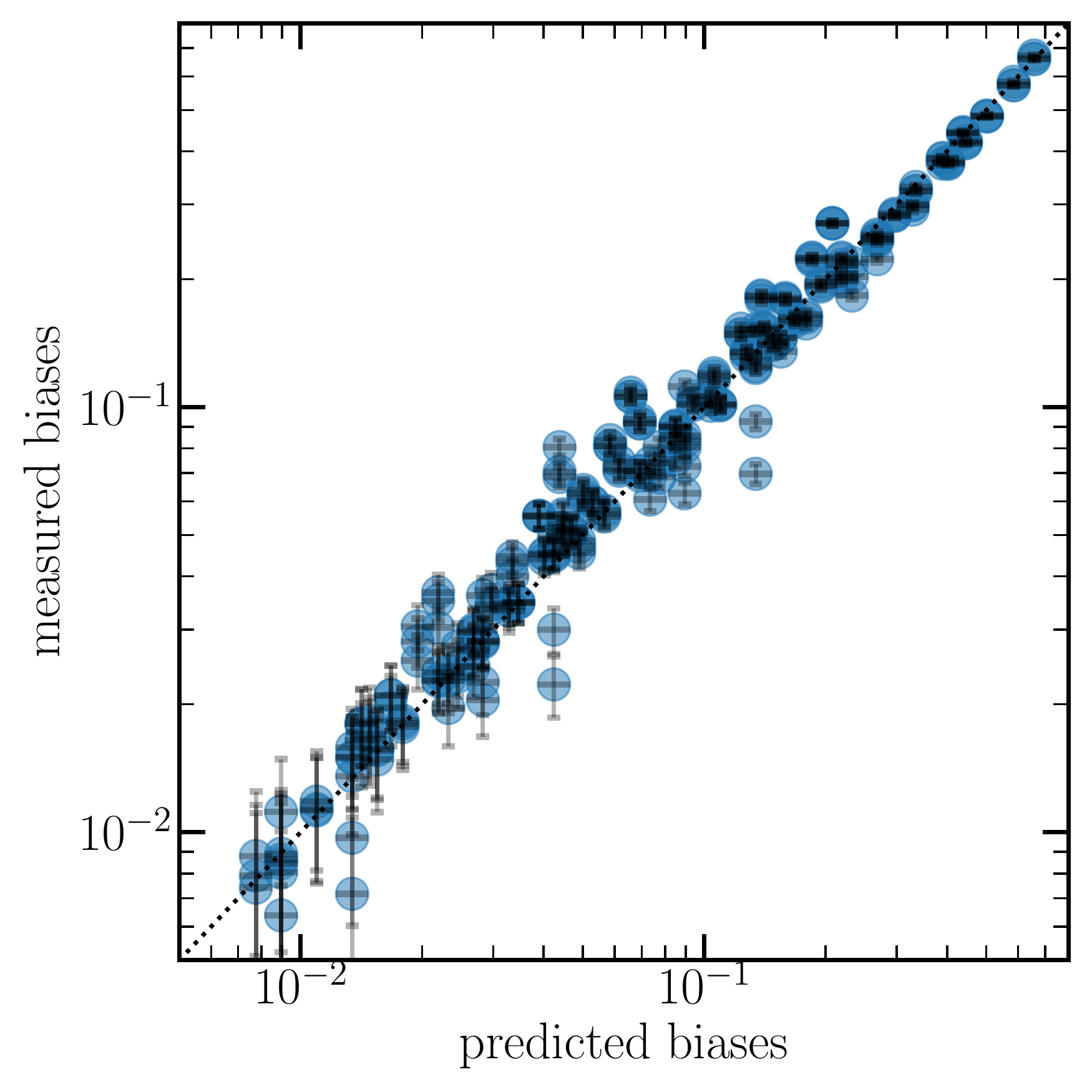}
    \caption{Predicted (Eq.~\ref{eq:bias_eq}) vs found bias from estimating expectation values using Prony's method.}
    \label{fig:bias_figure}
\end{figure}

When quantum phase estimation is used to estimate eigenvalues as well as amplitudes to sum together to give an expectation value (Eq.~\ref{eq:expval_from_eval}), finite sampling noise introduces a small bias to this estimation that may be cancelled.
This bias does not come from the QPE itself.
The sampling noise has a white spectrum which is invariant under a Fourier transform, so classical post-processing of a noisy spectrum yields a set of spurious eigenvalue/amplitude pairs evenly distributed around the circle.
However, in order to evaluate Eq.~\ref{eq:expval_from_eval}, we have to make a branch cut in this circle.
The resulting terms then average to bias the signal by a term $\Delta_{\mathrm{bias}}=\langle H\rangle - \overline{\langle H\rangle}$ towards the center of the resulting region.
(Here, $\langle H\rangle$ is the true expectation value, and $\overline{\langle H\rangle}$ that measured naively.)
For example, if we assume all eigenvalues $E_j\in [-\pi,\pi]$, this biases the signal towards zero.
This bias is dependent on both the number of steps $K$, and the number of samples $M$ used in the QPE process.
Numerically, we find (Fig.~\ref{fig:bias_figure}):
\begin{equation}
    \Delta_{\mathrm{bias}} = -\langle H\rangle \times(K-2)^{\frac{1}{2}}M^{-\frac{1}{2}}.\label{eq:bias_eq}
\end{equation}
Inverting this obtains
\begin{equation}
    \langle H\rangle = \overline{\langle H\rangle}\left[1+(K-2)^{\frac{1}{2}}M^{-\frac{1}{2}}\right]^{-1},
\end{equation}
which is used in the estimation in Sec.~\ref{sec:sampling_noise}.

\section{Demonstration of immunity to control noise in single-control VPE}

\begin{figure}[h!]
    \centering
    \includegraphics[width=\columnwidth]{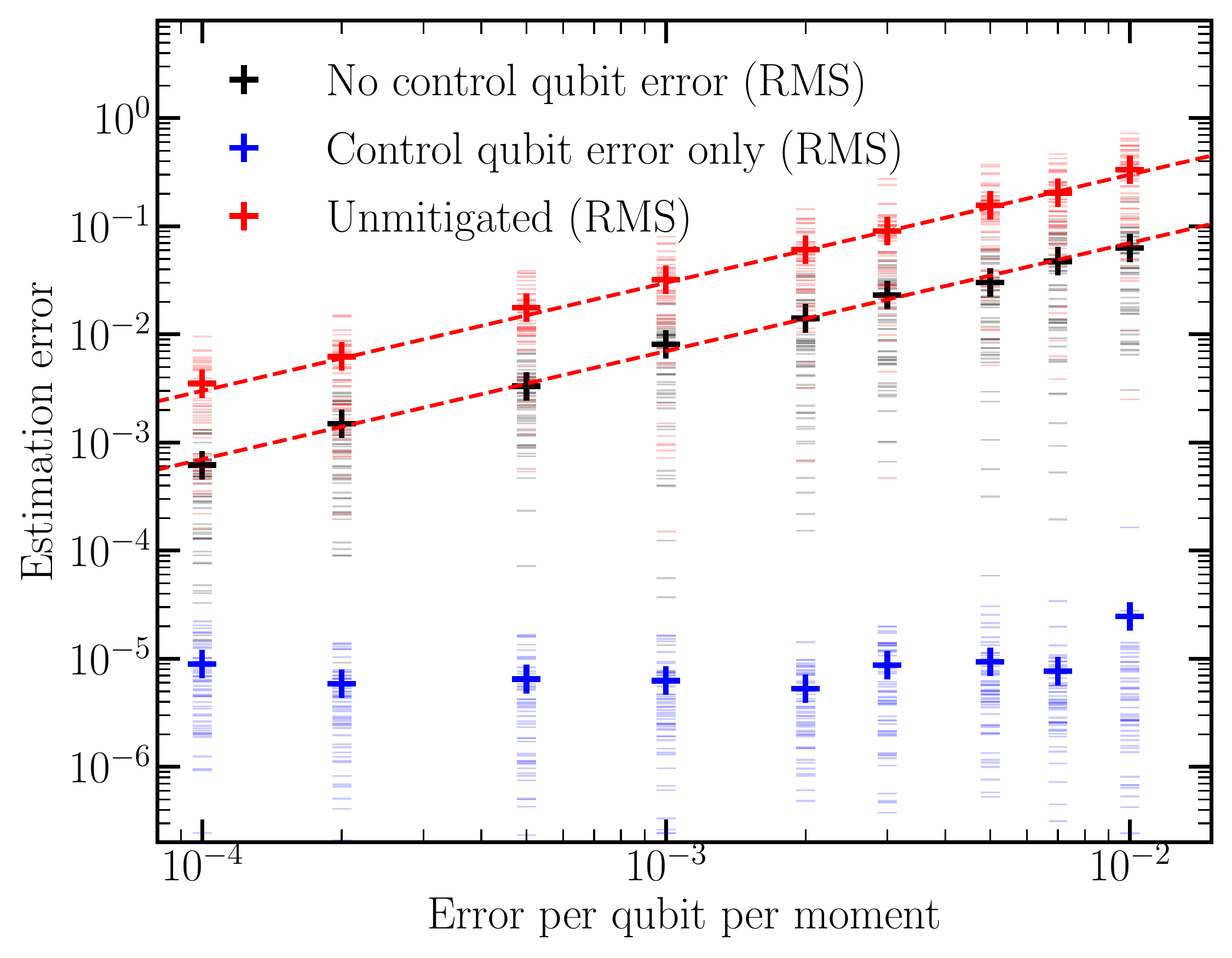}
    \caption{Mitigation of the same 4-qubit VHA circuit as in Fig.~\ref{fig:TFIM_expt_est}, but with either depolarizing noise on only the system register (black) or only on the control qubit (blue). This is compared with the error in estimation using partial state tomography instead of VPE (red). For each dataset, the RMS error (crosses) is plotted over $50$ different estimations for each error rate (with randomly-chosen ansatz parameters), and individual data points are plotted as dashes behind. For reference, dashed lines showing linear (red) dependence on the gate error rate are plotted.}
    \label{fig:TFIM_split_noise}
\end{figure}

One might expect that the discrepancy between the scaling of the error mitigation power of the control-free and single-control circuits seen throughout this work comes from accumulation of errors on the control qubit alone.
In this appendix, we show that this is not the case.
In Fig.~\ref{fig:TFIM_split_noise}, we see that removing all errors on the control qubit does little to reduce the total error rate (black crosses), whilst a model with noise only on the control qubit achieves an error limited by our use of Prony's method for post-processing.
In App.~\ref{app:noise_modelling} we argue that the increased error suppression from control-free VPE comes from the large separation between reference and starting states.
Errors will be removed by verification unless they maintain coherence between these states, which these error models fail to do.

\section{Use of a variational outer loop to mitigate constant unitary noise}
One of the main uses of expectation values $\langle H\rangle$ in quantum experiments is to use them as a cost function in a variational outer loop.
Optimizing the parameters $\vec{\theta}$ in a preparation unitary $U_{\mathrm{p}}(\vec{\theta})$ to minimize the expectation value of the prepared state $|\Psi(\vec{\theta})\rangle=U_{\mathrm{p}}(\vec{\theta})|0\rangle$ then gives an approximation for the true ground state of $H$.
The variational optimization process is itself known to be robust against certain types of error~\cite{Mcclean16Theory, Omalley16Scalable}, in particular control errors.
These occur when a signal meant to implement a gate $G(\theta)$ either drifts or is distorted and instead implements $G(\theta')$.
As this error is often repeated throughout an experiment, i.e. every instance of $G$ is miscalibrated by a similar amount, it will be repeated throughout the experiment.
Verification can only correct single errors, and as such is not targeted for this type of noise.
By contrast, the dominant source of errors in a VQE are often the incoherent errors that verification is designed to target.
As such, verification and variational optimization provide cumulative mitigation by targeting sources of error the other lets through.

\begin{figure}[h!]
    \centering
    \includegraphics[width=\columnwidth]{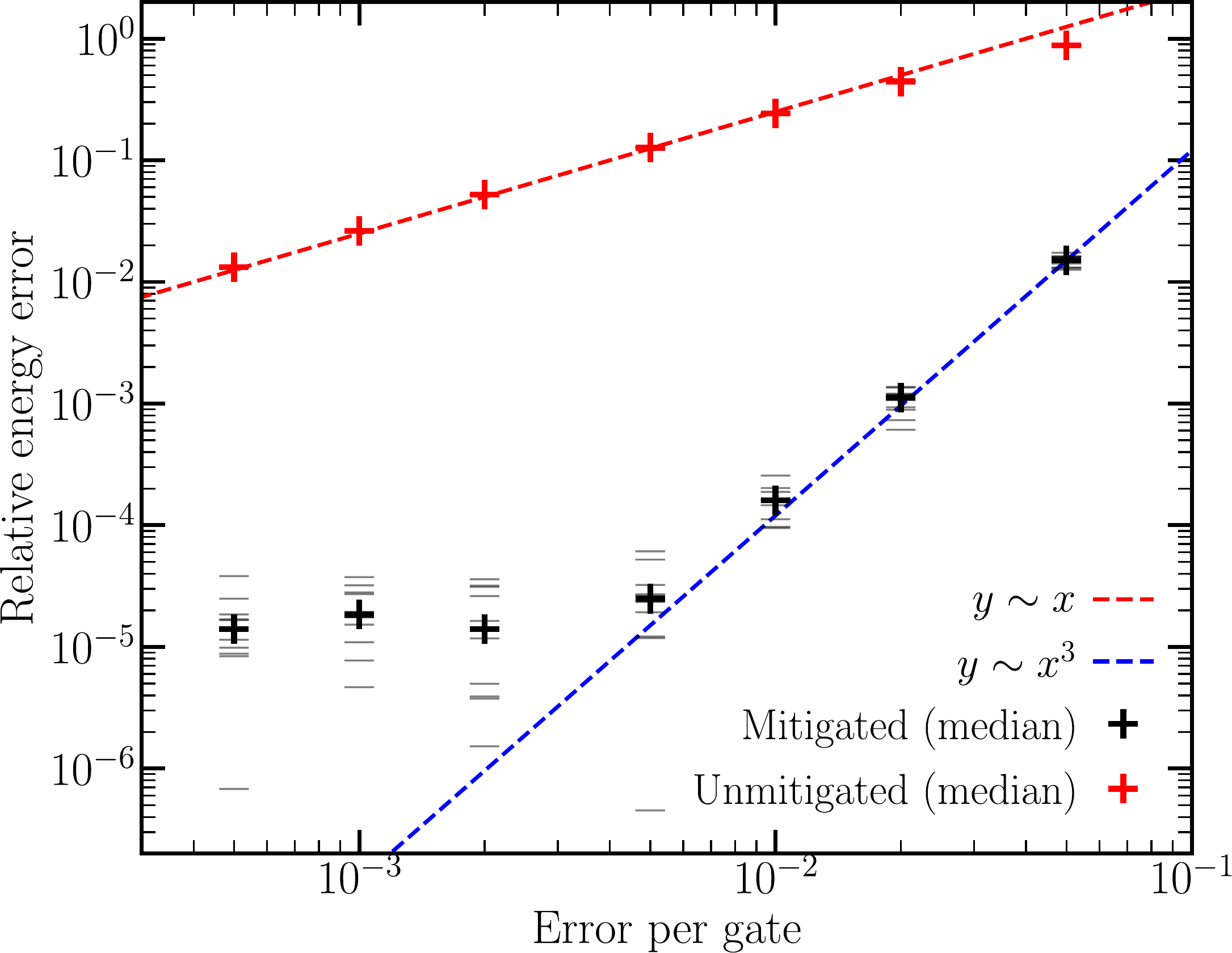}
    \caption{Error in estimating the ground state energy of a free-fermion system (Eq.~\ref{eq:free_fermion_chain}) of 4 fermions (on four qubits), using control-free verified phase estimation and a VQE. Noise model is a mixture of amplitude and phase damping and constant two-qubit control error (details in text). Median absolute errors for both verified estimation (black crosses) and standard partial state tomography (red crosses) are calculated over $10$ different optimization attempts. Individual simulations are plotted behind (faint dashes) Each optimization started from a different parameter set and had different control rates set. Linear (red dashed) and cubic (blue dashed) lines are shown as guides.}
    \label{fig:Givens_VQE}
\end{figure}

To demonstrate the combined mitigation effects, we use verified control-free phase estimation of a Givens rotation circuit in the inner loop of a variational quantum eigensolver.
In order to prevent oversimplifying the problem, we add a next-nearest-neighbour coupling and on-site potential to the Hamiltonian in Eq.~\ref{eq:free_fermion_chain}, yielding
\begin{align}
    H &= H_1 + H_2\\
    H_1 &= -t_1\sum_{j=1}^Nc^{\dag}_jc_{j+1} + \mathrm{h.c.}\\
    H_2 &= -t_2\left(\sum_{j=1}^Nc^{\dag}_jc_{j+1} + \mathrm{h.c.} + \sum_{j=1}^Nc^{\dag}_jc_j\right),
\end{align}
and estimate expectation values for $H_1$ and $H_2$ separately.
Here, we again take periodic boundary conditions for a $N=4$-site system (i.e. all sums in indices are taken modulo $4$), and fix $t_1=1$, $t_2=0.5$.
This ensures that the ground state of the system is neither a ground state of $H_1$ or $H_2$ (in which case the compiled variational ansatz and basis rotation would cancel to become an identity circuit).
For a simple model combining control error and incoherent noise, we fix $p$, draw a random offset $x_i\in [-\frac{p}{\pi},\frac{p}{\pi}]$ for each two-qubit ISWAP gate, and decompose the variational circuit into $\mathrm{ISWAP}^{1/2}$ gates.
(Though not terribly well-known, the $\mathrm{ISWAP}^{1/2}$ gate is a good native hardware gate for superconducting qubits, and decompositions of other gates into $\mathrm{ISWAP}^{1/2}$ gates are known~\cite{Google19Quantum}.)
Then, throughout the circuit, we implement $\mathrm{ISWAP}^{1/2+x_i}$ gates in place of $\mathrm{ISWAP}^{1/2}$ gate.
We additionally add amplitude and phase damping noise at a rate $\frac{p}{2}$.
In Fig.~\ref{fig:Givens_VQE}, we plot the result following optimization via the COBYLA algorithm implemented in scipy~\cite{scipy}, in the absence of sampling noise.
We see that the verification circuit is insensitive to the incoherent noise as expected, and behaves similarly to the effect of amplitude and phase damping alone (Fig.~\ref{fig:Givens_expt_est}, right).

\section{Term-wise comparison of VPE performance}\label{app:single_term_decomposition}

\begin{figure}[h!]
    \centering
    \includegraphics[width=\columnwidth]{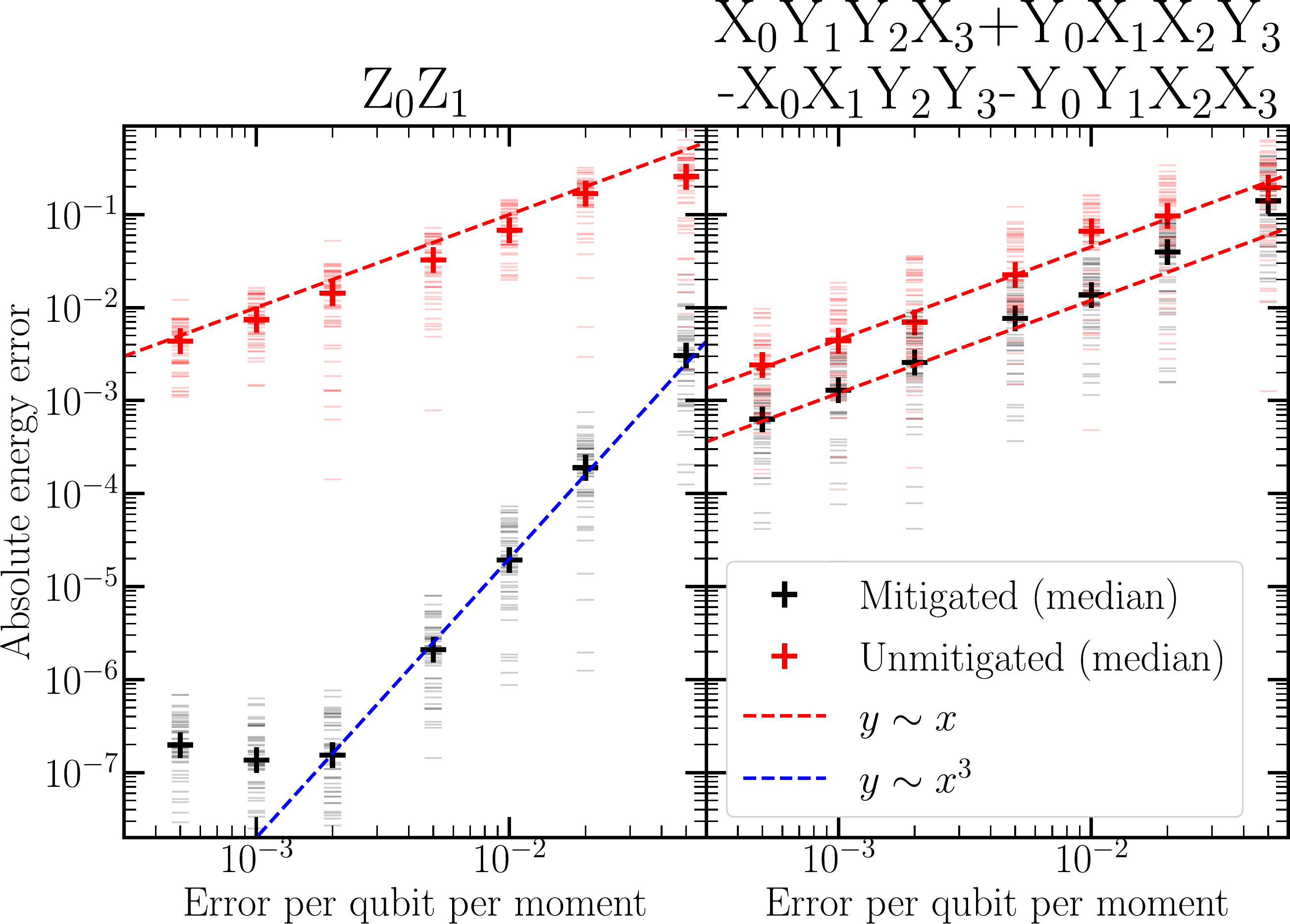}
    \caption{Expectation value estimation of two individual $H_s$ terms from the control-free number-conserving Pauli operator decomposition of the H$_2$ Hamiltonian studied in Fig.~\ref{sec:fsw_results} on states prepared by a fermionic swap network. The two terms here comprise part of the sum (Eq.~\ref{eq:expectation_value_sum}) for the expectation value of Fig.~\ref{fig:fsw_expt_est}(d) --- but are studied here without pre-factors (i.e. $\|H_s\|=1$). Each figure is labeled with the studied term, and guide-lines (dashed red and blue) are given to show observed scaling laws. Data presented is the median (crosses) over $50$ individual data points (faint dashes) of the absolute error in estimation using VPE (black) and standard partial state tomography (red).}
    \label{fig:single_term_fsw}
\end{figure}

To attempt to further understand the ability of VPE to mitigate errors, in this appendix we consider the effect of estimating different types of terms on the same preparation circuit.
We consider the fermionic swap network used in Sec.~\ref{sec:fsw_results} to prepare states for a H$_2$ Hamiltonian.
When this was split into number-conserving Pauli operator sums (Fig.~\ref{fig:fsw_expt_est}(c-f)), different circuits had to be used to estimate individual terms.
In Fig.~\ref{fig:single_term_fsw}, we show the result of estimating the expectation values of two of the individual terms used in the control-free Pauli operator decomposition under an amplitude-damping noise model (Fig.~\ref{fig:fsw_expt_est}(d)).
(Recall that this figure demonstrated first-order sensitivity to this error model, whilst the low-rank factorization demonstrated a third-order sensitivity to the same model.)
We see that the $H_s=Z_0Z_1$ term (left plot) shows the cubic dependence on error rate observed in previous amplitude-damping experiments, whilst the two-body scattering term (right plot)
\begin{equation}
    H_s = X_0Y_1Y_2X_3 + Y_0X_1X_2Y_3 - X_0X_1Y_2Y_3 - Y_0Y_1X_2X_3,
\end{equation}
does not.
This two-body scattering term is the only term contributing to the first-order decay of the VPE estimation observed in Fig.~\ref{fig:fsw_expt_est}(d) --- all other terms in the decomposition display similar decay to Fig.~\ref{fig:single_term_fsw}(left).
This indicates that the errors to which we are first-order sensitive occur during the circuit implementation of $e^{iH_st}$, and not the state preparation.
The circuit implementing $e^{iH_st}$ for the two-body scattering term is the only such circuit that does not conserve number throughout.
(Instead, this evolution is achieved in two steps: a basis transform of $XY,YX\rightarrow IZ, ZI$ on pairs of qubits, $ZZ$ rotations between the pairs and uncomputing, and then a basis transform of $XX,YY\rightarrow IZ,ZI$ on pairs of qubits, $ZZ$ rotations between the pairs, and uncomputing again.)
Finding decompositions of these circuits more amenable to VPE is a clear target for future work.

\end{document}